%% file: CEMW.tex
\documentclass[a4paper,11pt]{article}
\usepackage{jheppub,bm} % for details on the use of the package, please
\allowdisplaybreaks[4]

\usepackage{slashed}

\begin{document}

\title{The NNLO QCD soft function for $1$-jettiness} 

%% %simple case: 2 authors, same institution
\author[a]{John M. Campbell,}
\author[b]{R. Keith Ellis,}
\author[c]{Roberto Mondini,}
\author[c]{and Ciaran Williams}

\affiliation[a]{Fermilab, P.O.~Box 500, Batavia, IL 60510, USA}
\affiliation[b]{Institute for Particle Physics Phenomenology, Department of Physics, \\ University of Durham, Durham, DH1 3LE, UK}
\affiliation[c]{Department of Physics, University at Buffalo\\ The State University of New York, Buffalo, NY 14260, USA}

% e-mail addresses: one for each author, in the same order as the authors
\emailAdd{johnmc@fnal.gov}
\emailAdd{keith.ellis@durham.ac.uk}
\emailAdd{rmondini@buffalo.edu}
\emailAdd{ciaranwi@buffalo.edu}

\input macro.tex

\abstract{ 
We calculate the soft function for the global event variable $1$-jettiness
at next-to-next-to-leading order (NNLO) in QCD. We focus
specifically on the non-Abelian contribution, which, unlike the Abelian part, is
not determined by the next-to-leading order result.  The calculation uses the known
general forms for the emission of one and two soft partons and is performed using a
sector-decomposition method that is spelled out in detail.  Results are presented in
the form of numerical fits to the 1-jettiness soft function for LHC kinematics
(as a function of the angle between the incoming beams and the final-state jet)
and for generic kinematics (as a function of three independent angles). These fits 
represent one of the needed ingredients
for NNLO calculations that use the $N$-jettiness event variable to handle infrared
singularities.}

\preprint{
\noindent IPPP/17/76 \\
\hspace*{\fill} FERMILAB-PUB-17-493-T}
\maketitle
\flushbottom

%%%%%%%%%%%%%%%%%%%%%%%%%%%%%%%%%%%%%%%%%%%%%%%%%%%%%%%%%%%%%%%%%%

\section{Introduction}
\label{sec:intro}
\input{intro}

\section{Setup of the calculation}
\label{sec:calcsetup}
\input{calcsetup}

\section{Soft function and soft radiation}
\label{sec:calcsoft}
\input{calcsoft}

\section{Soft function at NLO}
\label{sec:calcnlo}
\input{calcnlo}

\section{Soft function at NNLO}
\label{sec:calcnnlo}
\input{calcnnlo}
\subsection{Double-real corrections}
\label{sec:calcrr}
\input{calcrr}
\section{Results}
\label{sec:results}
\input{results}

\section{Conclusions}
\label{sec:conclu}
\input{conclu}

\section*{Acknowledgments}
\label{acknow}
\input{acknowledgments}
\appendix
\input{Notation}

\input{Rotation}
\input{MEs}

\input{Integrals}

\input{Results0j}

\color{black}
\bibliographystyle{JHEP}
\bibliography{CEMW}

%%%%%%%%%%%%%%%%%%%%%%%%%%%%%%%%%%%%%%%%%%%%%%%%%%%%%%%%%%%%%%%%%%

\end{document}

%% file: macro.tex
\newcommand{\slsh}{\rlap{$\;\!\!\not$}}     % Feynman slash
\newcommand{\nn}{\nonumber}
\newcommand{\abs}[1]{\lvert#1\rvert}
\def\deltap{\delta^{+}}
\def\tS{\tilde{S}}
\def\rG{\mathrm{\Gamma}}
\def\bI{\boldsymbol{I}}
\def\bT{\boldsymbol{T}}
\def\WH{{$W\!H$}}
\def\ZH{{$Z\!H$}}
\def\VH{{$V\!H$}}
\def\Hbb{$H\to b{\bar b}$}
\def\MC@NLO{{\sc MC@NLO}}
\def\POWHEG{{\sc POWHEG}}
\def\PYTHIA8{{\sc PYTHIA8}}
\def\HERWIG++{{\sc HERWIG++}}
\newcommand{\vH}{\mathbf{H}}
\newcommand{\vW}{\mathbf{W}}
\newcommand{\vB}{\mathbf{B}}
\newcommand{\vD}{\mathbf{D}}
\newcommand{\LL}{\mathcal{L}}
\newcommand{\MET}{$\slashed{E}_T$}
\newcommand{\tr}[1]{\operatorname{tr}\left[#1\right]}
\newcommand{\off}{\rm off}
\newcommand{\vev}[1]{ \left\langle {#1} \right\rangle }
\def\bentarrow{\:\raisebox{1.3ex}{\rlap{$\vert$}}\!\rightarrow}
\def\dkp#1#2#3#4{
\begin{array}{r c l}
#1 & \rightarrow & #2#3 \\
 & & \phantom{\; #2}\bentarrow #4
\end{array}}
\def\bothdk#1#2#3#4#5{
\begin{array*}{r c l}
#1 & \rightarrow & #2#3 \\
 & & \:\raisebox{1.3ex}{\rlap{$\vert$}}\raisebox{-0.5ex}{$\vert$}
\phantom{#2}\!\bentarrow #4 \nonumber \\
 & & \bentarrow #5
\end{array*}
}
\def\bentarrow{\:\raisebox{1.3ex}{\rlap{$\vert$}}\!\rightarrow}
\def\bothdk#1#2#3#4#5{
\begin{equation}
\begin{array}{r c l}
#1 & \rightarrow & #2#3 \\
 & & \:\raisebox{1.3ex}{\rlap{$\vert$}}\raisebox{-0.5ex}{$\vert$}%
\phantom{#2}\!\bentarrow #4 \\
 & & \bentarrow #5
\end{array}
\end{equation}
}
\def\as{\alpha_s}
\def\kV{\kappa_V}
\def\kF{\kappa_F}
\def\spa#1.#2{\langle #1  #2\rangle}
\def\spab#1.#2.#3{\langle #1|#2|#3]}
\def\spb#1.#2{[ #1 #2]}
\def\kF{\kappa_F}
\def\beq{\begin{equation}}
\def\eeq{\end{equation}}
\def\beqn{\begin{eqnarray}}
\def\eeqn{\end{eqnarray}}
\def\bal{\begin{align}}
\def\eal{\end{align}}

% math abbreviations
\newcommand{\df}{\mathrm{d}}
\newcommand{\img}{\mathrm{i}}

\newcommand{\hH}{\widehat{H}}
\newcommand{\hS}{\widehat{S}}

% abbreviations for specific notations
\newcommand{\alem}{\alpha_\mathrm{em}}
\newcommand{\Ecm}{E_\mathrm{cm}}
\newcommand{\alphad}{\alpha_{2}}
\newcommand{\betad}{\beta_{2}}
\newcommand{\phiu}{\phi_{1}}
\newcommand{\phid}{\phi_{2}}
\newcommand{\alphaud}{\alpha_{12}}
\newcommand{\betaud}{\beta_{12}}
\newcommand{\phiud}{\phi_{12}}
\newcommand{\zud}{z_{12}}
\newcommand{\dotp}[2]{#1 \cdot #2}
\def\bom#1{{\mbox{\boldmath $#1$}}}
\newcommand{\la}{\langle}
\newcommand{\ra}{\rangle}
\def\tauN{{{\cal T}_N}}
\def\gs{g}
\newcommand{\id}{\mathbf{1}}
\newcommand{\hga}{\widehat{\gamma}}
\newcommand{\hZ}{\widehat{Z}}
\newcommand{\cf}{C_F}
\newcommand{\ca}{C_A}
\newcommand{\nf}{n_f}
\newcommand{\cfs}{C_F^2}
\newcommand{\Tcm}{\tau_{cm}}
\newcommand{\bare}{\mathrm{bare}}
\newcommand{\cut}{\mathrm{cut}}
\newcommand{\zero}{{(0)}}
\newcommand{\one}{{(1)}}
\newcommand{\hemi}{\mathrm{hemi}}
\newcommand{\cm}{{\mathcal M}}
\newcommand{\cU}{{\mathcal U}}
\newcommand{\cN}{{\mathcal N}}
\newcommand{\cL}{{\mathcal L}}
\newcommand{\cI}{{\mathcal I}}
\newcommand{\cJ}{{\mathcal J}}
\newcommand{\cS}{{\mathcal S}}
\newcommand{\cT}{{\mathcal T}}
\newcommand{\hp}{\hat{p}}
\newcommand{\hs}{y}
\def\rG{\mathrm{\Gamma}}
\def\bI{\boldsymbol{I}}
\def\bT{\boldsymbol{T}}
\def\WH{{$W\!H$}}
\def\ZH{{$Z\!H$}}
\def\VH{{$V\!H$}}
\def\Hbb{$H\to b{\bar b}$}
\def\MC@NLO{{\sc MC@NLO}}
\def\POWHEG{{\sc POWHEG}}
\def\PYTHIA8{{\sc PYTHIA8}}
\def\HERWIG++{{\sc HERWIG++}}
\newcommand{\e}{\epsilon}
\newcommand{\ep}{\epsilon}
\newcommand{\eps}{\epsilon}
\def\bentarrow{\:\raisebox{1.3ex}{\rlap{$\vert$}}\!\rightarrow}
\def\dkp#1#2#3#4{
\begin{array}{r c l}
#1 & \rightarrow & #2#3 \\
 & & \phantom{\; #2}\bentarrow #4
\end{array}}
\def\bothdk#1#2#3#4#5{
\begin{array*}{r c l}
#1 & \rightarrow & #2#3 \\
 & & \:\raisebox{1.3ex}{\rlap{$\vert$}}\raisebox{-0.5ex}{$\vert$}
\phantom{#2}\!\bentarrow #4 \nonumber \\
 & & \bentarrow #5
\end{array*}
}
\def\bentarrow{\:\raisebox{1.3ex}{\rlap{$\vert$}}\!\rightarrow}
\def\bothdk#1#2#3#4#5{
\begin{equation}
\begin{array}{r c l}
#1 & \rightarrow & #2#3 \\
 & & \:\raisebox{1.3ex}{\rlap{$\vert$}}\raisebox{-0.5ex}{$\vert$}%
\phantom{#2}\!\bentarrow #4 \\
 & & \bentarrow #5
\end{array}
\end{equation}
}
\def\as{\alpha_s}
\def\kV{\kappa_V}
\def\kF{\kappa_F}
\def\spa#1.#2{\langle #1  #2\rangle}
\def\spab#1.#2.#3{\langle #1|#2|#3]}
\def\spb#1.#2{[ #1 #2]}
\def\kF{\kappa_F}
\def\beq{\begin{equation}}
\def\eeq{\end{equation}}
\def\beqn{\begin{eqnarray}}
\def\eeqn{\end{eqnarray}}

\newcommand{\eq}[1]{Eq.~\eqref{eq:#1}}
\newcommand{\eqs}[2]{Eqs.~\eqref{eq:#1} and \eqref{eq:#2}}
\newcommand{\fig}[1]{Fig.~\ref{fig:#1}}

%% optional: allow align to break pages
\allowdisplaybreaks[4]
\newcommand{\two}{{(2)}}
\newcommand{\ztwo}{\zeta_2}
\newcommand{\zthree}{\zeta_3}

%%% Comment one or other of these out
% MW notation
\def\meas#1#2{n_#2 \cdot #1}
% CE notation
\def\meas#1#2{#1^{#2}}

\def\MF{\mathcal F}

%% file: intro.tex
The continued successful operation of the Large Hadron Collider (LHC)
has led to the accumulation of a very large data set with which to study
the Standard Model (SM) in unprecedented detail. The ever-increasing
precision of the experimental analyses has mandated a similar increase
in the precision of the corresponding theoretical predictions.  Over
the last few years a concerted effort has been made in the theoretical
community to provide predictions accurate to next-to-next-to-leading
order (NNLO) in QCD. Calculations that include colored final-state
radiation are particularly challenging.  Significant progress in this
direction has been made recently, including the NNLO calculation of the production of 
dijets~\cite{Currie:2014upa,Currie:2016bfm},
$V+j$~\cite{Ridder:2016rzm,Ridder:2015dxa,Ridder:2016nkl,Boughezal:2015ded,Boughezal:2016dtm,Campbell:2016lzl,Campbell:2017dqk},
$H+j$~\cite{Boughezal:2015dra,Boughezal:2015aha,Chen:2016zka}, $H+2~j$
in the vector boson fusion process~\cite{Cacciari:2015jma}, 
single top~\cite{Brucherseifer:2014ama,Berger:2016oht,Berger:2017zof}
and $t\overline{t}$~\cite{Czakon:2016ckf}.
 
A critical element in the completion of a NNLO calculation is a
manageable way to handle the copious infrared (IR) singularities
present in component pieces of the calculation. These singularities
occur in phase spaces of differing dimensionality, and
cancel only when combined in a suitably-inclusive, IR-safe
observable. At NLO the most widespread solutions use local
subtraction terms~\cite{Ellis:1980wv,Kunszt:1989km,Catani:1996vz}.
In these approaches one subtracts a user-defined
set of counter-terms from a given real-emission matrix element such
that the corresponding combination of matrix element plus
counter-terms is finite in all singly-unresolved IR limits. The
counter-terms are constructed in such a way as to be integrable
analytically over a single unresolved parton; the integrated
counter-terms can then be combined with the virtual one-loop matrix
elements, resulting in an analytic cancellation of IR poles. The two
phase spaces are then both manifestly finite and can be integrated
separately using Monte Carlo integration techniques. 

The construction of a similar subtraction scheme at NNLO accuracy is a
considerably more daunting task. This is primarily due to the presence
of one extra unresolved parton with respect to NLO,
yielding multiple overlapping singularities.  
Despite its difficulty, a number of subtraction schemes have
been developed and successfully applied to several LHC
processes~\cite{GehrmannDeRidder:2005cm,Somogyi:2006da,Czakon:2010td,GehrmannDeRidder:2012ja,Caola:2017dug}.

Alternatives to local subtraction schemes are possible. One such
method, based on a more global approach, is phase-space
slicing~\cite{Giele:1993dj}. These methods are simple to implement at
NNLO, particularly if the corresponding NLO process with one extra
parton in the final state is already known.  In slicing methods a
global parameter is used to divide the phase space into (at least) two
regions. At NNLO the two regions correspond to the region which
includes all of the doubly-unresolved emissions, and a region which
has at most one singly-unresolved parton. The IR structure in
the latter region is clearly akin to that obtained in a standard NLO
calculation, and hence this region is amenable to calculation using
existing NLO technology. The success of the method therefore relies on
the ability to calculate the region which contains doubly-unresolved
partons. To this end, factorization theorems are used to
calculate the cross section systematically in this region. The first
slicing method applied at NNLO~\cite{Catani:2007vq} used the
transverse momentum of the final state, $q_T$, as a slicing parameter,
and the Collins-Soper-Sterman factorization
theorem~\cite{Collins:1984kg} to compute the cross section in the
region of small $q_T$. As such, this method is applicable to final
states in which there is no $q_T$ associated with colored radiation,
i.e.~the production of color-singlet final states. A recently-developed 
method~\cite{Boughezal:2015dva,Gaunt:2015pea} uses the
$N$-jettiness event-shape variable
$\mathcal{T}_N$~\cite{Stewart:2010tn} rather than $q_T$, and a
factorization theorem from Soft-Collinear Effective Field theory
(SCET)~\cite{Bauer:2000ew,Bauer:2000yr,Bauer:2001ct,Bauer:2001yt,Bauer:2002nz}.
The theorem states that the cross section in the region of small $\mathcal{T}_N$ can be obtained from the following
convolution
\begin{eqnarray} \label{smalltauxsec}
\sigma(\mathcal{T}_N < \mathcal{T}_N^{{\rm{cut}}})  = 
\int  B \otimes B \otimes S \otimes H \otimes  \prod_{i=1}^{N} J_i + \mathcal{O}(\mathcal{T}_N^{\rm{cut}}) \, .
\end{eqnarray}
Here $B$ represents the beam function, which describes initial-state
collinear radiation, and $J$ the jet function, which describes
final-state collinear radiation. For both, expansions accurate to
$\mathcal{O}(\as^2)$ can be found in
Refs.~\cite{Gaunt:2014cfa,Gaunt:2014xga}.  The term $H$ denotes the
hard function, which is process-specific and finite. Finally, $S$
represents the soft function, which is the main focus of this
paper. The soft function is defined as the process-independent soft
limit of QCD amplitudes. Of the process-independent pieces of the
factorization theorem, it is by far the most complicated. For
color-singlet production processes (which for the LHC correspond to
zero jets in the final state), the soft function is reasonably simple and
analytic expressions are known. When three colored partons are present
(1-jet final states for LHC kinematics), the soft function is considerably more
intricate.  The calculation of the soft function at NLO was presented
in Ref.~\cite{Jouttenus:2011wh}.

A method to compute the 1-jettiness soft function
numerically to NNLO accuracy was presented in
Ref.~\cite{Boughezal:2015eha} and the results of this calculation have
since been used to compute several $V+j$ processes at
NNLO~\cite{Boughezal:2015ded,Boughezal:2016dtm}.  However, at present,
there is no publicly-available computation of the 1-jettiness soft
function presented in a form that can be implemented in an independent Monte Carlo
code. The main focus of Ref.~\cite{Boughezal:2015eha} is to provide a
methodology of computing the soft function numerically. Specific
results are only presented for the $qg \to q$ configuration, and only
in graphical form. Since the $gg \to g$ and
$q \bar{q} \to g$ configurations are absent, and due to the
nature of the result presented, it is currently impossible to
implement the 1-jettiness soft function at NNLO directly from the
literature. The primary aim of our paper is to provide this
information via an independent calculation of the 1-jettiness soft
function. 

Our paper proceeds as follows. In Section \ref{sec:calcsetup} we provide
a general overview of the calculation. We introduce the $N$-jettiness
variable $\mathcal{T}_N$ and present our parametrization of the phase space. 
In Section \ref{sec:calcsoft}
we present the formulae for soft parton emission at NLO and NNLO taken from the literature, 
paying particular attention to their color structures.
In Section \ref{sec:calcnlo} we validate the method by computing the NLO
soft function for the 1-jettiness case and compare the results
against known analytic formulae. In Section \ref{sec:calcnnlo} we discuss
our calculation for the NNLO 0- and 1-jettiness soft functions in
detail. In Section \ref{sec:results} we present the obtained results and
compare them against the known results in the literature. We draw our
conclusions in Section \ref{sec:conclu}.

%% file: calcsetup.tex
\subsection{$N$-jettiness variable $\mathcal{T}_N$}

For a parton
scattering event the $N$-jettiness variable $\mathcal{T}_N$
\cite{Stewart:2010tn} is defined as 
\bal \mathcal{T}_N = \sum_m
\text{min}_i \left\{ \frac{2 p_i \cdot q_m}{P_i} \right\} \, ,
\end{align}
where the subscript $N$ refers to the number of final-state jets in the
scattering event for Born-level kinematics. The momenta $p_i$ are the momenta
of the initial-state colored partons and final-state jets 
at Born level. For color-singlet production at the LHC, $N=0$ and $i \in \{1,2\}$,
while for LHC processes with one jet in the final state $N=1$ and
$i \in \{1,2,3\}$. The quantities $P_i$ are dimensionful normalization
factors that represent the hardness of the momenta $p_i$. The $q_m$ denote the momenta of final-state radiation.
For single-emission processes (NLO real corrections or NNLO real-virtual corrections) $m=1$, while for
double-emission processes (NNLO double-real corrections) the sum runs over two terms, $m \in \{1,2\}$.
For the calculation of the soft function the eikonal directions are given.
By defining dimensionless versions of the massless momenta $p_i$ through
$\hp_i = p_i/P_i$ and choosing $P_i = 2 E_i$, where $E_i$ is the energy of the parton, we can rewrite $\mathcal{T}_N$ as 
\bal \label{newdeftaun} \mathcal{T}_N = \sum_m
\text{min}_{i} \left\{ 2\hp_i \cdot q_m \right\} \, .
\end{align} 
In units where $\hbar=c=1$, $\mathcal{T}_N$ therefore has the units of mass.

\subsection{Sudakov decomposition}
\label{sec:Suddecomp}

A convenient way of parametrizing the momenta appearing in
the phase-space integrals that enter the calculation of the soft function is to use a Sudakov decomposition
of the momenta in terms of two of the momenta $\hp_i$ that appear in Eq.~(\ref{newdeftaun}). We first define a shorthand notation for the quantity that appears in Eq.~(\ref{newdeftaun}), namely the
projection of a vector $q$ along the direction of $\hp_i$,
\begin{equation}
q^x = 2\hp_x \cdot q \, .
\end{equation}
The parton momentum $q^\mu$ can then be expanded as
\begin{equation}
q^\mu  =  q^j\, \frac{\hp_i^\mu}{\hs_{ij}} +  q^i \frac{\hp_j^\mu}{\hs_{ij}}  +  q_{ij\perp}^\mu \, ,
\end{equation}
where $\hs_{ij}=2 \hp_i \cdot \hp_j$ and $q_{ij\perp}$ is transverse to the plane spanned
by $\hp_i$ and $\hp_j$.
The Sudakov expansion for $\hp_k$, which is not one of the Sudakov base vectors, is
\begin{equation}
\hp^\mu_k = \hp_i^\mu\frac{\hs_{jk}}{\hs_{ij}}+\hp_j^\mu\frac{\hs_{ik}}{\hs_{ij}}+\hp_{k\perp}^\mu \, .
\end{equation}
We can calculate $q^k$, the projection of $q$ on a non-Sudakov base vector, and obtain
\begin{equation}
q^k = 2 q \cdot \hp_k  =  q^j\, \frac{\hs_{ik}}{\hs_{ij}} +  q^i \frac{\hs_{jk}}{\hs_{ij}} 
 -2 \abs{q_{ij\perp}} \abs{\hp_{k\perp}}\cos \phi_{qk} \, ,
\end{equation}
where $\phi_{qk}$ is the angle in the transverse plane between $q$ and $\hp_k$, and
\beq
\hp_{k\perp}^2= \frac{\hs_{ik}\hs_{jk}}{\hs_{ij}}, \;\;\; q_{ij\perp}^2= \frac{q^i q^j}{\hs_{ij}} \, .
\eeq
The ratio of the projection along a non-Sudakov direction $k$ to the projection along
a Sudakov direction $i$ or $j$ is given by,
\beqn \label{Adefn}
\frac{q^k}{q^i} &=& \frac{\hs_{jk}}{\hs_{ij}} + x_{ji} \frac{\hs_{ik}}{\hs_{ij}}  
         - 2 \sqrt{\frac{x_{ji} \hs_{ik}\hs_{jk}}{\hs_{ij}^2}} \cos \phi_{qk} = A_{ji,k}(x_{ji},\phi_{qk})\, , \\
\frac{q^k}{q^j} &=& \frac{\hs_{ik}}{\hs_{ij}} 
         +  x_{ij}\frac{\hs_{jk}}{\hs_{ij}}
         - 2 \sqrt{\frac{x_{ij} \hs_{ik}\hs_{jk}}{\hs_{ij}^2}} \cos \phi_{qk} = A_{ij,k}(x_{ij},\phi_{qk}) \, ,
\eeqn
where $x_{ij}=q^i/q^j$.  For the case of hadronic collisions where two
of the directions, $\hp_1$ and $\hp_2$, are those of the beams we have,
\beq
\label{eq:beams}
\hp_1 = \frac{1}{2} (1, 0, 0, +1), \quad
\hp_2 = \frac{1}{2} (1, 0, 0, -1)\, ,
\eeq
so that $\hs_{12}=1$.

We note that this notation follows that of the calculation of the NLO soft function~\cite{Jouttenus:2011wh}.
It differs from that of Ref.~\cite{Boughezal:2015eha}, in which relevant quantities are expressed in terms
of pure directions, $n_i$.  Equivalent expressions can be obtained by making the replacement,
\begin{equation}
\hp_i \to \frac{n_i}{2}.
\end{equation}

\subsection{Measurement function}
\label{sec:measurement}
Written in terms of the projected momenta, the definition of the $N$-jettiness is
\bal \label{newdeftaunproj} \mathcal{T}_N = \sum_m
\text{min}_{i} \, \{ q_m^i \} \, .
\end{align}
The minimum over $i$ in this equation provides a natural
division of the phase space for extra emission into regions where each
projection is smallest.
For the case of $1$-jettiness we will label the three hard directions
as $i$, $j$ and $k$.  The single-emission phase space is then
partitioned by inserting a measurement function $F$ where,
\begin{equation} \label{Measureone}
F = F_i + F_j + F_k \, ,
\end{equation}
and $F_i$ corresponds to $q_1$ being closest to direction $i$ so that, for instance,
\begin{equation} \label{Measurement-oneparton}
F_i = \delta(\tauN-\meas{q_1}{i}) \, \theta(\meas{q_1}{j}-\meas{q_1}{i}) \theta(\meas{q_1}{k}-\meas{q_1}{i}).
\end{equation}
We note that in the original calculation of the NLO soft
function~\cite{Jouttenus:2011wh}, a further hemisphere decomposition of
the measurement function was used to separate the divergent and finite
parts of the calculation.  However, we will not pursue that method for
our NNLO calculation.  We can extend the decomposition of
$F$ to the double-emission case by writing the measurement function as,
\begin{equation} 
F = \sum_{a,b} F_{ab} \, ,
\label{eq:Fab}
\end{equation}
where the sum runs over the nine combinations of $a,b \in \{i,j,k\}$.  The notation $F_{ab}$ implies that
$q_1$ is closest to direction $a$ and $q_2$ is closest to $b$ with, for example,
\begin{equation} \label{Measurethree}
F_{ij} = \delta(\tauN-\meas{q_1}{i}-\meas{q_2}{j}) \,
 \theta(\meas{q_1}{j}-\meas{q_1}{i}) \theta(\meas{q_1}{k}-\meas{q_1}{i}) \,
 \theta(\meas{q_2}{i}-\meas{q_2}{j}) \theta(\meas{q_2}{k}-\meas{q_2}{j}).
\end{equation}
As it will be explicitly shown in Section \ref{sec:calcsoft}, the integrals that we have to evaluate in order to calculate the soft function have
an eikonal form in which there are two emitting directions $i$ and
$j$. We attach the labels of the emitters as superscripts to the
measurement functions. In this
notation we can write out the decomposition in Eq.~(\ref{eq:Fab})
explicitly as
\begin{align} \label{Measurement-twoparton}
F^{ij} &= F^{ij}_{ii} + F^{ij}_{jj} && \mbox{(case 1)} \nonumber \\
       &+ F^{ij}_{ij} + F^{ij}_{ji} &&  \mbox{(case 2)} \nonumber \\
       &+ F^{ij}_{ik} + F^{ij}_{jk} + F^{ij}_{ki} + F^{ij}_{kj} && \mbox{(case 3)} \nonumber \\
       &+ F^{ij}_{kk} \, . && \mbox{(case 4)}
\end{align}
Each term corresponds to one of four cases~\cite{Boughezal:2015eha}, as indicated:
\begin{enumerate}
\item Both $q_1$ and $q_2$ closest to the same emitting direction;
\item $q_1$ and $q_2$ closest to different emitting directions;
\item One of $q_1$ and $q_2$ closest to an emitter, while the other is closest to the non-emitting
direction $k$; 
\item Both $q_1$ and $q_2$ closest to the non-emitting direction $k$.
\end{enumerate}
In general, there is one more case where $q_1$ and $q_2$ are closest to different non-emitting directions $k,l$, but this  
is absent in the $1$-jettiness case.

\subsection{Phase space -- single emission}

Let us start by considering the phase space for a single emission:
\bal \label{mopps}
PS^{(1)} = \int \frac{\mathrm{d}^{d}q}{(2\pi)^{d-1}} \, \deltap(q^2) \, .
\end{align}
Using the Sudakov variables $q^i$ and $q^j$ defined in Section~\ref{sec:Suddecomp} above, we 
can rewrite the integration measure as
\bal
\mathrm{d}^{d}q = \frac{1}{2\hs_{ij}} \, \mathrm{d} q^i \, \mathrm{d} q^j \, \mathrm{d}^{d-2} q_{ij\perp} \, .
\end{align}
The phase space then becomes 
\bal
PS^{(1)}(i,j) 
 &= \frac{1}{(2\pi)^{d-1}} \frac{1}{2\hs_{ij}} \int \mathrm{d} q^i \, \mathrm{d} q^j \,
    \mathrm{d} \Omega_{(d-2)} \, \frac{\mathrm{d} q_{ij\perp}^2}{2 |q_{ij\perp}|} \left[q_{ij\perp}^2\right]^{\frac{d-3}{2}}
    \delta\left(\frac{q^i q^j}{\hs_{ij}} - q_{ij\perp}^2\right) \notag \\
 &= \frac{1}{(2\pi)^{d-1}} \frac{1}{4\hs_{ij}} \int \mathrm{d} q^i \, \mathrm{d} q^j \, \mathrm{d} \Omega_{(d-2)} \,
    \left[\frac{q^i q^j}{\hs_{ij}}\right]^{\frac{d}{2}-2} \, ,
\end{align}
where $\mathrm{d} \Omega_{(d-2)}$ is the $(d-2)$-dimensional angular
measure.  Setting $d=4-2\eps$ and using the standard expression for the angular 
measure after integrating over unconstrained angles given in Eq.~(\ref{eq:PS1angles}),
we obtain
\begin{equation}
PS^{(1)}(i,j) = \frac{\pi^\ep}{16\pi^3} \frac{\Gamma(1-\ep)} {\Gamma(1-2 \ep)} \frac{1}{\hs_{ij}^{1-\ep}} 
    \int \mathrm{d} q^i \, \mathrm{d} q^j \, \left[q^i q^j\right]^{-\eps} \,
    \int_0^{\pi} d \phi \, \sin^{-2 \ep}\phi .
\end{equation}
It is convenient to normalize the remaining angular integration so that it integrates to one using,
\begin{equation}
N_\phi = \int_0^{\pi} d \phi \, \sin^{-2 \ep}\phi = 4^\ep \pi \frac{\Gamma(1-2\ep)}{\Gamma(1-\ep)^2}.
\end{equation}
The final expression for the single-emission phase space is then,
\begin{equation}
PS^{(1)}(i,j) = \left[\frac{1}{16\pi^2} \frac{(4\pi)^\ep}{\Gamma(1-\ep)}\right] \frac{1}{\hs_{ij}^{1-\ep}} 
    \int \mathrm{d} q^i \, \mathrm{d} q^j \, \left[q^i q^j\right]^{-\eps} \,
    \int_0^{\pi} \frac{d \phi}{N_\phi} \, \sin^{-2 \ep}\phi \, .
\label{PhaseSpace1}
\end{equation}

\color{black}
\subsection{Phase space -- double emission}

For the double-emission phase space we employ the same Sudakov decomposition as the single-emission case and, following the same steps as above, we
have
\beqn
&&PS^{(2)}(i,j)=\int \frac{d^d q_1}{(2 \pi)^{d-1}} \frac{d^d q_2}{(2 \pi)^{d-1}} 
\deltap(q_1^2) \deltap(q_2^2) \nonumber \\
&=& 
\frac{1}{(2\pi)^{2d-2}} \frac{1}{16\hs_{ij}^2} \int \mathrm{d} q_1^i \, \mathrm{d} q_1^j
  \, \mathrm{d} \Omega_{(d-2)}^{(q_1)} \, \left[\frac{q_1^i q_1^j}{\hs_{ij}}\right]^{\frac{d}{2}-2}
  \int \mathrm{d} q_2^i \, \mathrm{d} q_2^j
  \, \mathrm{d} \Omega_{(d-2)}^{(q_2)} \, \left[\frac{q_2^i q_2^j}{\hs_{ij}}\right]^{\frac{d}{2}-2} .
\eeqn
The integral over the transverse space for $q_1$ can be performed just
as in the single-emission case, with the result given in
Eq.~(\ref{eq:PS1angles}).  The integral over the transverse space for
$q_2$ is more complicated since one of the angles cannot be integrated
out; the form of the integral is given in Eq.~(\ref{eq:PS2angles}).
Combining these expressions we arrive at the final form for the phase
space,
\beqn \label{PhaseSpace2}
PS^{(2)}(i,j)&=&\frac{1}{2^8 \pi^4} \left[\frac{(4 \pi)^\ep}{\Gamma(1-\ep)}\right]^2 \Big(\frac{1}{\hs_{ij}}\Big)^{2-2\ep}
\int dq_1^i dq_1^j dq_2^i dq_2^j \, \left[q_1^i q_1^j q_2^i q_2^j\right]^{-\ep} \nonumber \\
&\times& \int_0^\pi \frac{d \phiu}{N_{\phiu}} \sin^{-2 \ep}\phiu
         \int_0^\pi \frac{d \phid}{N_{\phid}} \sin^{-2 \ep}\phid
         \int_0^\pi \frac{d \beta }{N_{\beta }} \sin^{-1 -2 \ep}\beta,
\eeqn
where
\beq
N_\beta = -\frac{1}{\ep} \sqrt{\pi} \frac{\Gamma(1-\ep)}{\Gamma(\frac{1}{2}-\ep)},\;\;\;
N_{\phiu} = N_{\phid} = 4^\ep \pi \frac{\Gamma(1-2\ep)}{\Gamma(1-\ep)^2}
  = \sqrt{\pi} \frac{\Gamma(\frac{1}{2}-\ep)}{\Gamma(1-\ep)}.
\eeq

%% file: calcsoft.tex
\color{black}

\subsection{Soft function}

We define the unrenormalized $N$-jettiness soft function $\tS(\mathcal{T}_N)$ as a perturbative series in powers of the bare strong coupling $\overline{\alpha}_s$:
\bal
\tS(\mathcal{T}_N) = \tS^{(0)}(\mathcal{T}_N) + \Big[\frac{\overline{\alpha}_s}{2 \pi}\Big] \tS^{(1)}(\mathcal{T}_N) + \Big[\frac{\overline{\alpha}_s}{2 \pi}\Big]^2 \tS^{(2)}(\mathcal{T}_N) + \mathcal{O}(\overline{\alpha}_s^3) \, .
\end{align}
We renormalize the coupling constant by performing the replacement 
\bal \label{ccrenor}
\overline{\alpha}_s \to \as \, Z_{\alpha} \, .
\end{align}
The renormalization factor is given by, 
\beq 
Z_{\alpha} = 1 - \Big[\frac{\as}{2 \pi}\Big] \frac{\beta_0}{2 \eps} + \mathcal{O}(\as^2), 
\hspace{0.5cm} \beta_0 = \frac{11}{3} C_A -\frac{4}{3} T_R N_F
\eeq
with $C_A=3$, $T_R=\frac{1}{2}$, $N_F=5$, and $\as \equiv \as(\mu)$ at the renormalization scale $\mu$. The coefficients of the perturbation series of the renormalized soft function $S(\mathcal{T}_N)$ in terms of the unrenormalized ones then read:
\bal \label{rensoftdef}
S^{(0)}(\mathcal{T}_N) &= \tS^{(0)}(\mathcal{T}_N) \notag \\
S^{(1)}(\mathcal{T}_N) &= \tS^{(1)}(\mathcal{T}_N) \notag \\
S^{(2)}(\mathcal{T}_N) &= -\frac{\beta_0}{2 \eps} \, \tS^{(1)}(\mathcal{T}_N) + \tS^{(2)}(\mathcal{T}_N) \, .
\end{align}
The leading-order contribution is simply
$S^{(0)}(\mathcal{T}_N) = \delta(\mathcal{T}_N)$, since at leading order there is no emitted radiation. 

\subsection{Soft radiation at NLO}
We start by computing the soft function at NLO, which allows us to
illustrate the main features of the method as well. The result for the
NLO soft function is known analytically~\cite{Jouttenus:2011wh} and
can be used to validate our numerical evaluation. The NLO corrections
to the leading-order soft function are made up of two different
contributions: Born-type processes with one-loop corrections
(``virtual'' corrections) and tree-level processes with the emission
of one additional parton (``real'' corrections). The former only
contribute at $\mathcal{T}_N=0$. Since we are considering corrections
on massless eikonal lines, there is no dimensionful quantity to carry
the dimension of the one-loop integrals; their contribution therefore vanishes
in dimensional regularization.  The only contribution at NLO is
therefore the one from real radiation.

The form of the squared amplitude representing the emission of a single soft gluon is well known.  Using the same notation
that will be employed at NNLO we can write the factorization at ${\cal O}(\gs^2)$ as,
\beq
\label{ccfact}
| \cm^{(0)}(q,p_1,\dots,p_m) |^2 \simeq
- \gs^2 \,S_\ep \, \mu^{2\ep} \,2 \,\sum_{i,j=1}^m\, {\cal S}_{ij}(q)
\;| \cm^{(0)}_{(i,j)}(p_1,\dots,p_m) |^2 \;\;,
\eeq
c.f. Eq.~(12) of Ref.~\cite{Catani:2000pi}.  The eikonal function ${\cal S}_{ij}(q)$ is given by
\begin{equation}
\label{eikfun}
{\cal S}_{ij}(q) = \frac{p_i \cdot p_j}{2 (p_i \cdot q)\, (p_j\cdot q)}  = \frac{ \hs_{ij}}{q^i q^j} \;\;.
\end{equation}
Note that here we have introduced the normal $\overline{MS}$ factor $S_\ep$,
\beq \label{Sepdefinition}
S_\ep = \left(\frac{e^{\gamma_E}}{4 \pi}\right)^{\ep},
\eeq
where $\gamma_E$ is the Euler-Mascheroni constant.
The emission of a soft gluon produces color correlations that are indicated in Eq.~(\ref{ccfact})
by the subscripts $i$ and $j$ in $\cm^{(0)}_{(i,j)}$,
\beq
\label{colam}
| \cm^{(0)}_{(i,j)}(p_1,\dots,p_m) |^2 \equiv
\la \,\cm^{(0)}(p_1,\dots,p_m) \,|
\,{\bT}_i \cdot {\bT}_j
\,|\,\cm^{(0)}(p_1,\dots,p_m)\rangle.
\eeq
Factoring out the leading-order amplitude squared we thus have a simple expression for the
soft-gluon approximation,
\beqn
|M^{(1)}|^2 &=& - 4g^2 \mu^{2 \ep}  S_\ep \, \sum_{i < j}\, \bT_i\cdot  \bT_j \, \frac{\hs_{ij} }{q^i q^j} \nn \\
 & \equiv & \sum_{i<j} \, \bT_i\cdot  \bT_j \, |M^{(1)}_{ij}|^2\,,  \label{lowestordermatrixelement}
\eeqn
where we have limited the sum that appears in Eq.~(\ref{ccfact}) to $i < j$ and added the consequent factor of two.
Introducing the measurement function of Eq.~\eqref{Measureone}, the total result for the soft function at NLO is then,
\beqn 
\label{NLOsoft}
\Big[\frac{\as}{2 \pi}\Big] \tS^{(1)} &=& \sum_{i<j} \, \bT_i\cdot  \bT_j \, |M_{ij}^{(1)}|^2 \; PS^{(1)} \; F \nonumber \\
            &=& \sum_{i<j} \, \bT_i\cdot  \bT_j \, |M_{ij}^{(1)}|^2 \; PS^{(1)} \; \Big[F_i+F_j+F_k\Big] \label{eq:soft1} \\
            &=& \sum_{i<j} \, \bT_i\cdot  \bT_j \, |M_{ij}^{(1)}|^2 \Biggl\{ PS^{(1)}(i,j) \; \Big[F^{ij}_i+F^{ij}_j\Big]
               + PS^{(1)}(k,i) \; \Big[F^{ij}_k\Big] \Biggr\}, \nonumber
\eeqn
where, in the last line, we have explicitly indicated the choice of momenta in the Sudakov decomposition in the
phase-space.  The explicit expressions for the quantities in curly braces 
in Eq.~(\ref{NLOsoft}) are given in
Eqs.~(\ref{Measurement-oneparton}) and~(\ref{PhaseSpace1}), respectively.
When we evaluate the different contributions in Eq.~\eqref{eq:soft1} we will have
two different cases, corresponding to $F^{ij}_i$ and $F^{ij}_k$. The case $F^{ij}_j$ can be obtained by relabelling 
since $S_{ij}(q)$ is symmetric under $i \leftrightarrow j$.

\subsection{Color conservation}
\label{sec:color}
The eikonal expressions given above are written using color-space notation~\cite{Catani:1996vz}. 
Using color conservation we have,
\begin{equation}\label{eq:colcons}
\sum_j \bT_j | {\cal M}\rangle= 0\, .
\end{equation}
Thus for the case of $0$-jettiness ($j \in \{1,2\}$) we find that $\bT_1^2 = \bT_2^2 =- \bT_1 \cdot \bT_2\,$,
whereas for the case of $1$-jettiness ($j \in \{1,2,3\}$) we have
$\bT_1^2=-\bT_1 \cdot \bT_2 - \bT_1 \cdot \bT_3 \nonumber $ and cyclic permutations. 
For the $1$-jettiness case we can write,
\beqn
\bT_1 \cdot \bT_2&=& \frac{1}{2} \Big[\bT_3^2-\bT_1^2-\bT_2^2  \Big] \nonumber \\
\bT_2 \cdot \bT_3&=& \frac{1}{2} \Big[\bT_1^2-\bT_2^2-\bT_3^2  \Big] \nonumber \\
\bT_3 \cdot \bT_1&=& \frac{1}{2} \Big[\bT_2^2-\bT_3^2-\bT_1^2  \Big]\, .
\label{ColorIdentities}
\eeqn
All products $\bT_i \cdot \bT_j$ can therefore be expressed in terms of sums of Casimirs $\bT_i^2$
(and vice versa) with 
\beq
\bT_i^2 = C_F=\frac{4}{3}~\mbox{for}~i=q,\bar{q} \quad \mbox{and} \quad \bT_i^2 = C_A=3~\mbox{for}~i=g \,.
\eeq

\subsection{Soft radiation at NNLO: real-virtual}
\label{sec:RV}

As earlier discussed, the virtual diagrams do not contribute because of scaling arguments, so that at NNLO we are left with 
the real-virtual contribution and the double-real contribution (to be considered in the following two 
sections).  Since the real-virtual corrections involve only one real emission, the
calculation follows closely the NLO case. 

The one-loop contribution to the
soft-gluon current has been given in Ref.~\cite{Catani:2000pi}.
Combining Eqs.~(23) and~(26) of Ref.~\cite{Catani:2000pi}, the ${\cal O}(\gs^4)$
real-virtual contribution is,
\begin{eqnarray}
&& - (\gs_{\rm bare} \, \mu^\ep)^4
\left[
\la \,\cm^{(0)}(\{p\}) \,| \, {\bom J}_{\mu}^{(0)}(q)  \cdot
{\bom J}^{\mu \,(1)}(q, \ep) \, | \,\cm^{(0)}(\{p\}) \,\ra + \;{\rm c.c.} \right] =
 \nn \\ &&
 \frac{g^4}{4\pi^2}\; S_\ep^2 \; \mu^{4 \ep} \frac{(4\pi)^\ep}{\ep^2} \;\frac{\Gamma(1-\ep)^3
\,\Gamma(1+\ep)^2}{\Gamma(1-2\ep)} \Bigg\{ C_A \; \cos (\pi \ep) \;{\sum_{i,j}}^{\prime} 
\left[ {\cal S}_{ij}(q) \right]^{1+\ep}
\; | \cm^{(0)}_{(i,j)}(\{p\}) |^2 \nn \\ && \qquad \qquad
 + 2 \sin (\pi \ep) \;{\sum_{i,j,k}}^{\prime} {\cal S}_{ki}(q) \,
\left[ {\cal S}_{ij}(q)\right]^{\ep} \left(\lambda_{ij} - \lambda_{iq}
-\lambda_{jq} \right) 
\; | \cm^{(0)}_{(k,i,j)}(\{p\}) |^2 \Bigg\} \;,
\label{1loopcolcor}
\end{eqnarray}
where $S_{ij}(q)$ is given in Eq.~(\ref{eikfun}).
The notation $\sum^\prime$ represents a sum over values
of the indices that are distinct (for instance for the second term this explicitly means $i\neq j, j\neq k, k \neq i$).  The calculation of $1$-jettiness
does not permit such a contribution from the second term and therefore we do not consider it further. 

The real-virtual contribution is thus,
\beqn \label{NLOMsq} |M^{(RV)}|^2 
&=&\frac{g^4}{2 \pi^2} \mu^{4 \ep} S_\ep^2 \frac{( 4 \pi)^\ep}{\ep^2} \frac{\Gamma(1-\ep)^3 \Gamma(1+\ep)^2}{\Gamma(1-2\ep)}
C_A \cos(\pi \ep)  \sum_{i < j}\, \bT_i\cdot \bT_j\, \Big(\frac{\hs_{ij}}{q^i q^j}\Big)^{1+\ep} \nonumber \\
&=&\frac{g^4}{2 \pi^2} \mu^{4 \ep} C_A S_\ep^2 \frac{( 4 \pi)^\ep}{\ep^2} 
\frac{\Gamma(1-\ep)^4 \Gamma(1+\ep)^3}{\Gamma(1-2\ep)^2 \Gamma(1+2\ep)}
\sum_{i < j}\, \bT_i\cdot \bT_j\,  \Big(\frac{\hs_{ij}}{q^i q^j}\Big)^{1+\ep} \nonumber \\
&=&
2 \left( \frac{\as} {2 \pi}\right)^2
\mu^{4 \ep} \left[ 16 \pi^2 \, \frac{\Gamma(1-\ep)}{(4\pi)^\ep}\right]
\frac{C_A}{\ep^2}  B_{RV} \frac{\Gamma(1-2\ep)}{\Gamma(1-\ep)^2}
\sum_{i < j}\, \bT_i\cdot \bT_j\,  \Big(\frac{\hs_{ij}}{q^i q^j}\Big)^{1+\ep} \nonumber \\
& \equiv & \sum_{i<j} \, \bT_i\cdot  \bT_j \, |M^{(RV)}_{ij}|^2\,,
\eeqn
where we have used
\begin{equation}
\cos (\pi \ep) = \frac
{\Gamma(1-\ep) \Gamma(1+\ep)}
{\Gamma(1-2 \ep) \Gamma(1+2 \ep)}
\end{equation}
and extracted an overall factor
\beq
B_{RV} =\frac{e^{2 \gamma_E \ep}\Gamma(1-\ep)^5 \Gamma(1+\ep)^3}{\Gamma(1-2\ep)^3 \Gamma(1+2\ep)} 
= 1- \frac{2\pi^2}{3}\ep^2 -\frac{14 \zeta_3}{3} \ep^3 + \frac{\pi^4}{15}\ep^4 + O(\ep^5).
\eeq
We have also identified a factor (shown in square brackets in the third line of Eq.~(\ref{NLOMsq})) that will be naturally cancelled
by a corresponding one in the phase space, c.f. Eq.~(\ref{PhaseSpace1}).

Finally, we note that the method for calculating this
contribution will be very similar to the one used for the NLO soft function, after the replacement of the integrand factor
$\hs_{ij}/q^i q^j$ by $[\hs_{ij}/q^i q^j]^{1+\ep}$. For the real-virtual contribution to the NNLO soft function we therefore have
\beq
\Big[\frac{\as}{2 \pi}\Big]^2 \tS^{(2,RV)} = \sum_{i<j} \, \bT_i\cdot  \bT_j \, |M^{(RV)}_{ij}|^2 \Biggl\{  PS^{(1)}(i,j) \; \Big[F^{ij}_i+F^{ij}_j\Big]
               + PS^{(1)}(k,i) \; \Big[F^{ij}_k\Big] \Biggr\}.
\eeq

\subsection{Soft radiation at NNLO: $q\overline{q}$ emission}
\label{sec:RRqq}

Turning now to the double-real emission, we first consider the radiation of a soft $q \bar{q}$ pair.
The factorization of QCD amplitudes in the double-soft limit has been given in Refs.~\cite{Campbell:1997hg,Catani:1999ss}.
From Eq.~(95) of Ref.~\cite{Catani:1999ss} we have,
\begin{equation} \label{eq:QCDfac1}
|{\cal M}^{(0)}(p_1,\dots,p_m,q_1,q_2)|^2 = g^4 \mu^{4 \ep} \, S_\ep^2 \, T_R \, \langle {\cal M} | \left( \sum_{i,j=1}^m {\cI}_{ij}\, \bT_i \cdot \bT_j  \right)| {\cal M}\rangle \, .
\end{equation}
Using the color identities of Eq.~(\ref{ColorIdentities}), we can rewrite the result for the matrix element as,
\begin{equation} \label{TRmatrixelementsq}
|{\cal M}^{(0)}(p_1,\dots,p_m,q_1,q_2)|^2 = - T_R \, \langle {\cal M} | \left( \sum_{i<j} {\cal U}_{ij}\, \bT_i \cdot \bT_j  \right)| {\cal M}\rangle \, ,
\end{equation}
where the quark eikonal result always appears in  the special combination~\cite{Boughezal:2015eha}
\beq \label{Udef}
\cU_{ij}=g^4 \mu^{4 \ep} S_\ep^2 \Big[ \cI_{ii}+\cI_{jj}-2 \cI_{ij}\Big] \, .
\eeq
From Eq.~(96) of Ref.~\cite{Catani:1999ss} the soft quark pair production result for $\cI_{ij}$ is 
\beq
\cI_{ij}=\frac{(p_i.q_1 \,p_j.q_2+p_j.q_1 \,p_i.q_2-p_i.p_j \,q_1.q_2)}{(q_1.q_2)^2 \, p_i.(q_1+q_2) \, p_j.(q_1+q_2)}\, .
\eeq
Inserting this result into Eq.~(\ref{Udef}),
the result for quark pair emission can be written in terms of two functions
\beq
\cU_{ij}=g^4 \mu^{4 \ep} S_\ep^2 \Big[ \cJ^{I}_{ij}+\cJ^{II}_{ij} \Big]\, ,
\label{CUdefn}
\eeq
where
\beqn 
\label{CJIdefn}
\cJ^{I}_{ij}&=&-2 \frac{(p_i.q_1 \,p_j.q_2-p_j.q_1 \,p_i.q_2)^2}{(q_1.q_2)^2\,[p_i.(q_1+q_2)]^2\,[p_j.(q_1+q_2)]^2}\,, \\
\label{CJIIdefn}
\cJ^{II}_{ij}&=&2 \frac{p_i.p_j}{q_1.q_2 \, p_i.(q_1+q_2) \, p_j.(q_1+q_2)} \, . 
\eeqn
Using Eq.~(\ref{TRmatrixelementsq}), we see that the contribution to the soft function due to the emission of $N_F$ flavors of light quarks is given by,
\beq
\left[ \frac{\as} {2 \pi}\right]^2 \tS^{(2)}_{qq} = -T_R \,N_F \; \sum_{i<j} \bT_i \cdot \bT_j \; \cU_{ij} \; PS^{(2)} \; F^{ij} \, ,
\eeq
where $F^{ij}$ is the measurement function, Eq.~(\ref{Measurement-twoparton}).

\subsection{Soft radiation at NNLO: $gg$ emission}
\label{sec:RRgg}

The case of two-gluon emission gives rise to both Abelian and non-Abelian contributions.  
The Abelian two-gluon matrix element squared is given by the product of two single-gluon currents 
weighted by a factor of $\frac{1}{2}$. The integrations over the two emitted momenta factorize,
so that the Abelian two-gluon emission result is determined by the NLO result, and we will not consider it further.

The result for non-Abelian soft radiation has been given in Eq.~(108) of Ref.~\cite{Catani:1999ss}
and is proportional to 
\begin{equation} \label{eq:QCDfac2}
|{\cal M}^{(0)}(p_1,\dots,p_m,q_1,q_2)|^2 = -g^4 \mu^{4 \ep} \, S_\ep^2 \, C_A \langle {\cal M} |
 \left( \sum_{i,j=1}^m {\cS}_{ij}(q_1,q_2) \, \bT_i \cdot \bT_j  \right)| {\cal M}\rangle \, .
\end{equation}
The two-gluon soft function is given in Eq.~(109) of Ref.~\cite{Catani:1999ss},
\beqn
\cS_{ij}(q_1,q_2)&=&[1-\ep] \frac{(p_i.q_1 \,p_j.q_2+p_i.q_2 \,p_j.q_1)}{(q_1.q_2)^2 \, p_i.(q_1+q_2) \, p_j.(q_1+q_2)} \nonumber \\
 &-&\frac{(p_i.p_j)^2}{ 2 \, p_i.q_1 \, p_j.q_2 \, p_i.q_2 \, p_j.q_1} \Bigg[ 2-\frac{(p_i.q_1 \,p_j.q_2+p_i.q_2 \,p_j.q_1)}
{p_i.(q_1+q_2) \, p_j .(q_1+q_2)} \Bigg]\nonumber \\
 &+&\frac{p_i.p_j}{ 2 \, q_1.q_2} 
  \Bigg[\frac{2}{p_i.q_1 \, p_j.q_2}+\frac{2}{p_j.q_1 \, p_i.q_2}-\frac{1}{p_i.(q_1+q_2) \, p_j.(q_1+q_2)} \nonumber \\
 &\times& \Bigg(4+\frac{(p_i.q_1 \,p_j.q_2+p_i.q_2 \,p_j.q_1)^2}{p_i.q_1 \, p_j.q_2 \, p_i.q_2 \, p_j.q_1}\Bigg)\Bigg] .
\eeqn
Using color conservation, Eq.~(\ref{ColorIdentities}), it is clear that we only require the combination,
\beq
\cT_{ij}=g^4 \mu^{4 \ep} S_\ep^2 \Big[ \cS_{ii}+\cS_{jj}-2 \cS_{ij} \Big] \, .
\eeq
This result can further be decomposed as~\cite{Boughezal:2015eha},
\beq
\cT_{ij}=g^4 \mu^{4 \ep} S_\ep^2 \Big [(1-\ep) \cJ^{I}_{ij}+2 \cJ^{II}_{ij}+\cJ^{III}_{ij} \Big]
\label{CTdefn}
\eeq
where $\cJ^{I}_{ij},\cJ^{II}_{ij}$ are given in Eqs.~\eqref{CJIdefn} and \eqref{CJIIdefn} and 
\beqn \label{CJIIIdefn}
\cJ^{III}_{ij}&=&\left(\frac{p_i.q_1 p_j.q_2+p_j.q_1 p_i.q_2}{p_i.(q_1+q_2) \, p_j.(q_1+q_2)}-2\right) S^{(s.o.)}(p_1,p_2) \\
\cS^{(s.o.)}(p_i,p_j)&=&\frac{p_i.p_j}{q_1.q_2} \left(\frac{1}{p_i.q_1 \, p_j.q_2}+\frac{1}{p_j.q_1 \, p_i.q_2}\right)
-\frac{(p_i.p_j)^2}{p_i.q_1 \, p_i.q_2 \, p_j.q_1 \, p_j.q_2}.
\eeqn
Thus the final result for this contribution to the soft function is
\beq
\left[ \frac{\as} {2 \pi}\right]^2  \tS^{(2)}_{gg} = C_A \; \sum_{i<j} \bT_i \cdot \bT_j\;\cT_{ij}\; PS^{(2)} \;S_f \; F^{ij} 
\eeq
where $S_f=1/2$ is the statistical factor for the two final-state gluons, 
and $F^{ij}$ is the measurement function, Eq.~(\ref{Measurement-twoparton}).

%% file: calcnlo.tex
\color{black}

In this section we will make the contributions shown in
Eq.~(\ref{eq:soft1}) explicit and then assemble the complete NLO soft function.

\subsection{Phase-space sector $F^{ij}_i$}
To evaluate this contribution we first perform the change of variables,
\beq  \label{xidef}
q^i = \tauN \xi,\;\;\; q^j = \tauN \xi/x_{ij}.
\eeq
The measurement function defined in Eq.~(\ref{Measurement-oneparton}) then becomes,
\begin{eqnarray}
F_i &=& \delta(\tauN(1-\xi)) \, \theta(\tauN(1/x_{ij}-1)) \, \theta(q^j(A_{ij,k}(x_{ij},\phi_{qk})-x_{ij}) \nn \\
 &=& \frac{1}{\tauN} \delta(1-\xi) \theta(1-x_{ij}) \, \theta(A_{ij,k}(x_{ij},\phi_{qk})-x_{ij})
\end{eqnarray}
Parameterizing the phase space given in Eq.~(\ref{PhaseSpace1}) using the same set of variables, Eq.~(\ref{xidef}), and
combining the two gives,
\beqn
PS^{(1)}(i,j) F_i^{ij} = 
&=& \frac{1}{16 \pi^2} \frac{(4 \pi)^\ep}{\Gamma(1-\ep)} \; \tauN^{1-2\ep} \frac{1}{\hs_{ij}^{1-\ep}} \; 
\int \; d\xi \; \xi^{1-2\ep} \; \int_0^1 dx_{ij} \; x_{ij}^{-2+\ep} \nonumber \\
&\times & \frac{1}{N_\phi} \int_0^\pi d \phi_{qk}\; \sin^{-2\ep}\phi_{qk} \;\;
\delta(1-\xi) \; \theta(A_{ij,k}(x_{ij},\phi_{qk})-x_{ij})
\label{eq:FPS}
\eeqn
The matrix element from Eq.~(\ref{lowestordermatrixelement}) can be written as,
\beq
|M^{(1)}_{ij}|^2=- 4g^2 \; \mu^{2\ep} S_\ep \; \frac{\hs_{ij}}{q^i q^j}
 = - 4g^2  \; \mu^{2\ep} S_\ep \;\frac{\hs_{ij}}{\tauN^2 \xi^2} \, x_{ij} 
\eeq
This contribution to the NLO soft function is then,
\beqn
|M^{(1)}_{ij}|^2 PS^{(1)}(i,j) F_i^{ij}
 &=&-\Big[\frac{\as}{2 \pi}\Big] \frac{e^{\gamma_E \ep}}{\Gamma(1-\ep)} 
\frac{2}{\tauN} \, \Big[\frac{\tauN}{\mu \sqrt{\hs_{ij}}}\Big]^{-2\ep} \,
\int_0^1 \; dx_{ij} \; x_{ij}^{-1+\ep} \nonumber \\
& \times & \frac{1}{N_\phi} \int_0^\pi d \phi_{qk}\; \sin^{-2\ep}\phi_{qk}
 \; \theta(A_{ij,k}(x_{ij},\phi_{qk})-x_{ij})
\label{eq:Fiji}
\eeqn

\subsection{Phase-space sector $F^{ij}_k$}
For this contribution we use the change of variables,
\beq
q^k = \tauN \xi,\;\;\; q^i = \tauN \xi/x_{ki},
\eeq
and note that the projection that enters the matrix element ($q^j$) can be related to
these through $q^j = q^i A_{ki,j}(x_{ki},\phi_{qj})$.
The combination of measurement function and phase space is trivially related to the
expression in Eq.~(\ref{eq:FPS}) by cyclic permutation of the labels $i, j$ and $k$. The matrix
element is,
\beq
|M^{(1)}_{ij}|^2=- 4g^2 \; \mu^{2\ep} S_\ep \; \frac{\hs_{ij}}{q^i q^j}
 = - 4g^2  \; \mu^{2\ep} S_\ep \;\frac{\hs_{ij}}{\tauN^2 \xi^2}
  \, \frac{x_{ki}^2}{A_{ki,j}(x_{ki},\phi_{qj})}
\eeq
which yields the expression for this contribution,
\beqn
|M^{(1)}_{ij}|^2 PS^{(1)}(k,i) F_k^{ij}
 &=&-\Big[\frac{\as}{2 \pi}\Big] \frac{e^{\gamma_E \ep}}{\Gamma(1-\ep)} 
\frac{2}{\tauN} \, \Big[\frac{\tauN}{\mu \sqrt{\hs_{ij}}}\Big]^{-2\ep} \,
\Big(\frac{\hs_{ij}}{\hs_{ki}}\Big)^{1-\ep}
\int_0^1 dx_{ki} \; x_{ki}^{\ep} \nonumber \\ 
& \times & \frac{1}{N_\phi} \int_0^\pi d \phi_{qj}\;
\frac{\sin^{-2\ep}\phi_{qj}}{ A_{ki,j}(x_{ki},\phi_{qj})}
\theta(A_{ki,j}(x_{ki},\phi_{qj})-x_{ki}).
\label{eq:Fkik}
\eeqn

\subsection{$0$-jettiness}
The NLO soft function for $0$-jettiness is straightforward to compute
analytically and we need not resort to the numerical methods that we
will employ throughout the rest of this paper.
A short description of this calculation is given in
Appendix~\ref{sec:Results0j} for completeness.

\subsection{$1$-jettiness}

In the $1$-jettiness case, the calculation of the soft function at NLO
is more involved since one of the $\theta$-functions depends on an angle 
and therefore we have to resort to numerical integrations.
In the $1$-jettiness case we can have three
different leading-order configurations, $gg \to g$, $q\bar{q} \to g$,
and $qg \to q$, where the configuration determines the color factors that appear
in Eq.~(\ref{lowestordermatrixelement}) through the relations presented in Section~\ref{sec:color}.
For the LHC kinematics we define $\hp_1$ and $\hp_2$ as the directions of
the initial-state partons, as in Eq.~(\ref{eq:beams}), and $\hp_3$ as the direction of the
final-state parton.

The soft function is given by Eq.~(\ref{eq:soft1}), where each of the contributions is
evaluated using either Eq.~(\ref{eq:Fiji}) or Eq.~(\ref{eq:Fkik}).
The integrals are evaluated using the sector decomposition
approach~\cite{Anastasiou:2003gr,Czakon:2010td,Boughezal:2011jf}. The term $x_{ij}^{-1+\eps}$
in Eq.~\eqref{eq:Fiji} is expanded in terms of delta and plus distributions by means of the
relation
\begin{align} 
x^{-1-k\epsilon} = -\frac{\delta(x)}{k\,\epsilon} + \sum_{n=0}^\infty \frac{(-k\,\epsilon)^n}{n!} \Big[\frac{\ln^n(x)}{x}\Big]_+ \label{deltaplusx}
\end{align}
where, given a sufficiently-smooth function $f(x)$, the plus distributions are defined as
\begin{align}
\int_0^1 \mathrm{d}x \, \Big[\frac{\ln^n(x)}{x}\Big]_{\!+} f(x) = \int_0^1 \mathrm{d}x \, \frac{\ln^n(x)}{x} \big[f(x)-f(0)\big] \, .
\label{eq:plusdef}
\end{align}
After using the expansion and expanding the integrals as Laurent series in $\eps$,
the coefficients of the series are obtained by numerical integration.
This integration is straightforward and has been performed with a Fortran code
using double precision accuracy without, in this case, any additional cuts for numerical safety.\footnote{
As an additional check of the methodology and numerical integration, all of the results
in this paper have been cross-checked with two independent codes.} 
The final expression for the soft function is obtained by expanding the overall factor
$\mathcal{T}_1^{-1-2\eps}$ in terms of delta and plus distributions,
\begin{align} 
\mathcal{T}_N^{-1-k\epsilon} = -\frac{\delta(\mathcal{T}_N)}{k\,\epsilon}
 + \sum_{n=0}^\infty \frac{(-k\,\epsilon)^n}{n!} \mathcal{L}_n(\mathcal{T}_N)
\label{deltaplus}
\end{align}
where $\mathcal{L}_n(\mathcal{T}_N) = \Big(\frac{\ln^n(\mathcal{T}_N)}{\mathcal{T}_N}\Big)_+$.
Since the expansion of the term $\mathcal{T}_1^{-1-2\eps}$ using Eq.~\eqref{deltaplus} starts at
order $\eps^{-1}$, we note that it is necessary to compute all integrals up to
$\mathcal{O}(\eps)$. For our purposes, namely the factorized cross section defined in Eq.~\eqref{smalltauxsec},
we require only the $\mathcal{O}(\eps^0)$ part of the
renormalized soft function. The result is then given as the coefficients of
$\delta(\mathcal{T}_1)$ and $\mathcal{L}_n(\mathcal{T}_1)$ with $n=0,1$.

\begin{figure}
\centerline{\includegraphics[scale=0.7]{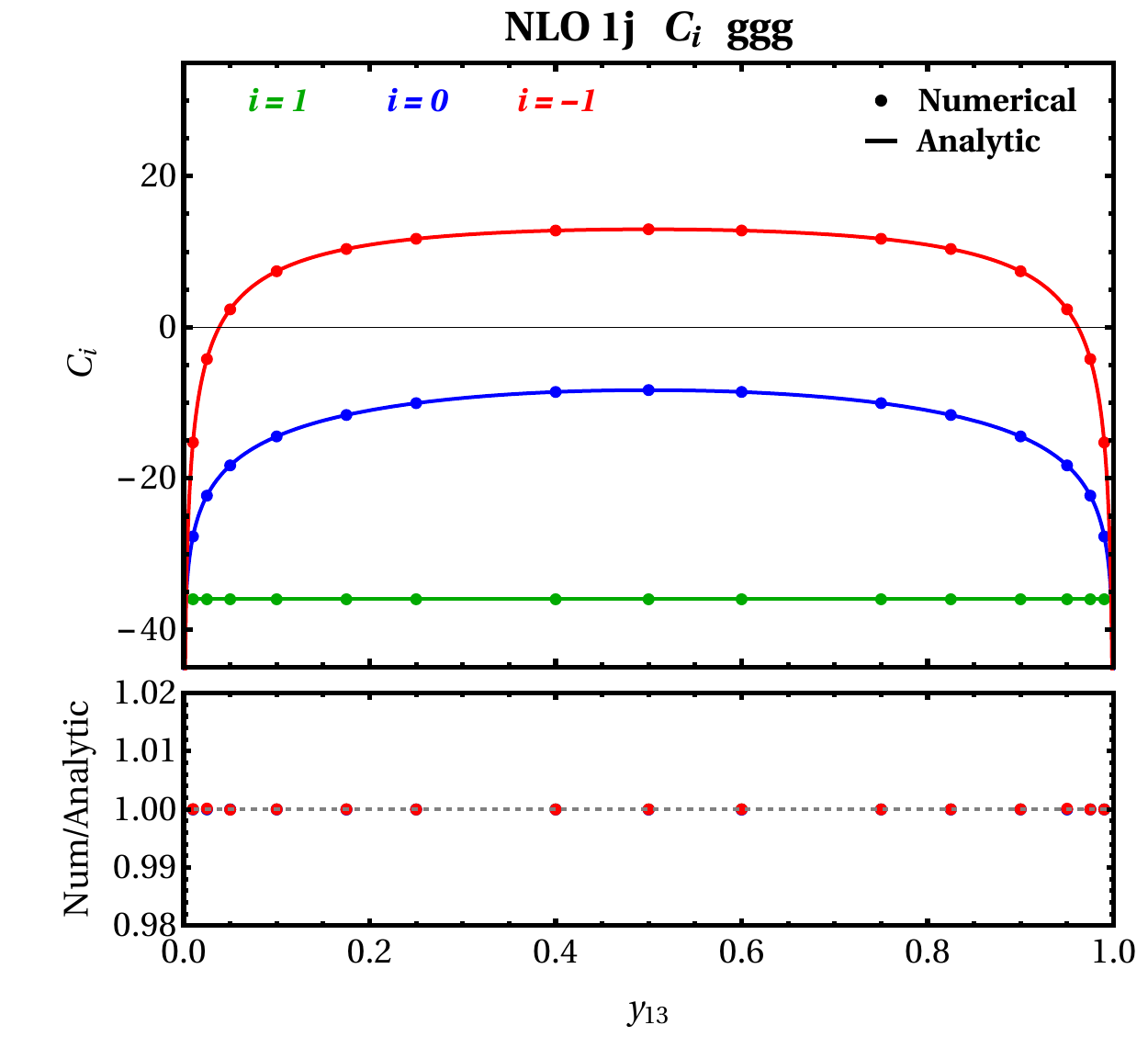} \includegraphics[scale=0.7]{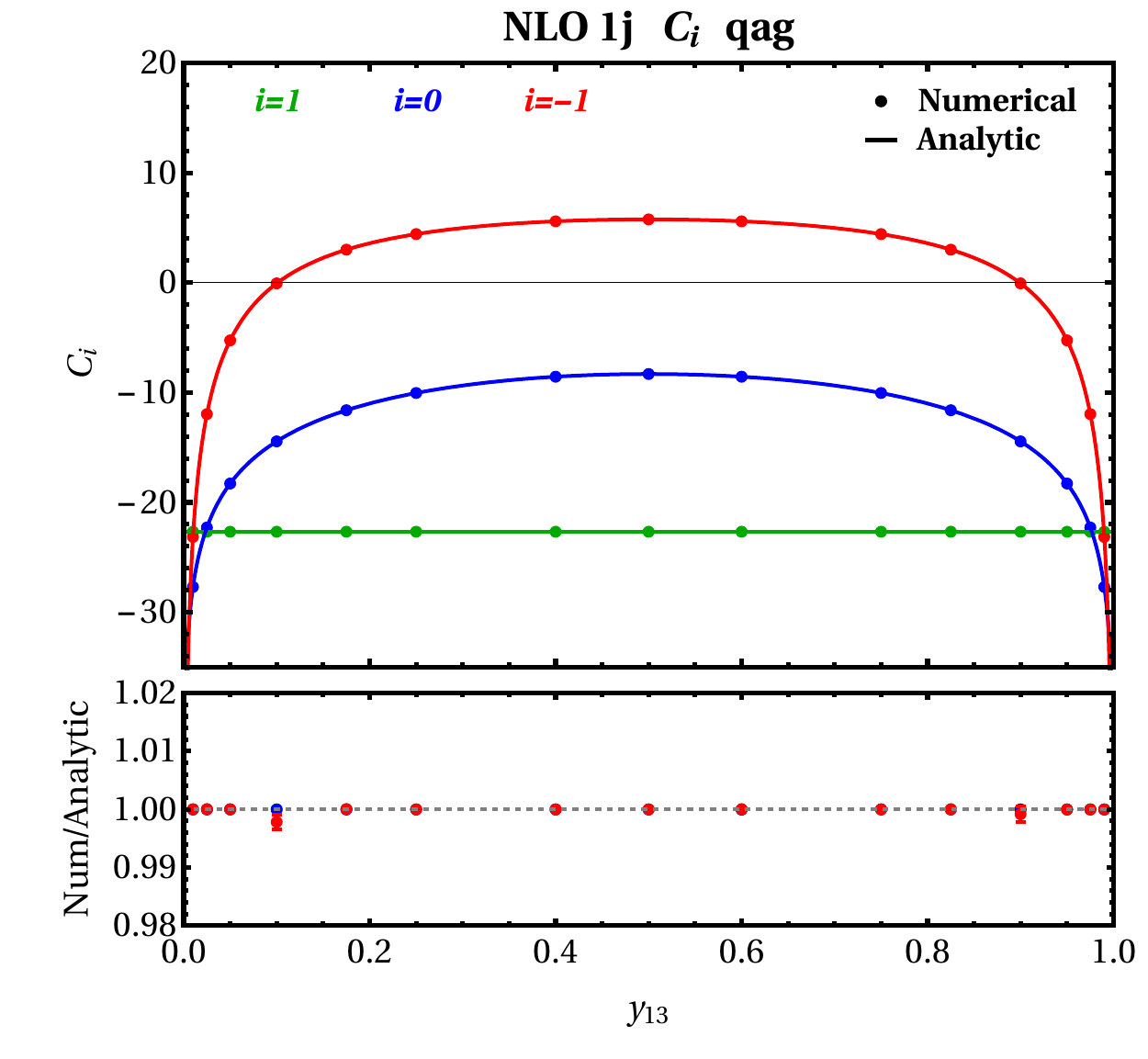}}
\centerline{\includegraphics[scale=0.7]{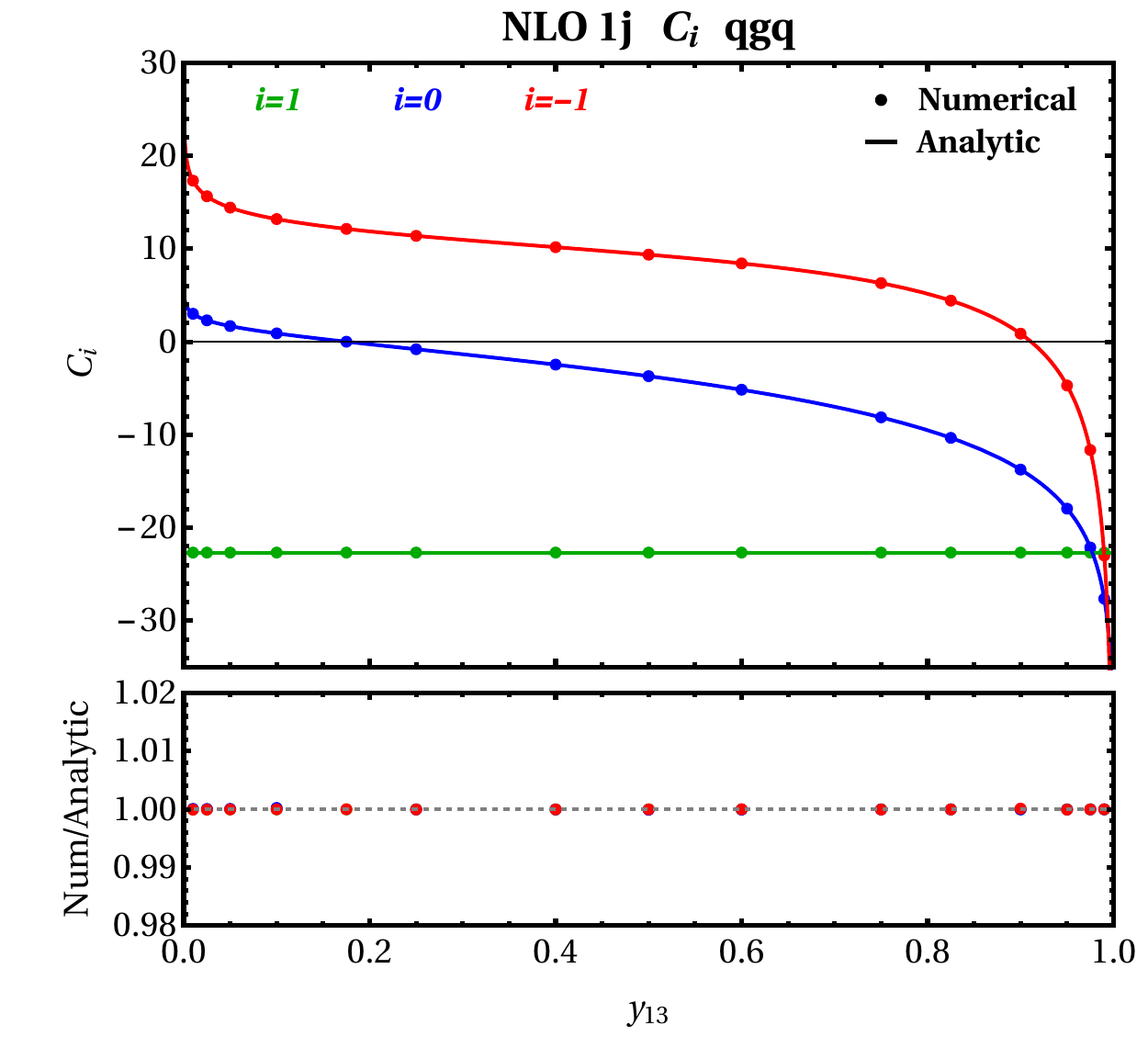}}
\caption{Numerical results for the $1$-jettiness soft function at NLO for the partonic channels $gg \to g$, $q\bar{q} \to g$, and $qg \to q$. We plot the coefficients of $\delta(\mathcal{T}_1)$, $\mathcal{L}_0(\mathcal{T}_1)$, and $\mathcal{L}_1(\mathcal{T}_1)$ as functions of $y_{13} \in [0,1]$. The known analytic results are taken from Ref.~\cite{Jouttenus:2011wh}.}
\label{NLO1j}
\end{figure}

We now present the numerical results for the $1$-jettiness NLO soft function.
We specialize our calculation for LHC kinematics, where we have $y_{12}=1$ and momentum conservation
gives $y_{23}=1-y_{13}$ so that the result is a function of $y_{13}$ alone.
We compute the soft function for the three channels ($gg \to g$, $q\bar{q} \to g$, and
$qg \to q$) with 15 different values of $y_{13}$:
\bal
\{0.01,0.025,0.05,0.1,0.175,0.25,0.4,0.5,0.6,0.75,0.825,0.9,0.95,0.975,0.99\} \, .
\end{align}
For each channel we compare the coefficients of $\delta(\mathcal{T}_1)$,
$\mathcal{L}_0(\mathcal{T}_1)$, and $\mathcal{L}_1(\mathcal{T}_1)$  with the
known analytic results in the literature~\cite{Jouttenus:2011wh}. 
We find excellent agreement with the known results for all channels and all
coefficients, as demonstrated in Fig.~\ref{NLO1j}.  The agreement is at the
level of $0.2\%$ or better.

\begin{figure}
\centerline{\includegraphics[scale=0.7]{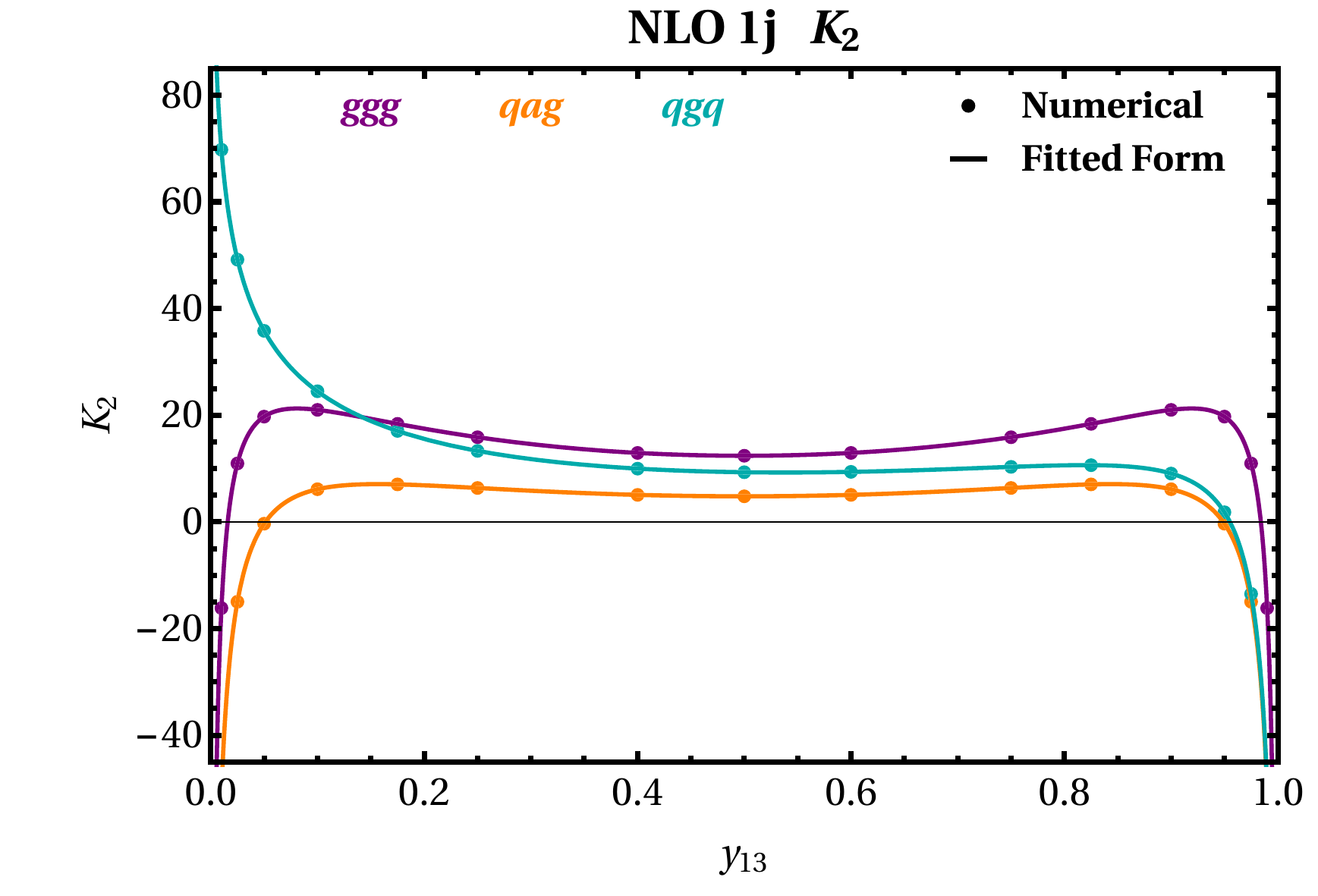}}
\caption{Numerical results for the $1$-jettiness soft function at NLO for the partonic channels $gg \to g$, $q\bar{q} \to g$, and $qg \to q$. We plot the $\mathcal{O}(\eps^2)$ coefficient of the $\epsilon$-expansion of the soft function as a function of $y_{13} \in [0,1]$. Included (but not visible) in the plot are the fitting uncertainties at the 95\% confidence level.}
\label{NLO1jeps}
\end{figure}

In Fig.~\ref{NLO1jeps} we plot the $\mathcal{O}(\eps^2)$ coefficient of the series
expansion in $\epsilon$ of the soft function before the term
$\mathcal{T}_1^{-1-2\eps}$ has been written out using Eq.~\eqref{deltaplus}.
The $\mathcal{O}(\epsilon^{2})$ term does not contribute at NLO once $\mathcal{T}_1^{-1-2\eps}$ is expanded, but instead enters the
coefficient of $\delta(\mathcal{T}_1)$ in the renormalization contribution to
the NNLO soft function, as shown in Eq.~(\ref{rensoftdef}) and explained in detail in
the next section. Since this contribution is not available in the literature,
we perform a fit of our numerical results and present the fit here for completeness:
\bal \label{fitformep2nlo}
K_{2,\text{fit}}(y_{13}) = \sum_{m,n = 0}^3 k_{(m,n)} \left[\ln{(y_{13})}\right]^m \left[\ln{(1-y_{13})}\right]^n \, .
\end{align}
The coefficients $k_{(m,n)}$ are collected in Table \ref{tablefitep2nlo}.
For the channels $gg \to g$ and $q\bar{q} \to g$ we set $k_{(m,n)} = k_{(n,m)}$ (with $m \neq n$)
since these channels are symmetrical in the initial state.

%%%%%%%%%%%%%%%%%%%%%

% NNLO1j fit

\begin{table}[ht]
\centering
\begin{tabular}{|c|r|r|r|}
\hline
\rule{0pt}{2ex} & $gg \to g$ & $q\bar{q} \to g$ & $qg \to q$\\
\hline
$k_{(0,0)}$ & $-7.991 \pm 0.056$ & $-10.594 \pm 0.053$ & $ -2.168 \pm 0.213$\\
%$k_{(0,0)}$ & $-7.99085 \pm 0.05554$ & $-10.59398 \pm 0.05258$ & $ -2.16786 \pm 0.21314$\\
\hline
$k_{(1,0)}$ & $-14.163 \pm 0.045$ & $-11.348 \pm 0.047$ & $ -6.467 \pm 0.196$\\
%$k_{(1,0)}$ & $-14.16314 \pm 0.04500$ & $-11.34817 \pm 0.04744$ & $ -6.46658 \pm 0.19600$\\
\hline
$k_{(2,0)}$ & $1.532 \pm 0.016$ & $0.350 \pm 0.015$ & $ 2.099 \pm 0.066$\\
%$k_{(2,0)}$ & $1.53228 \pm 0.01617$ & $0.34986 \pm 0.01542$ & $ 2.09862 \pm 0.06576$\\
\hline
$k_{(3,0)}$ & $1.086 \pm 0.002$ & $1.022 \pm 0.002$ & $ 0.025 \pm 0.007$\\
%$k_{(3,0)}$ & $1.08575 \pm 0.00175$ & $1.02160 \pm 0.00169$ & $ 0.02524 \pm 0.00745$\\
\hline
$k_{(0,1)}$ & $-14.163 \pm 0.045$ & $-11.348 \pm 0.047$ & $ -8.790 \pm 0.190$\\
%$k_{(0,1)}$ & $-14.16314 \pm 0.04500$ & $-11.34817 \pm 0.04744$ & $ -8.79038 \pm 0.19004$\\
\hline
$k_{(0,2)}$ & $1.532 \pm 0.016$ & $0.350 \pm 0.015$ & $ 0.494 \pm 0.062$\\
%$k_{(0,2)}$ & $1.53228 \pm 0.01617$ & $0.34986 \pm 0.01542$ & $ 0.49410 \pm 0.06168$\\
\hline
$k_{(0,3)}$ & $1.086 \pm 0.002$ & $1.022 \pm 0.002$ & $ 1.009 \pm 0.007$\\
%$k_{(0,3)}$ & $1.08575 \pm 0.00175$ & $1.02160 \pm 0.00169$ & $ 1.00936 \pm 0.00680$\\
\hline
\end{tabular}
\caption{Non-zero coefficients of the numerical fit of Eq.~\eqref{fitformep2nlo} for the three partonic configurations $gg \to g$, $q\bar{q} \to g$, and $qg \to q$. Coefficients not shown here are understood to be zero.}
\label{tablefitep2nlo}
\end{table}

%%%%%%%%%%%%%%%%%%%%%

%% file: calcnnlo.tex
From Eq.~\eqref{rensoftdef} the
renormalized $N$-jettiness soft function at NNLO, $S^{(2)}(\mathcal{T}_N)$, is
\bal \label{softnnlo}
S^{(2)}(\mathcal{T}_N) &= -\frac{\beta_0}{2 \eps} \, \tS^{(1)}(\mathcal{T}_N) + \tS^{(2)}(\mathcal{T}_N) \, .
\end{align}
We have already discussed the calculation of the first term in
Eq.~\eqref{softnnlo} in the previous section, so we 
now study the second term in more
detail. At NNLO three types of corrections arise: two-loop virtual
corrections, one-loop corrections with one real gluon emission
(``real-virtual''), and double-real corrections, consisting of the
emission of $q \bar{q}$ and $gg$ pairs. In the latter case, the
emission of two gluons is made up of two diagrammatic contributions:
an Abelian contribution, i.e.~two single-gluon currents, and a
non-Abelian contribution, i.e.~emission of a $gg$ pair through a
three-gluon vertex. In light of this, and since the virtual two-loop
corrections vanish in dimensional regularization, we can rewrite the
term $\tS^{(2)}(\mathcal{T}_N)$ as the sum of four
non-vanishing contributions:
\bal \label{abnabsoftfunc}
\tS^{(2)}(\mathcal{T}_N) = \tS^{(2)}_{ab}(\mathcal{T}_N) + \tS^{(2)}_{RV}(\mathcal{T}_N)
 + \tS^{(2)}_{gg}(\mathcal{T}_N) + \tS^{(2)}_{q\overline{q}}(\mathcal{T}_N)
\end{align}
with $\tS^{(2)}_{gg}(\mathcal{T}_N)$ indicating the non-Abelian
part of the double-real $gg$ corrections. The Abelian term
$\tS^{(2)}_{ab}(\mathcal{T}_N)$ can be obtained directly from
the NLO soft function thanks to well-known exponentiation
theorems~\cite{Frenkel:1984pz,Gatheral:1983cz}.  
The calculation of the real-virtual contribution $\tS^{(2)}_{RV}(\mathcal{T}_N)$ is conceptually
identical to the calculation of the NLO soft function and has been described in Section~\ref{sec:RV}.
We therefore spend the remainder of this section discussing the calculation of the double-real corrections
represented by $\tS^{(2)}_{gg}(\mathcal{T}_N)$ and $ \tS^{(2)}_{q\overline{q}}(\mathcal{T}_N)$.
The decomposition of the relevant eikonal matrix elements into a set of basis integrals is given in
Sections~\ref{sec:RRqq} and~\ref{sec:RRgg} for the $q\bar q$ and $gg$ cases respectively.
We will detail the transformations necessary to render the basis integrals,
obtained by combining the double-emission phase space with the eikonal integrands, amenable
to numerical evaluation.  We will provide a number of illustrative examples
of integrals, leaving a complete enumeration of all contributions to Appendix~\ref{DRME}.

%% file: calcrr.tex
We first identify a common overall factor that is associated with the total angular volume, 
in the phase-space parametrization of Eq.~(\ref{PhaseSpace2}), 
\beq \label{Nidentity}
\frac{1}{N_\beta N_{\phiu} N_{\phid}} = -\frac{\ep}{2^{2 \ep} \pi^2} \frac{\Gamma(1-\ep)^2}{\Gamma(1-2 \ep)}   = 
-\frac{\ep}{2^{2 \ep} \pi^2} 
\Big[\frac{\Gamma(1-\ep)}{(4 \pi)^\ep}\Big]^2\frac{B_{RR}}{S_\ep^2} \, ,
\eeq
where
\begin{equation} \label{BRR}
B_{RR} = \frac{e^{2 \ep \gamma_E}}{\Gamma(1-2 \ep)}
=1 - \frac{\pi^2}{3} \epsilon^2-\frac{8}{3} \zeta_3 \epsilon^3+\frac{\pi^4}{90} \epsilon^4 + O(\ep^5) \, ,
\end{equation}
and $S_\ep$ is given in Eq.~(\ref{Sepdefinition}).
We will always be able to trivially rescale one of the angular integrals
so that the range of integration is between zero and one,
\beq
\int_0^\pi d\phi \, \sin^{-2\ep}\phi = \pi \int_0^1 dx_\phi \, \sin^{-2\ep}\phi\, ,
\label{eq:phirescaling}
\eeq
with $\phi=\pi x_\phi$, and the integral over $\beta$ can be recast in similar
fashion,
\beq \label{betaintegral}
\int_0^\pi d \beta \, \sin^{-1-2\ep}\beta
 =\frac{1}{2^{1+2\ep}} \int_0^1 dx_\beta \; x_\beta^{-1-\ep}\, (1-x_\beta)^{-1-\ep} \, ,
\eeq
by performing the change of variable $\cos{\beta} = 1-2 x_{\beta}$.
We thus rewrite Eq.~(\ref{PhaseSpace2}) as,
\beqn \label{PhaseSpace2simp}
PS^{(2)}(i,j)&=&-\frac{\ep}{2^{9+4\ep} \, \pi^4} \frac{B_{RR}}{S_\ep^2} \Big(\frac{1}{\hs_{ij}}\Big)^{2-2\ep}
\int dq_1^i dq_1^j dq_2^i dq_2^j \, \left[q_1^i q_1^j q_2^i q_2^j\right]^{-\ep} \nonumber \\
&\times& \int_0^1 dx_\beta \; x_\beta^{-1-\ep}\, (1-x_\beta)^{-1-\ep}
         \int_0^1 d x_{\phiu} \sin^{-2 \ep}\phiu
         \int_0^\pi \frac{d \phid}{\pi} \sin^{-2 \ep}\phid \, ,
\eeqn
where we have written the remaining $\phid$ integral in a form that
anticipates the Jacobian factor that will eventually emerge as in
Eq.~(\ref{eq:phirescaling}).  Finally, in our eventual evaluation of
the $x_\beta$ integral we will replace this form with a simpler one in
which singularities can only appear at $x_\beta = 0$,\footnote{An
alternative procedure~\cite{Anastasiou:2005qj} is to split the
integration range in half, into $[0,\frac{1}{2})$ and
$[\frac{1}{2},1]$, and then remap the second range so that all
singularities are at $x_\beta = 0$.  This method has been adopted in
one of our codes.}
\beqn
\int_0^1 dx_\beta \, x_\beta^{-1-\ep} (1-x_\beta)^{-1-\ep} f(x_\beta)
 &=& \int_0^1 dx_\beta \, \Big [x_\beta^{-1-\ep} (1-x_\beta)^{-\ep} +x_\beta^{-\ep} (1-x_\beta)^{-1-\ep}\Big] f(x_\beta)
\nonumber \\
 &=& \int_0^1 \frac{dx_\beta}{x_\beta} \, \Big [x_\beta (1-x_\beta)\Big]^{-\ep}
\Big[f(x_\beta)+f(1-x_\beta)\Big] \, .
\label{eq:xbetatrick}
\eeqn
As before, we compute the integrals numerically using sector decomposition and
the plus-distribution expansion given in Eq.~(\ref{deltaplusx}).

As in the NLO case, the choice of the phase-space parametrization will
depend upon which case is picked out by the measurement function.  We
will discuss each of the cases described in
Section~\ref{sec:measurement} in turn.  The combination of the matrix
elements, phase space, and measurement function for each case will
expressed as,
\beq
\MF \, \cJ^X_{ij} \, PS^{(2)} \, F^{ij}_{ab} \equiv \cN_{ij} \, I^{(2),ij,X}_{ab} \, ,
\label{eq:Idefinition}
\eeq
where the coupling-associated factor $\MF$ is defined in Eq.~(\ref{MFdefn}).  In this equation $X$ labels the
division of the eikonal approximations, Eqs.~\eqref{CJIdefn}, \eqref{CJIIdefn}, and \eqref{CJIIIdefn}, and 
$a$, $b$ the different cases.
We have extracted an overall factor that will be universal across all contributions,
\beq \label{Normalizationdefinition}
\cN_{ij}=\Big(\frac{\as}{2 \pi}\Big)^2 \, \frac{1}{\tauN}
 \, \Big[\frac{\tauN}{\mu \sqrt{\hs_{ij}}}\Big]^{-4\ep} \, .
\eeq
\subsubsection{Case 1: $F^{ij}_{ii}$}
The appropriate change of variables for this case is,
\beq
q_1^i=\tauN \xi,\;
q_1^j=\frac{\tauN \xi}{s},\;
q_2^i=\tauN (1-\xi),\;
q_2^j=\frac{\tauN (1-\xi)}{t} \, .
\label{Denominators1}
\eeq
We then have,
\beqn
\int dq_1^i dq_1^j dq_2^i dq_2^j \, F^{ij}_{ii} &=& \int dq_1^i dq_1^j dq_2^j \,
 \theta(\meas{q_1}{j}-\meas{q_1}{i}) \theta(\meas{q_1}{k}-\meas{q_1}{i}) \,
 \theta(\meas{q_2}{j}-\meas{q_2}{i}) \theta(\meas{q_2}{k}-\meas{q_2}{i})\\
 &=& \tauN^3 \int d\xi ds dt \, \frac{\xi(1-\xi)}{s^2t^2} \,
 \theta(A_{ij,k}(s,\phiu)-s) \, \theta(A_{ij,k}(t,\phid)-t) \, , \nonumber
\eeqn
so that the combination of the phase space in Eq.~(\ref{PhaseSpace2simp}) with the measurement
function $F^{ij}_{ii}$ yields,
\beqn \label{Twophasespacecase1untransformed}
PS^{(2)}(i,j) \, F^{ij}_{ii} &=&
-\frac{\ep}{2^{9+4\ep} \pi^4} \tauN^{3-4\ep} \frac{B_{RR}}{S_\ep^2} \Big(\frac{1}{\hs_{ij}}\Big)^{2-2\ep}
\int_0^1 d\xi ds dt [\xi (1-\xi)]^{1-2 \ep} \frac{1}{s^{2-\ep} t^{2-\ep}} \nonumber \\
&\times& \int_0^1 d x_\beta \; x_\beta^{-1-\ep}\; (1-x_\beta)^{-1-\ep}
         \int_0^1 d x_{\phiu} \; \sin^{-2 \ep}\phiu
         \int_0^\pi \frac{d\phid}{\pi} \; \sin^{-2 \ep}\phid \nonumber \\
&\times& \theta(A_{ij,k}(s,\phiu)-s)\theta(A_{ij,k}(t,\phid)-t) \, ,
\eeqn
where $\phiu=\pi x_{\phiu}$.  In the simplest case we will be
able to perform a rescaling of $\phid$ as in Eq.~(\ref{eq:phirescaling}).  However, in general
this is not true.

The reason for the additional complication is the presence of the factor $q_1 \cdot q_2$ in
the eikonal factors $\cT_{ij}$ and $\cU_{ij}$.  Although the Sudakov decomposition is very
convenient for describing the other dot products, the expression for this one is considerably
more complicated,
\beq
2 q_1.q_2 = \frac{1}{\hs_{ij}} \frac{\tauN^2 \xi  (1-\xi)}{s t}
  \Big[(\sqrt{s}-\sqrt{t})^2 +4 \zud \sqrt{s t}\Big] \, ,
\eeq
where $\zud=\frac{1}{2}(1-\cos\phiud)$ and $\phiud$ is the angle between $q_1$ and $q_2$.
A method for handling this denominator has been outlined in
Refs.~\cite{Boughezal:2015eha,Boughezal:2013uia} and we follow this strategy here.
We map $\zud$ to a new variable $\lambda$ through the relation,
\beq \label{zud}
\zud = \frac{ (\sqrt{s}-\sqrt{t})^2 (1-\lambda)}{(\sqrt{s}-\sqrt{t})^2 +4 \lambda \sqrt{st}}\,,\quad\quad
1-\zud = \frac{ (\sqrt{s}+\sqrt{t})^2 \lambda}{(\sqrt{s}-\sqrt{t})^2 +4 \lambda \sqrt{st}}\, ,
\eeq
or the inverse,
\beq
\lambda = \frac{(\sqrt{s}-\sqrt{t})^2\;(1-\zud)}{(\sqrt{s}-\sqrt{t})^2+4 \zud\sqrt{st}}\, , \quad\quad
\lambda(1-\lambda)=\frac{((s-t)^2\;\zud (1-\zud)}{[(\sqrt{s}-\sqrt{t})^2+4 \zud \sqrt{st} ]^2} \, .
\eeq
The Jacobian associated with the transformation from $\zud$ to $\lambda$ is, 
\beq
\frac{d \zud}{d \lambda} = -\frac{(s-t)^2}{[(\sqrt{s}-\sqrt{t})^2 +4 \lambda \sqrt{st}]^2}\, .
\eeq
In terms of the new variable $\lambda$ we have,
\beq \label{q1Dq2_transformed}
(\sqrt{s}-\sqrt{t})^2 +4 \zud \sqrt{st}
 = \frac{(s-t)^2}{(\sqrt{s}-\sqrt{t})^2 +4 \lambda \sqrt{st}} \, ,
\eeq
so that
\beq
2 q_1.q_2 = \frac{1}{\hs_{ij}} \frac{\tauN^2 \xi(1-\xi)}{st}
 \frac{(s-t)^2}{(\sqrt{s}-\sqrt{t})^2 +4 \lambda \sqrt{st}} \, .
\eeq
Further, from Eq.~(\ref{zud}) we have that,
\beq
\sin^2 \phiud = 4 \zud (1-\zud)
 = 4 \frac{(s-t)^2 \lambda(1-\lambda)}{[(\sqrt{s}-\sqrt{t})^2 +4 \lambda \sqrt{st}]^2} \, .
\label{eq:phiud}
\eeq
We observe that by means of this change of variable the collinear singularity
$q_1 \cdot q_2 \to 0$ has been mapped to $s \to t$. 
In the presence of this denominator (and associated factor of $\zud$) we must be careful to ensure
that the relevant angles are handled appropriately.  It is convenient to perform the integration in
a rotated frame, c.f. Appendix~\ref{sec:rotation}.  Following that logic, we replace the measure
$d\beta \, d\phid$ with $d\betaud \, d\phiud$ and, from Eq.~(\ref{eq:anglerelation}),
relate the angle $\phid$ to $\phiu$, $\betaud$, and $\phiud$ through,
\beq
\cos\phid=\cos\phiu \cos\phiud-\sin\phiu \sin\phiud\cos\betaud \, .
\label{eq:phid}
\eeq
The final integral in Eq.~(\ref{Twophasespacecase1untransformed}) then becomes,
\beqn \label{phiudintegral}
\int_0^\pi \frac{d\phiud}{\pi} \, \sin^{-2\ep}\phiud &=& \frac{1}{\pi}
 \int_{-1}^1 d(\cos\phiud) \, [ \sin^2\phiud ]^{-\frac{1}{2}-\ep} \nn \\
 &=& \frac{2}{\pi} \int_{0}^1 d\zud \left[ 4\zud(1-\zud) \right]^{-\frac{1}{2}-\ep} \nn \\
 &=& \frac{2}{\pi} \int_0^1 d\lambda \, \frac{(s-t)^2}{[(\sqrt{s}-\sqrt{t})^2 +4 \lambda \sqrt{st}]^2}
  \left[ \frac{4(s-t)^2 \lambda(1-\lambda)}{[(\sqrt{s}-\sqrt{t})^2 +4 \lambda \sqrt{st}]^2} \right]^{-\frac{1}{2}-\ep} \nn \\
 &=& \frac{2^{-2\ep}}{\pi} \int_0^1 d\lambda \, \left[ \lambda(1-\lambda) \right]^{-\frac{1}{2}-\ep} 
  \, \frac{|s-t|^{1-2\ep}}{[(\sqrt{s}-\sqrt{t})^2 +4 \lambda \sqrt{st}]^{1-2\ep}} \nn \\
 &=& 2^{-2\ep} \int_0^1 dx_{\phid} \, \left[ \lambda(1-\lambda) \right]^{-\ep} 
  \, \frac{|s-t|^{1-2\ep}}{[(\sqrt{s}-\sqrt{t})^2 +4 \lambda \sqrt{st}]^{1-2\ep}},
\eeqn
where $\lambda=\sin^2(\pi x_{\phid}/2)$.   Thus the final form for the phase space is,
\beqn \label{Twophasespacecase1transformed}
PS^{(2)}(i,j) \, F^{ij}_{ii} &=&
-\frac{\ep}{2^{9+6\ep} \pi^4} \tauN^{3-4\ep} \frac{B_{RR}}{S_\ep^2} \Big(\frac{1}{\hs_{ij}}\Big)^{2-2\ep}
\int_0^1  d\xi ds dt [\xi (1-\xi)]^{1-2 \ep} \frac{1}{s^{2-\ep} t^{2-\ep}} \nonumber \\
&\times&\int_0^1 dx_{\phiu} \sin^{-2 \ep}\phiu
         \int_0^1 \, d x_{\betaud} \; x_{\betaud}^{-1-\ep}\; (1-x_{\betaud})^{-1-\ep} \nonumber \\
&\times&\int_0^1 dx_{\phid} \, (\lambda (1-\lambda))^{-\ep}
	 \frac{|s-t|^{1-2\ep}}{[(\sqrt{s}-\sqrt{t})^2 +4 \lambda \sqrt{st}]^{1-2\ep}} \nn \\
&\times&\theta(A_{ij,k}(s,\phiu)-s)\theta(A_{ij,k}(t,\phid)-t) \, ,
\eeqn
where $\phiu=\pi x_{\phiu}$, $\cos\betaud=1-2x_{\betaud}$, $\phiud$ is defined through Eq.~(\ref{eq:phiud}),
and $\phid$ can be obtained from Eq.~(\ref{eq:phid}).

The matrix elements that must be evaluated to compute the double-real contributions for this case
are given in Appendix~\ref{MEcase1}.  They are expressed in terms of the new variables
$\xi$, $s$, and $t$ and to aid in the evaluation of contribution $III$ they have been
further subdivided into integrals that are simpler to compute.  

To illustrate a further
complication that arises for these integrals, consider the evaluation of contribution $I$.
Combining the phase space from Eq.~(\ref{Twophasespacecase1transformed}) with the matrix element 
in Eq.~(\ref{eq:MEcase1I}) we obtain,
\beqn
I^{(2),ij,I}_{ii} &=& B_{RR} \, 2^{-6\ep} \ep
\int_0^1  d\xi ds dt [\xi (1-\xi)]^{1-2 \ep} \; (st)^{\ep} |s-t|^{-1-2\ep}\nonumber \\
&\times&\int_0^1 dx_{\phiu} \sin^{-2 \ep}\phiu
         \, \int_0^1 \, d x_\beta \; x_\beta^{-1-\ep}\; (1-x_\beta)^{-1-\ep} \nonumber \\
&\times& \int_0^1 dx_{\phid} (\lambda (1-\lambda))^{-\ep} \, \frac{[(\sqrt{s}-\sqrt{t})^2 +4 \lambda \sqrt{st}]^{1+2 \ep}}
 {[\xi t +(1-\xi) s ]^2}\nonumber \\
&\times& \theta(A_{ij,k}(s,\phi_{1})-s)\theta(A_{ij,k}(t,\phi_{2})-t),
\eeqn
where $B_{RR}$ is given in Eq.~(\ref{BRR}) and we have removed an overall factor of $\cN_{ij}$
in the definition of $I^{(2),ij,I}_{ii}$, c.f. Eq.~(\ref{eq:Idefinition}).
In order to be able to perform the integration, we have to handle the line singularity
associated with the $|s-t|^{-1-2\eps}$ term. Following Ref.~\cite{Boughezal:2015eha},
we do so by partitioning the integral into two contributions by means of the identity,
\bal
1 = \theta(s-t) + \theta(t-s) \,.
\label{eq:stpartition}
\end{align}
In the $s>t$ sector we then perform the change of variables
\bal
s = x_2, \hspace{2cm} t = x_2 \,(1-x_3) \, ,
\label{eq:sGTt}
\end{align}
while in the $t>s$ sector we have
\bal
t = x_2, \hspace{2cm} s = x_2 \,(1-x_3) \,,
\label{eq:tGTs}
\end{align}
such that, in general, all singularities are located at $x_2 = 0$ and $x_3 = 0$.
The only complication is that, for the integrand $\cJ^{IIIb}_{ij}$, this procedure also
yields singularities at $x_3 = 1$.  However, these are simple to handle by using a
simplification similar to the one in Eq.~(\ref{eq:xbetatrick}), but carried out for $x_3$.
These transformations are sufficient to treat all of the singularities in this case.

\subsubsection{Case 2: $F^{ij}_{ij}$}
The change of variables for this case is,
\beq
q_1^i=\tauN \xi,\;
q_1^j=\frac{\tauN \xi}{s},\;
q_2^j=\tauN (1-\xi),\;
q_2^i=\frac{\tauN (1-\xi)}{t} \, .
\label{Denominators2}
\eeq
In combination with the measurement function, this results in a phase-space
parametrization that is very similar to the simplest one for case 1,
\beqn \label{Twophasespacecase2untransformed}
PS^{(2)}(i,j) \, F^{ij}_{ij} &=&
-\frac{\ep}{2^{9+4\ep} \pi^4} \tauN^{3-4\ep} \frac{B_{RR}}{S_\ep^2} \Big(\frac{1}{\hs_{ij}}\Big)^{2-2\ep}
\int_0^1 d\xi ds dt [\xi (1-\xi)]^{1-2 \ep} \frac{1}{s^{2-\ep} t^{2-\ep}} \nonumber \\
&\times& \int_0^1 d x_\beta \; x_\beta^{-1-\ep}\; (1-x_\beta)^{-1-\ep}
         \int_0^1 d x_{\phiu} \; \sin^{-2 \ep}\phiu
         \int_0^1 d x_{\phid} \; \sin^{-2 \ep}\phid \nonumber \\
&\times& \theta(A_{ij,k}(s,\phiu)-s)\theta(A_{ji,k}(t,\phid)-t) \, ,
\eeqn
where $\phiu=\pi x_{\phiu}$ and $\phid=\pi x_{\phid}$.  The remaining invariant
that enters the matrix elements (given in Appendix~\ref{MEcase2}) becomes,
\beq
2 q_1.q_2 = \frac{\tauN^2 \xi  (1-\xi)}{s t \hs_{ij}} \Big[(1-\sqrt{st})^2 +4 \zud \sqrt{s t}\Big],
\eeq
where $\zud=\frac{1}{2}(1-\cos\phiud)$ and $\phiud$ is obtained from, c.f.
Eq.~(\ref{eq:anglerelation12}),
\beq
\cos\phiud=\cos\phiu \cos\phid +\sin\phiu \sin\phid\cos\beta \, .
\eeq
This does not require any further reparametrization of the phase space because factors of $1/q_1 \cdot q_2$
do not lead to singularities as $s, t \to 0$.

\subsubsection{Case 3: $F^{ij}_{ik}$}
This case uses the following change of variables, based on a Sudakov expansion with respect to $i$ and $k$,
\beq
q_1^i=\tauN \xi,\;
q_1^k=\frac{\tauN \xi}{s},\;
q_2^k=\tauN (1-\xi),\;
q_2^i=\frac{\tauN (1-\xi)}{t} \, .
\label{Denominators3}
\eeq
The phase-space parametrization becomes,
\beqn \label{Twophasespacecase3untransformed}
PS^{(2)}(i,k) \, F^{ij}_{ik} &=&
-\frac{\ep}{2^{9+4\ep} \pi^4} \tauN^{3-4\ep} \frac{B_{RR}}{S_\ep^2} \Big(\frac{1}{\hs_{ik}}\Big)^{2-2\ep}
\int_0^1 d\xi ds dt [\xi (1-\xi)]^{1-2 \ep} \frac{1}{s^{2-\ep} t^{2-\ep}} \nonumber \\
&\times& \int_0^1 d x_\beta \; x_\beta^{-1-\ep}\; (1-x_\beta)^{-1-\ep}
         \int_0^1 d x_{\phiu} \; \sin^{-2 \ep}\phiu
         \int_0^1 d x_{\phid} \; \sin^{-2 \ep}\phid \nonumber \\
&\times& \theta(A_{ik,j}(s,\phiu)-s)\theta(A_{ki,j}(t,\phid)-t) \, ,
\eeqn
where $\phiu=\pi x_{\phiu}$ and $\phid=\pi x_{\phid}$.  Since one of the emitting
lines is no longer one of the vectors in the Sudakov expansion, more invariants must be defined.  The
remaining quantities are,
\beqn
  q_1^j &=&\frac{\tauN \xi}{s} \, A_{ik,j}(s,\phiu)\, ,  \quad \quad
  q_2^j =\frac{\tauN (1-\xi)}{t} \, A_{ki,j}(t,\phid)  \nonumber \\
2 q_1.q_2&=&\frac{1}{\hs_{ik}}\frac{\tauN^2 \xi  (1-\xi)}{s t} \Big[(1-\sqrt{st})^2 +4 \zud \sqrt{s t}\Big],
\eeqn
where $\zud=\frac{1}{2}(1-\cos\phiud)$ and $\phiud$ is again defined through Eq.~(\ref{eq:anglerelation12}).
There are no further singularities to disentangle in this case, for which all matrix elements are
specified in Appendix~\ref{MEcase3}.

\subsubsection{Case 4: $F^{ij}_{kk}$}
We perform the Sudakov expansion with respect to base vectors $i$ and $k$ and then the following
change of variables,
\beq
q_1^k=\tauN \xi,\;
q_1^i=\frac{\tauN \xi}{s},\;
q_2^k=\tauN (1-\xi),\;
q_2^i=\frac{\tauN (1-\xi)}{t} \, .
\label{Denominators4}
\eeq
After this transformation the phase space parametrization is,
\beqn \label{Twophasespacecase4untransformed}
PS^{(2)}(i,k) \, F^{ij}_{kk} &=&
-\frac{\ep}{2^{9+4\ep} \pi^4} \tauN^{3-4\ep} \frac{B_{RR}}{S_\ep^2} \Big(\frac{1}{\hs_{ik}}\Big)^{2-2\ep}
\int_0^1 d\xi ds dt [\xi (1-\xi)]^{1-2 \ep} \frac{1}{s^{2-\ep} t^{2-\ep}} \nonumber \\
&\times& \int_0^1 d x_\beta \; x_\beta^{-1-\ep}\; (1-x_\beta)^{-1-\ep}
         \int_0^1 d x_{\phiu} \; \sin^{-2 \ep}\phiu
         \int_0^1 d x_{\phid} \; \sin^{-2 \ep}\phid \nonumber \\
&\times& \theta(A_{ki,j}(s,\phiu)-s)\theta(A_{ki,j}(t,\phid)-t) \, .
\eeqn
The other invariants that enter the matrix elements are,
\beqn
q_1^j &=&\frac{\tauN \xi }{s} \, A_{ki,j}(s,\phiu)\,,\quad \quad  
q_2^j =\frac{\tauN (1-\xi)}{t} \, A_{ki,j}(t,\phid)  \nonumber \\
2 q_1.q_2&=&\frac{1}{\hs_{ik}}\frac{\tauN^2 \xi  (1-\xi)}{s t} \Big[(\sqrt{s}-\sqrt{t})^2 +4 \zud \sqrt{s t}\Big].
\eeqn
Since this denominator takes the same form as the one considered in case 1, it must be handled in a similar
manner.  Explicitly we have,
\beqn \label{Twophasespacecase4transformed}
PS^{(2)}(i,k) \, F^{ij}_{kk} &=&
-\frac{\ep}{2^{9+6\ep} \pi^4} \tauN^{3-4\ep} \frac{B_{RR}}{S_\ep^2} \Big(\frac{1}{\hs_{ik}}\Big)^{2-2\ep}
\int_0^1  d\xi ds dt [\xi (1-\xi)]^{1-2 \ep} \frac{1}{s^{2-\ep} t^{2-\ep}} \nonumber \\
&\times&\int_0^1 dx_{\phiu} \sin^{-2 \ep}\phiu
         \int_0^1 \, d x_{\betaud} \; x_{\betaud}^{-1-\ep}\; (1-x_{\betaud})^{-1-\ep} \nonumber \\
&\times&\int_0^1 dx_{\phid} \, (\lambda (1-\lambda))^{-\ep}
	 \frac{|s-t|^{1-2\ep}}{[(\sqrt{s}-\sqrt{t})^2 +4 \lambda \sqrt{st}]^{1-2\ep}} \nn \\
&\times&\theta(A_{ki,j}(s,\phiu)-s)\theta(A_{ki,j}(t,\phid)-t)\, ,
\eeqn
where the definitions of all the angles are taken over from case 1.

The matrix elements for this case are given in Appendix~\ref{MEcase3}.  In the calculation of
the contribution $\cJ^I_{ij}$ an additional subtlety arises.  Consider the evaluation of the $s>t$ sector
that appears after the partitioning of Eq.~(\ref{eq:stpartition}).  Following the change of variables
in Eq.~(\ref{eq:sGTt}) we have,
\beqn
I^{(2),ij,I}_{kk} &=& B_{RR} \, 2^{-6\ep} \left(\frac{\hs_{ik}}{\hs_{ij}}\right)^{2\ep} \ep
\int_0^1  d\xi dx_2 dx_3 [\xi (1-\xi)]^{1-2 \ep} \; x_2^{2\ep} x_3^{-1-2\ep} (1-x_3)^{2+\ep}\nonumber \\
&\times&\int_0^1 dx_{\phiu} \sin^{-2 \ep}\phiu
         \, \int_0^1 \, d x_\beta \; x_\beta^{-1-\ep}\; (1-x_\beta)^{-1-\ep} \,
         \int_0^1 dx_{\phid} (\lambda (1-\lambda))^{-\ep} \nonumber \\
&\times& \frac{[(1-\sqrt{1-x_3})^2+4\lambda\sqrt{1-x_3}]^{1+2 \ep}}
 {(1-\xi x_3)^2 \, [\xi(1-x_3)A_{ki,j}(x_2,\phi_{1})+(1-\xi)A_{ki,j}(x_2(1-x_3),\phi_{2})]^2} \nonumber \\
&\times& \frac{[ A_{ki,j}(x_2,\phiu)- A_{ki,j}(x_2(1-x_3),\phid)]^2}{x_2 x_3^2} \nonumber \\
&\times& \theta[A_{ki,j}(x_2,\phi_{1})-x_2] \, \theta[A_{ki,j}(x_2(1-x_3),\phi_{2})-x_2(1-x_3)]\, ,
\eeqn
The penultimate line of this equation appears to indicate that a plus-distribution expansion is required for $x_2$ and
that the $x_3$ integration is too singular as $x_3 \to 0$ (i.e. $s \to t$ in the original variables).  However, a careful
analysis reveals that neither of these is true.  Instead, these additional denominator factors
are actually regulated by the numerator, $[ A_{ki,j}(x_2,\phiu)- A_{ki,j}(x_2(1-x_3),\phid)]^2$.
To see that this is the case we write out the expressions for $A_{ki,j}$ explicitly and use the the relation between
the angles in Eq.~(\ref{eq:phid}) to find,
\beqn
&& A_{ki,j}(x_2,\phiu)- A_{ki,j}(x_2(1-x_3),\phid) \nn \\
 && = \frac{\hs_{ij}}{\hs_{ik}} x_2 x_3
  - 2 \sqrt{x_2 \frac{\hs_{ij} \hs_{jk}}{\hs_{ik}^2}}
  \left(\cos\phiu - \sqrt{1-x_3} \cos\phid \right) \\
 &&= \frac{\sqrt{x_2}}{\hs_{ik}} \left[ \hs_{ij} \sqrt{x_2} x_3
  - 2 \sqrt{\hs_{ij} \hs_{jk}}
  \left((1-\sqrt{1-x_3} \cos\phiud) \cos\phiu
   + \sqrt{1-x_3} \sin\phiu \sin\phiud\cos\betaud \right) \right]. \nn
\eeqn
This expression makes it clear that the $x_2$ denominator factor is harmless.  Moreover,
we observe that in the limit $x_3 \to 0$,
\beq
 \cos\phiud \to 1 - \frac{1-\lambda}{8\lambda} x_3^2 \,, \qquad
 \sin\phiud \to \frac{\sqrt{1-\lambda}}{2\sqrt{\lambda}} x_3 \,,
\eeq 
so that,
\beqn
&& A_{ki,j}(x_2,\phiu)- A_{ki,j}(x_2(1-x_3),\phid) \nn \\
&\to & \frac{x_3 \sqrt{x_2}}{\hs_{ik}} \left[ \hs_{ij} \sqrt{x_2}
  - \sqrt{\hs_{ij} \hs_{jk}}
  \left(\cos\phiu
   + \sin\phiu \cos\alphaud \sqrt{\frac{1-\lambda}{\lambda}} \right) \right]
\eeqn
The use of this limit is essential in order to obtain the correct subtraction of the singularity at $x_3=0$.

%% file: results.tex
We first make a few observations regarding our numerical integration procedure.
Each contributing integral is computed using VEGAS in double precision.  Since many
of the integrals contain square-root singularities, we routinely perform remappings
to remove such factors and improve the convergence of the numerical integration,
\beq
\int_0^1 \frac{dx}{\sqrt x} \, f(x) = 2 \int_0^1 du \, f(u^2) \,.
\eeq
Moreover, in order to avoid numerical instability when evaluating the double-real integrands
extremely close to singularities, we impose a tiny cut on every integration range:
\beq
\int_0^1 dx \; \longrightarrow \; \int_\delta^1 dx \,.
\eeq
We do not observe any sensitivity of our results to reasonable variations
of $\delta$, within Monte Carlo uncertainties, and choose
$\delta=10^{-12}$ for the final results presented below. We have additionally checked that running the code in quadruple precision with a cutoff
reduced to $\delta=10^{-22}$ does not alter the results.  In order to
provide the reader with an example of our raw results, and a point of
comparison for an independent implementation, we present the numerical
value of all double-real integrals at a single phase-space point in
Appendix~\ref{sec:RRintegrals}.  Each integral is typically evaluated
with an uncertainty that is far smaller than one percent, but which
can be at the percent level for a few contributions where the absolute value is very small.
This accuracy is more than sufficient to obtain the NNLO soft
function at the level necessary for phenomenology. 

Having described all of the necessary ingredients to perform the calculations, we can now present our numerical results for the
$0$- and $1$-jettiness soft functions at NNLO. In particular, we focus on the non-Abelian contribution to the soft functions.  
Following the notation of Sec.~\ref{sec:calcnnlo}, we define the non-Abelian part of
the NNLO soft function as the sum of four different contributions:
\bal
S^{(2)}_{nab}(\mathcal{T}_N,\ep) = -\frac{\beta_0}{2 \eps} \, \tS^{(1)}(\mathcal{T}_N)  + \tS^{(2)}_{RV}(\mathcal{T}_N) + \tS^{(2)}_{gg}(\mathcal{T}_N) + \tS^{(2)}_{q\overline{q}}(\mathcal{T}_N) \, .
\end{align}
Each individual contribution is computed as explained in the previous sections. After performing the sum, the non-Abelian soft function
corresponds to the $\mathcal{O}(\eps^0)$ contribution to the total.  This is written as
\bal
S^{(2)}_{nab}(\mathcal{T}_N) = C_{-1} \,\delta(\mathcal{T}_N) + C_{0} \,\mathcal{L}_0(\mathcal{T}_N) + C_{1} \,\mathcal{L}_1(\mathcal{T}_N) + C_{2} \,\mathcal{L}_2(\mathcal{T}_N) + C_{3} \,\mathcal{L}_3(\mathcal{T}_N) \, ,
\end{align}
where $C_{n}$ with $n=-1,\dots,3$ are numerical coefficients, functions of the invariants $y_{ij}$.

\subsection{$0$-jettiness}
\label{0jetresults}

Although the calculation of the $1$-jettiness soft function is the focus of
this paper, recomputing the $0$-jettiness soft function provides a useful
check of a subset of the integrals and of the assembly of the final result.
In this case the result is known analytically~\cite{Kelley:2011ng,Monni:2011gb},
which we can use to provide a robust check of our calculation.

\begin{figure}
\centerline{\includegraphics[scale=0.9]{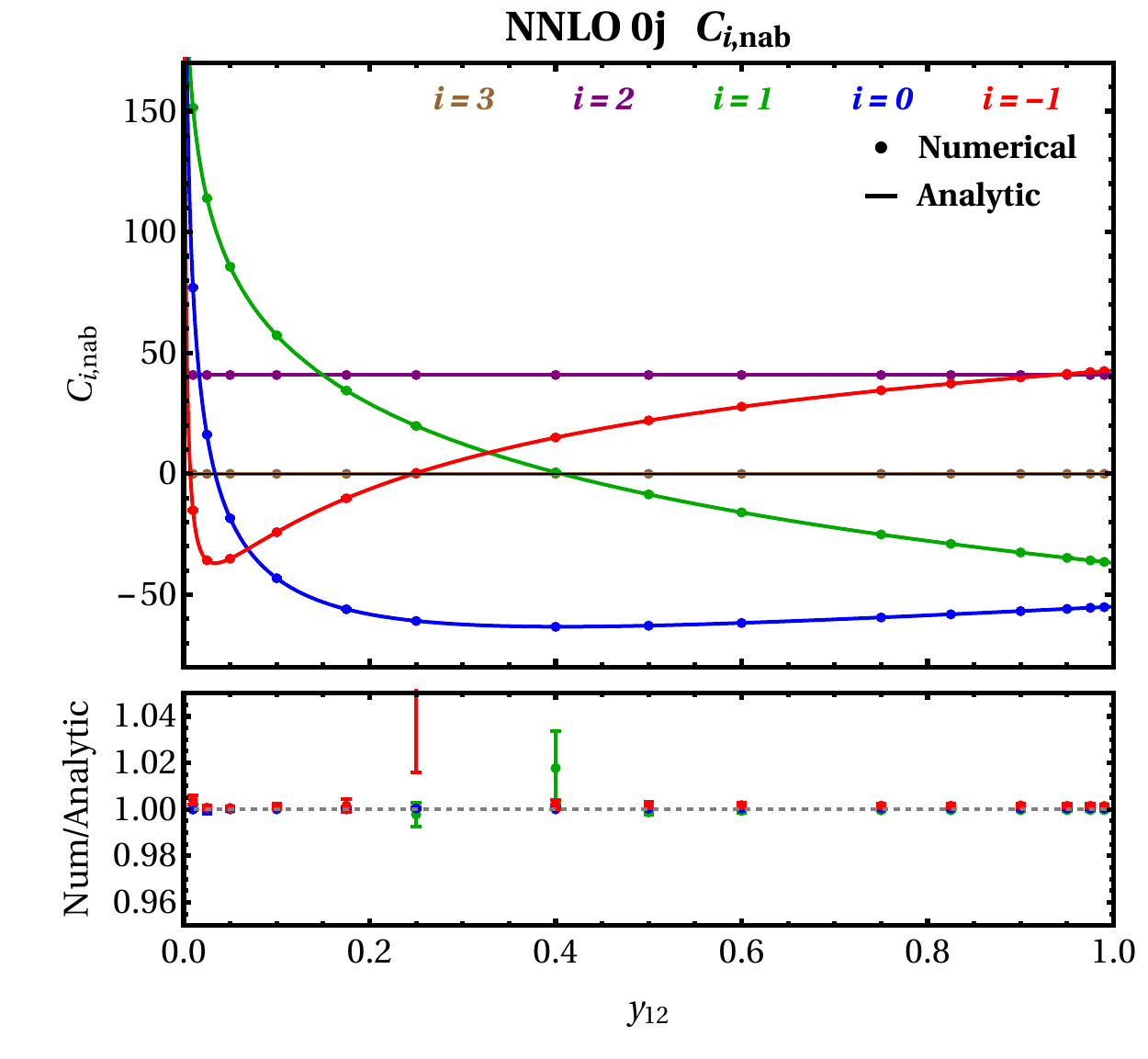}}
\caption{Numerical results for the non-Abelian part of the $0$-jettiness soft function at NNLO for the partonic channel $q\bar{q}$. We plot the coefficients of $\delta(\mathcal{T}_0)$ and $\mathcal{L}_n(\mathcal{T}_0)$ with $n=0,3$ as functions of $y_{12} \in [0,1]$. The analytic results are taken from Ref.~\cite{Boughezal:2015eha}.}
\label{plotsnnlo0j}
\end{figure}

We choose to present our results using the color factor $C_F$ (i.e. the partons present at leading order are a quark and an anti-quark)
and compute the coefficients $C_{n}$ for 15 different values of $y_{12}$ (the only invariant in the case of $0$-jettiness):
\bal
\{0.01,0.025,0.05,0.1,0.175,0.25,0.4,0.5,0.6,0.75,0.825,0.9,0.95,0.975,0.99\} \, .
\end{align}
The $gg$ channel is obtained by simply rescaling by a factor
$C_A/C_F$.   A comparison of our numerical evaluation of the NNLO soft function with the analytic
results, for the values of $y_{12}$ above, is shown in Fig.~\ref{plotsnnlo0j}.
Apart from regions where the soft functions are very close to zero, the numerical and analytic
results agree perfectly, at the level of a few per-mille or better.

\subsection{$1$-jettiness}
\label{1jetresults}

\begin{figure}
\centerline{\includegraphics[scale=0.7]{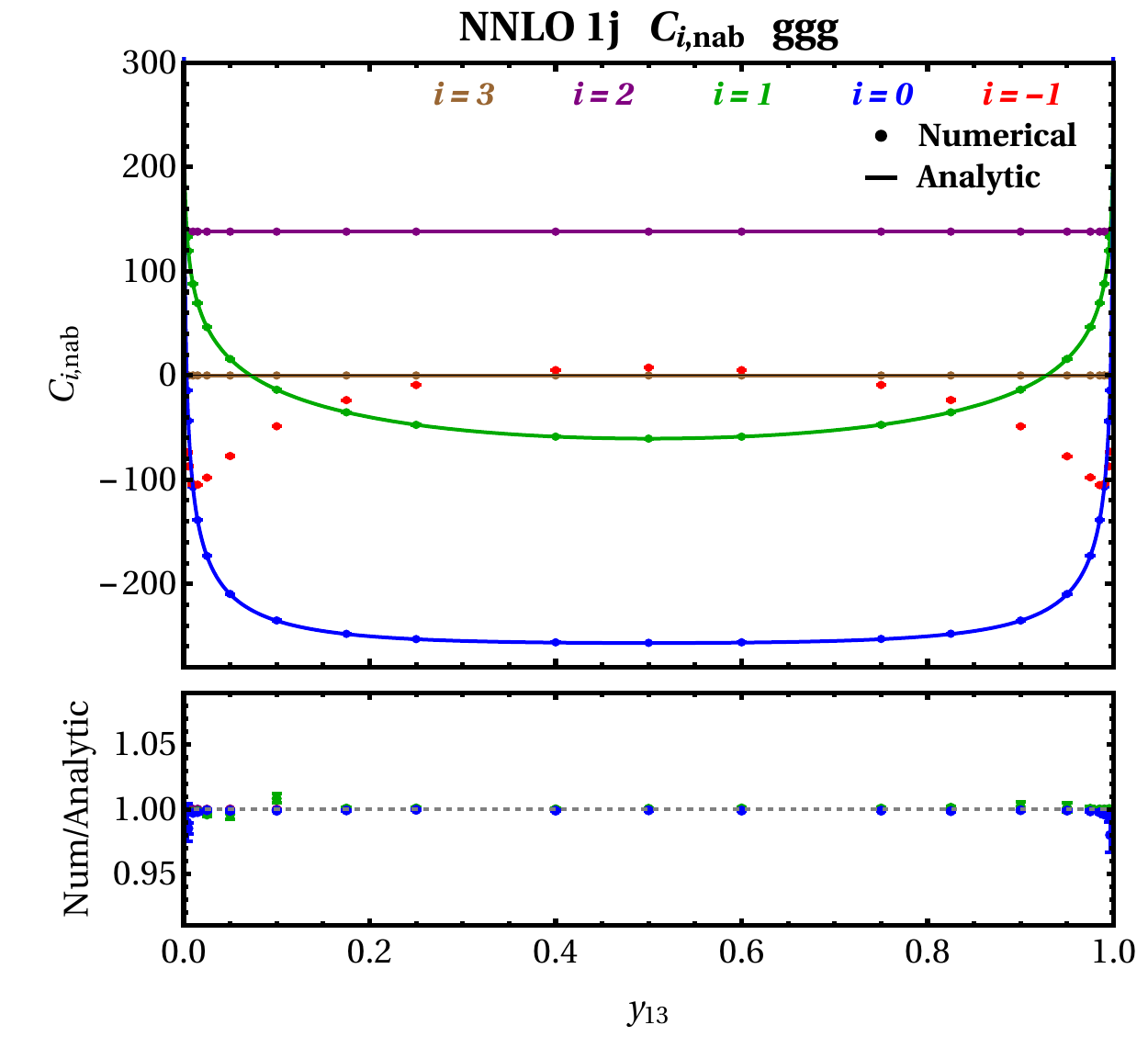} \includegraphics[scale=0.7]{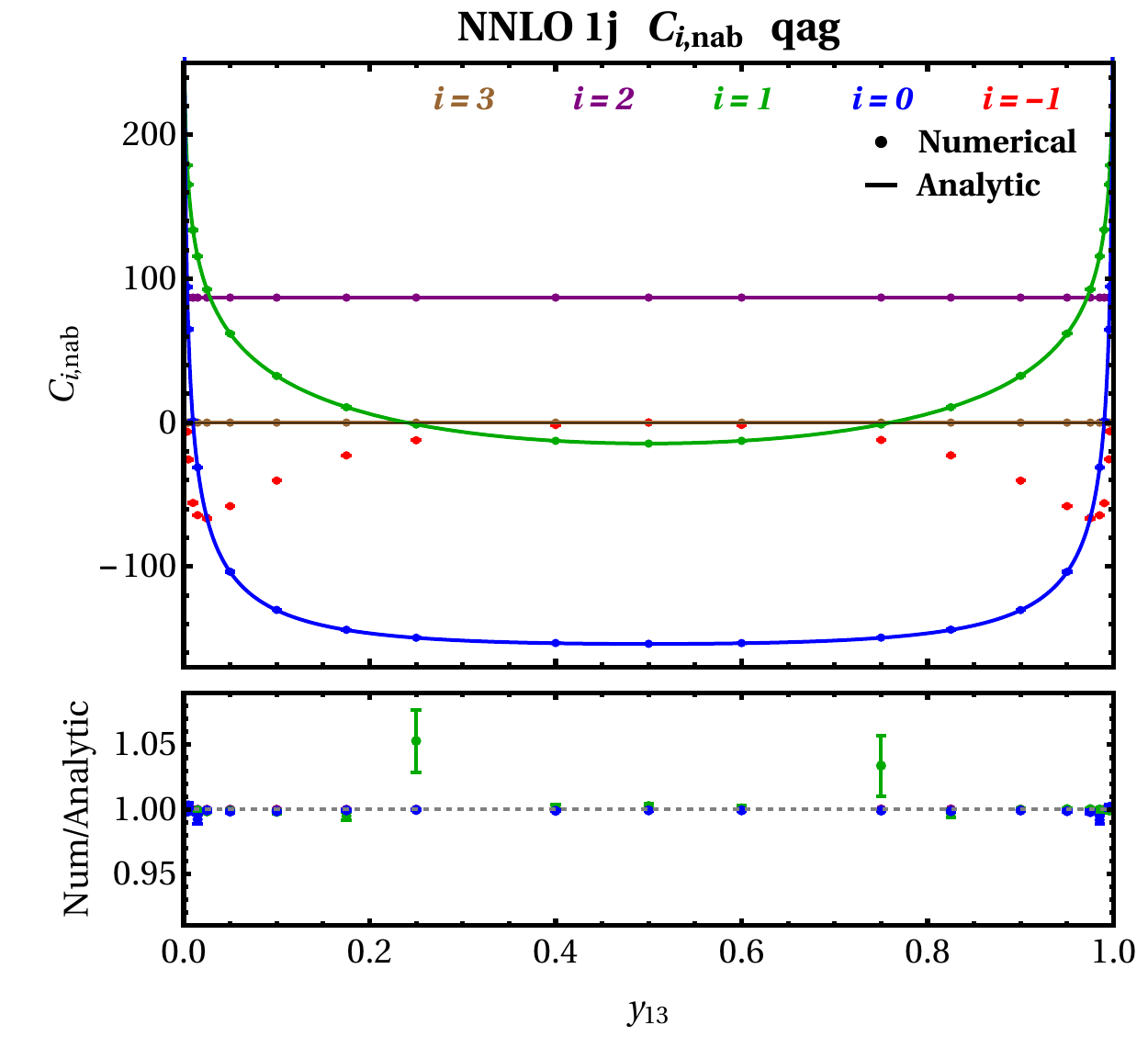}}
\vspace{0.2cm}
\centerline{\includegraphics[scale=0.7]{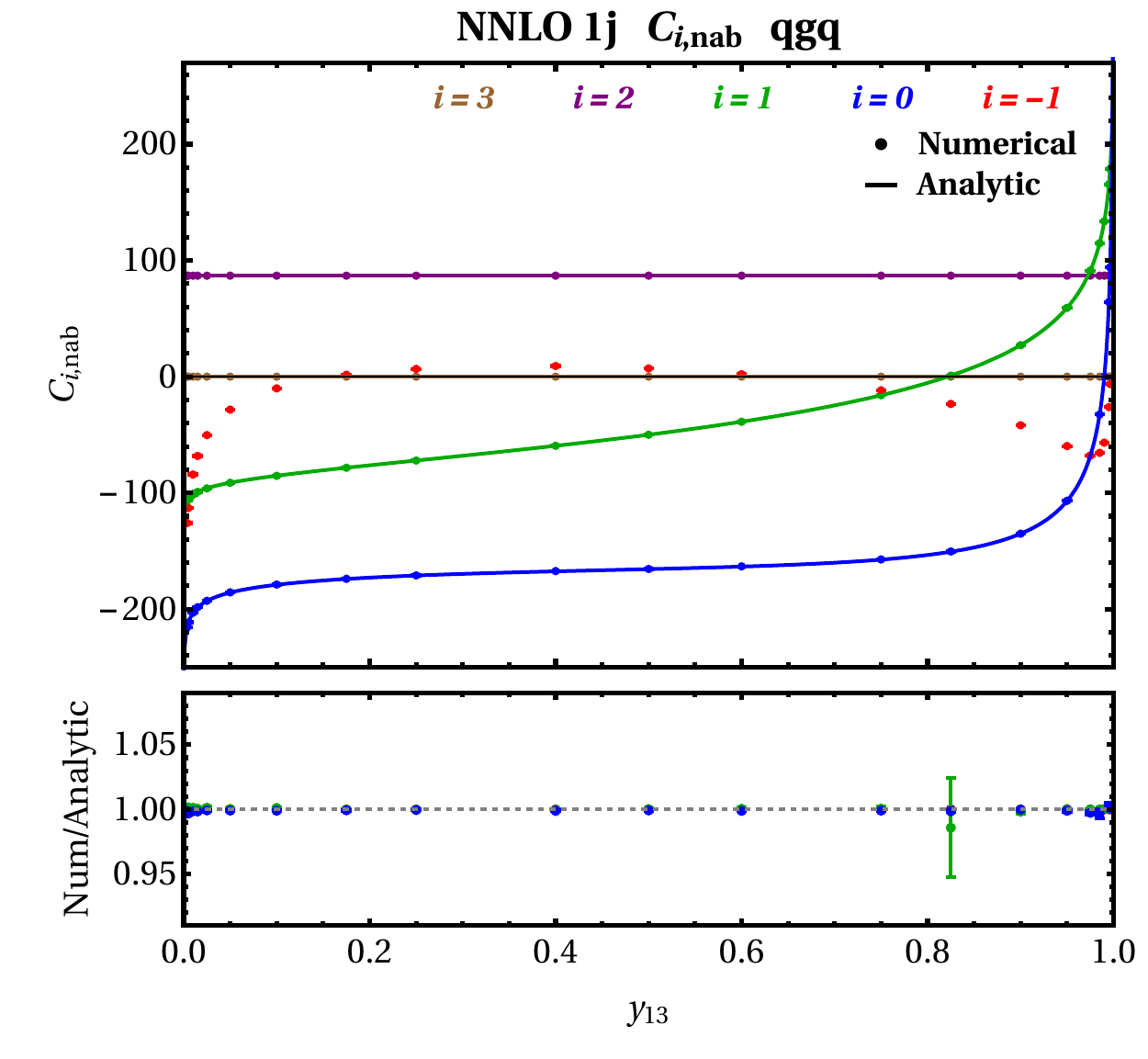}}
\vspace{0.2cm}
\caption{Numerical results for the non-Abelian part of the $1$-jettiness soft
function at NNLO for the partonic channels $gg \to g$, $q \bar{q} \to
g$, and $qg \to q$. We plot the coefficients of
$\delta(\mathcal{T}_1)$ and $\mathcal{L}_n(\mathcal{T}_1)$ with
$n=0,3$ as functions of $y_{13} \in [0,1]$. The analytic results are
taken from Refs.~\cite{Jouttenus:2011wh,Boughezal:2015eha,Gaunt:2015pea}.}
\label{plotsnnlo1j}
\end{figure}

We present the non-Abelian contribution to the $1$-jettiness soft function. As in the NLO case, we specialize our calculation for LHC kinematics. The coefficients $C_{n}$ are functions of $y_{13}$ only since $y_{12} = 1$ and $y_{23} = 1 - y_{13}$.
We compute the coefficients $C_{n}$ for the three different partonic configurations ($gg \to g$, $q\bar{q} \to g$, and $qg \to q$) and for 21 different values of $y_{13}$:
\bal
\{0.00375,0.005,0.01,0.015,0.025,0.05,0.1,0.175,0.25,0.4,0.5, \notag \\ 0.6,0.75,0.825,0.9,0.95,0.975,0.985,0.99,0.995,0.99625\} \, .
\end{align}
The coefficients $C_{0}$--$C_{3}$ are known analytically and can be derived,
for instance, from renormalization group constraints~\cite{Jouttenus:2011wh,Boughezal:2015eha,Gaunt:2015pea}.  They therefore provide
a useful check of our calculation, particularly in the case of $C_0$, which receives contributions from all the basis integrals and expansions that
must be performed for $C_{-1}$. Our results are shown in Fig.~\ref{plotsnnlo1j} and indicate that, as at NLO, the
numerical integration is accurate to the per-mille level when compared with the analytic results for $C_{0}$--$C_{3}$.

\begin{figure}
\centerline{\includegraphics[scale=0.7]{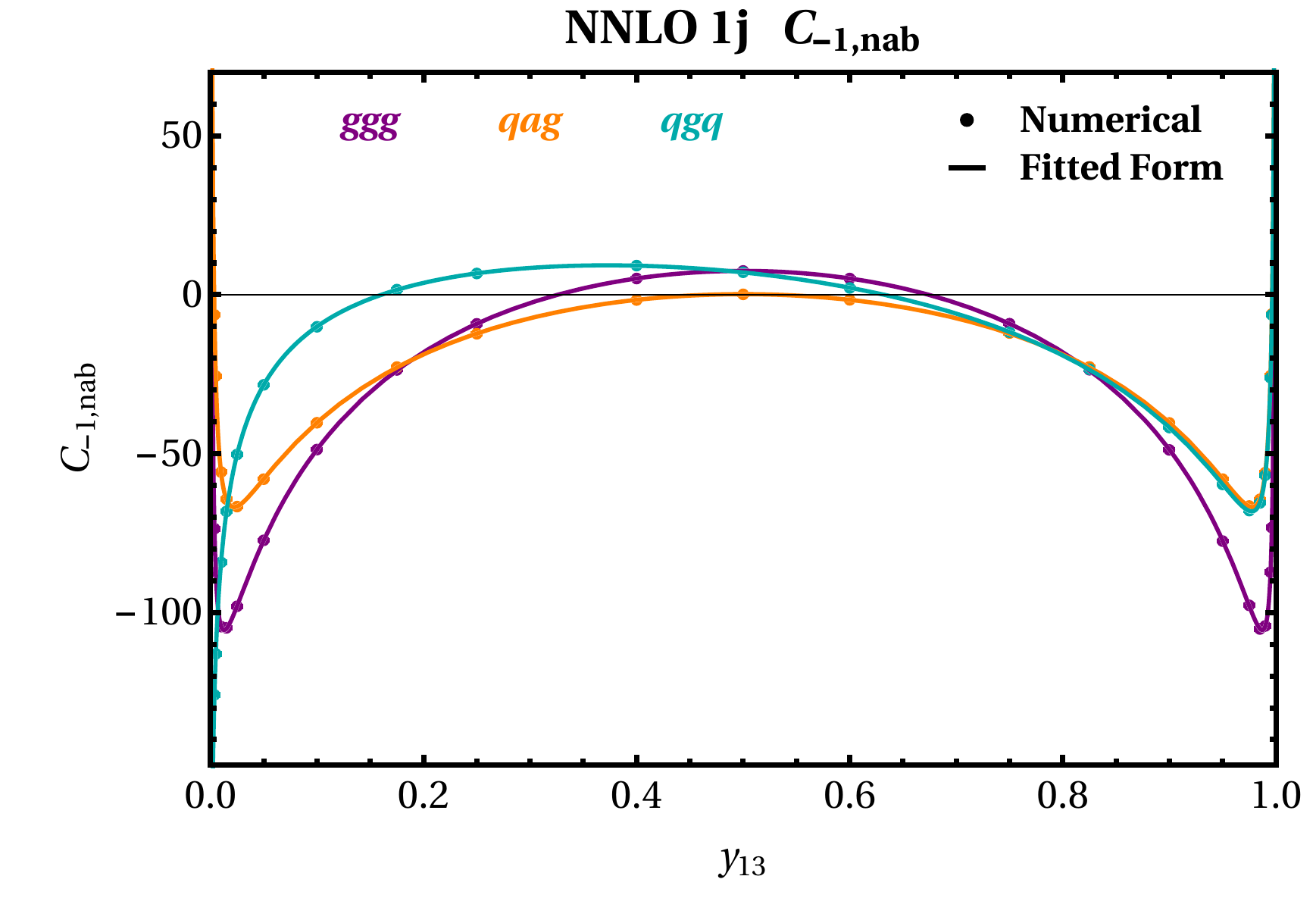}}
\caption{Numerical results and fitted form for the coefficient of $\delta(\mathcal{T}_1)$ in the three partonic configurations as a function of $y_{13} \in [0,1]$. Included in the plot are the fitting uncertainties at the 95\% confidence level.}
\label{plotsnnlo1jdelpiece}
\end{figure}

The calculation of the endpoint contribution $C_{-1}$ is the central result of this paper. Our results for this contribution are shown
separately in Fig.~\ref{plotsnnlo1jdelpiece}, for each of the three
configurations.  We have performed fits to our results using the
functional form,
\bal \label{fitform}
C_{-1,\text{fit}}(y_{13}) = \sum_{m,n = 0}^3 c_{(m,n)} \left[\ln{(y_{13})}\right]^m \left[\ln{(1-y_{13})}\right]^n \,,
\end{align}
and these fits are also shown in the figure. Uncertainties on the fits are estimated at the 95\% confidence level, and are plotted in the figure (they essentially correspond to the line thickness).  The fits provide a
very good description of the endpoint coefficient and are far more efficient
for subsequent evaluation of the soft function. The values of the fit
coefficients of Eq.~\eqref{fitform} for the three cases are collected
in Table~\ref{tablefit}. For the channels $gg \to g$ and $q\bar{q} \to g$ we set $c_{(m,n)} = c_{(n,m)}$ (with $m \neq n$).
We have tested the fits using randomly-generated phase-space points for comparison and within their respective MC uncertainties every event agrees with the prediction from the fits. When the MC uncertainties are small the agreement is within $0.5\%$, while for points for which the coefficients are close to zero the MC uncertainties are significantly larger and the agreement is at the few percent level.

% NNLO1j fit table

\begin{table}
\centering
\begin{tabular}{|c|r|r|r|}
\hline
\rule{0pt}{2ex} & $gg \to g$ & $q\bar{q} \to g$ & $qg \to q$\\
\hline
$c_{(0,0)}$ & $63.187 \pm 0.903$ & $42.357 \pm 0.786$ & $39.101 \pm 0.698$\\
%$c_{(0,0)}$ & $63.1867 \pm 0.9032$ & $42.3565 \pm 0.7855$ & $39.1008 \pm 0.6981$\\
\hline
$c_{(1,0)}$ & $33.599 \pm 0.779$ & $25.158 \pm 0.678$ & $13.726 \pm 0.615$\\
%$c_{(1,0)}$ & $33.5992 \pm 0.7794$ & $25.1578 \pm 0.6775$ & $13.7263 \pm 0.6150$\\
\hline
$c_{(2,0)}$ & $-11.056 \pm 0.227$ & $-9.100 \pm 0.197$ & $-2.737 \pm 0.186$\\
%$c_{(2,0)}$ & $-11.0558 \pm 0.2265$ & $-9.0998 \pm 0.1967$ & $-2.7374 \pm 0.1860$\\
\hline
$c_{(3,0)}$ & $-2.273 \pm 0.021$ & $-2.158 \pm 0.019$ & $0.016 \pm 0.018$\\
%$c_{(3,0)}$ & $-2.27305 \pm 0.0213$ & $-2.15775 \pm 0.0185$ & $0.0164 \pm 0.0183$\\
\hline
$c_{(0,1)}$ & $33.599 \pm 0.779$ & $25.158 \pm 0.678$ & $25.591 \pm 0.602$\\
%$c_{(0,1)}$ & $33.5992 \pm 0.7794$ & $25.1578 \pm 0.6775$ & $25.5908 \pm 0.6019$\\
\hline
$c_{(0,2)}$ & $-11.056 \pm 0.227$ & $-9.100 \pm 0.197$ & $-8.749 \pm 0.177$\\
%$c_{(0,2)}$ & $-11.0558 \pm 0.2265$ & $-9.0998 \pm 0.1967$ & $-8.7492 \pm 0.1773$\\
\hline
$c_{(0,3)}$ & $-2.273 \pm 0.021$ & $-2.158 \pm 0.019$ & $-2.126 \pm 0.017$\\
%$c_{(0,3)}$ & $-2.27305 \pm 0.0213$ & $-2.15775 \pm 0.0185$ & $-2.1255 \pm 0.0170$\\
\hline
\end{tabular}
\caption{Non-zero coefficients of the numerical fit of Eq.~\eqref{fitform} for the three partonic configurations $gg \to g$, $q\bar{q} \to g$, and $qg \to q$. Coefficients not shown here are understood to be zero.}
\label{tablefit}
\end{table}

\begin{figure}
\centerline{\includegraphics[scale=0.53]{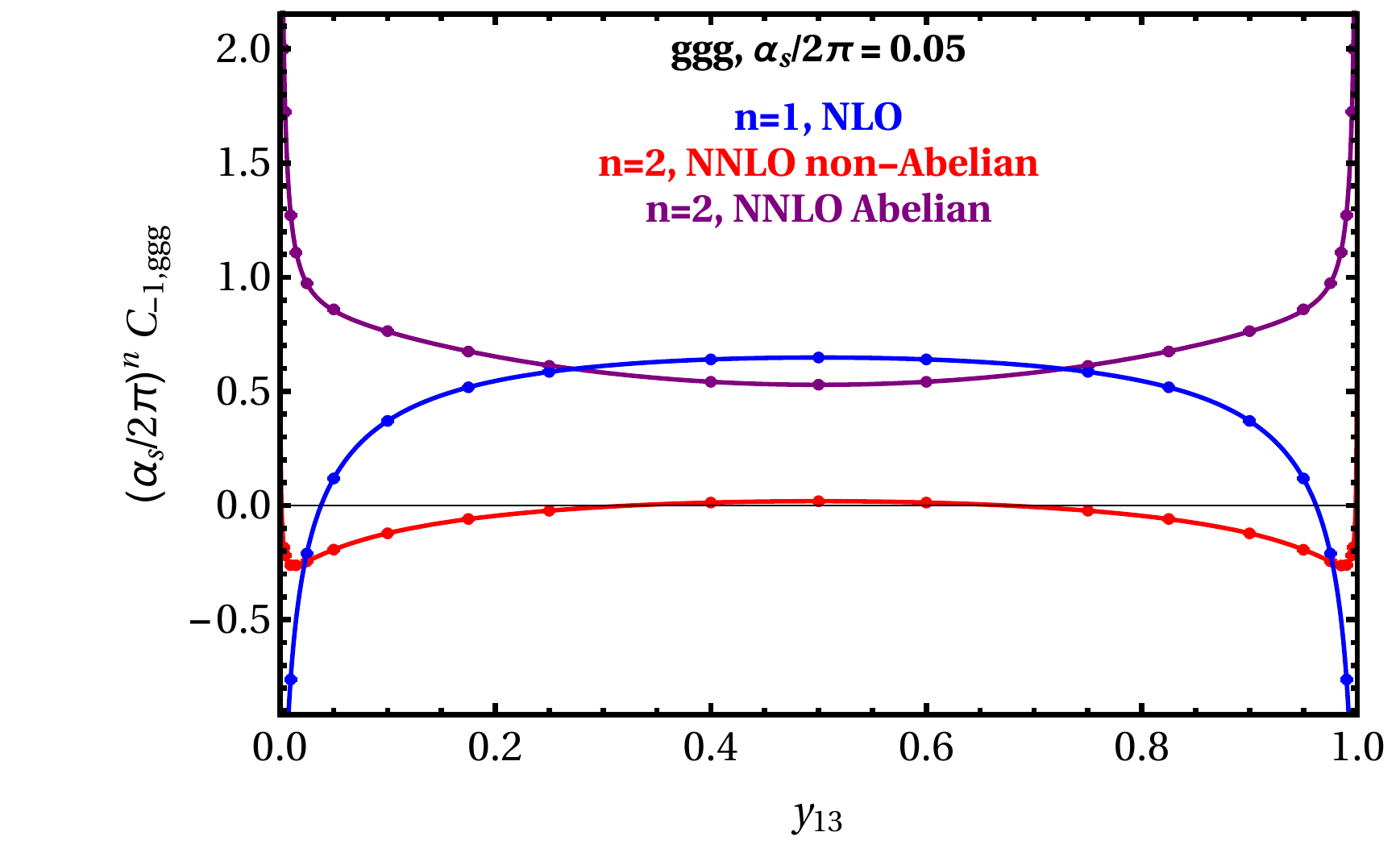}\includegraphics[scale=0.53]{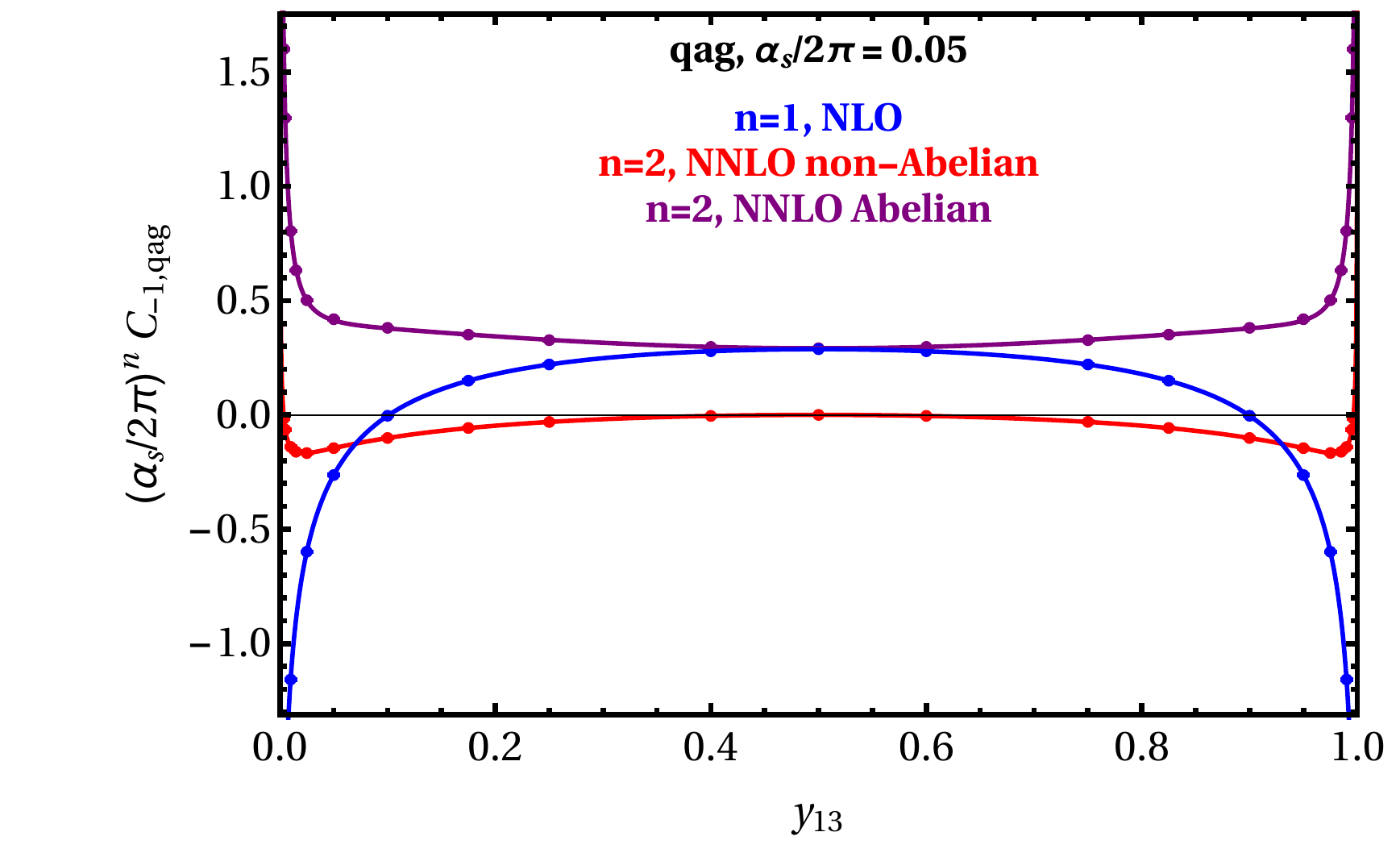}}
\centerline{\includegraphics[scale=0.53]{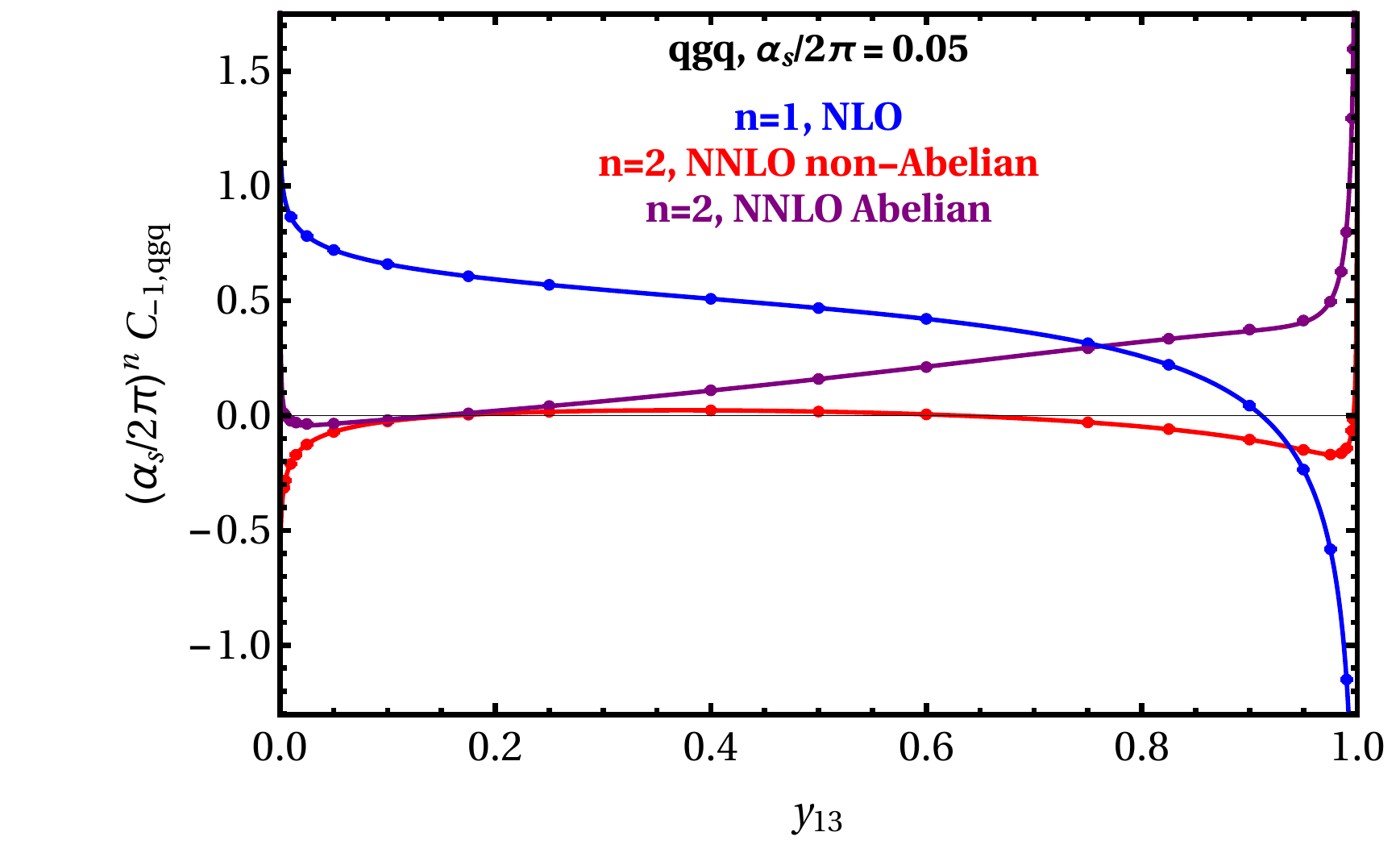}}
\caption{The NLO and NNLO contributions to the soft function, shown for a notional
value of $\as/(2 \pi)=0.05$.}
\label{del_total}
\end{figure}

To assess the overall impact of the corrections to the soft function, 
in Fig.~\ref{del_total} we show the NLO contribution along with the 
Abelian and non-Abelian pieces of the NNLO contribution. It is clear that the Abelian pieces are significantly larger than the 
non-Abelian pieces over the entire phase space for each partonic configuration. This can be readily understood from the analytic structures of the two pieces. 
The $\delta({\cal{T}}_1)$ coefficient of the Abelian part arises from the convolution of the NLO result with itself, and therefore it has two contributing parts. Firstly, there is 
the ``pure" $\delta({\cal{T}}_{1})$ piece, which arises from the $\delta({\cal{T}}_1)$ coefficient of the NLO soft function squared. This term is comparable to the non-Abelian 
part in size. Secondly, there are mixed contributions which arise from convolutions of the form 
\begin{eqnarray}
(\mathcal{L}_m \otimes \mathcal{L}_n)({\cal{T}}) \equiv  \int d {\cal{T}'} {\cal{L}}_{m}({\cal{T}'}-{\cal{T}}){\cal{L}}_{n}({\cal{T}'}) = V^{mn}_{-1}\delta({\cal{T}}) + \sum_{k=0}^{m+n+1} V^{mn}_{k} {\cal{L}}_{k}({\cal{T}}) 
\end{eqnarray}
where the coefficients $V_{k}^{mn}$ can be found, for instance, in Table 1 of Ref.~\cite{Gaunt:2015pea}, and are roughly $\mathcal{O}(1)$. After both contributions are combined together the dominant part of the total coefficient is determined by the $V^{01}_{-1}C_0 C_1$ term. Such a term is not present in the non-Abelian calculation. 

\subsection{Generic kinematics}
Finally, we compute the non-Abelian part of the 1-jettiness soft function at NNLO with generic kinematics, i.e. for the case where the scattering process has three colored partons at Born level, but where two of them do not represent the directions of incoming beams (as is the case for LHC kinematics). We parametrize the three directions $\hp_1$, $\hp_2$, and $\hp_3$ as
\bal
\hp_1 &= \frac{1}{2} (1,0,0,1) \notag \\
\hp_2 &= \frac{1}{2} (1,0,\sin{\theta_2},\cos{\theta_2}) \notag \\
\hp_3 &= \frac{1}{2} (1,\sin{\phi}\sin{\theta_3},\cos{\phi}\sin{\theta_3},\cos{\theta_3}) \, ,
\end{align}
with $\theta_2,\theta_3 \in [0,\pi]$ and $\phi \in [0,2\pi]$. For the invariants $y_{ij}$ (with $i,j \in \{1,2,3\}$) we therefore explicitly have
\bal
y_{12} &= \frac{1}{2} (1-\cos{\theta_2}) \notag \\
y_{13} &= \frac{1}{2} (1-\cos{\theta_3}) \notag \\
y_{23} &= \frac{1}{2} (1-\cos{\phi}\sin{\theta_2}\sin{\theta_3}-\cos{\theta_2}\cos{\theta_3}) \, .
\end{align}
We observe that with this parametrization the LHC limit is recovered by setting $\phi = 0$ and $\theta_2 = \pi$. We compute the coefficients $C_{n}$ of the non-Abelian part of the soft function for the three different partonic configurations ($ggg$, $q\bar{q}g$, and $qgq$) by choosing 200 random values for $\theta_2$, $\theta_3$, and $\phi$. We then perform numerical fits to our results for the coefficient of $\delta(\mathcal{T}_1)$, $C_{-1}$. The functional form of the fits is taken to be
\bal \label{fitformgenkin}
C^{\rm{gen}}_{-1,\text{fit}}(y_{12},y_{13},y_{23}) = \sum_{k,m,n = 0}^3 c_{(k,m,n)} \left[\ln{(y_{12})}\right]^k \left[\ln{(y_{13})}\right]^m  \left[\ln{(y_{23})}\right]^n \, .
\end{align}
In order to obtain accurate fits, we retain all 64 coefficients in Eq.~\eqref{fitformgenkin} for each partonic channel. The values of the coefficients $c_{(k,m,n)}$ for the three cases are collected in ancillary files that we include in the arXiv submission.

\begin{figure}
\centerline{\includegraphics[scale=0.68]{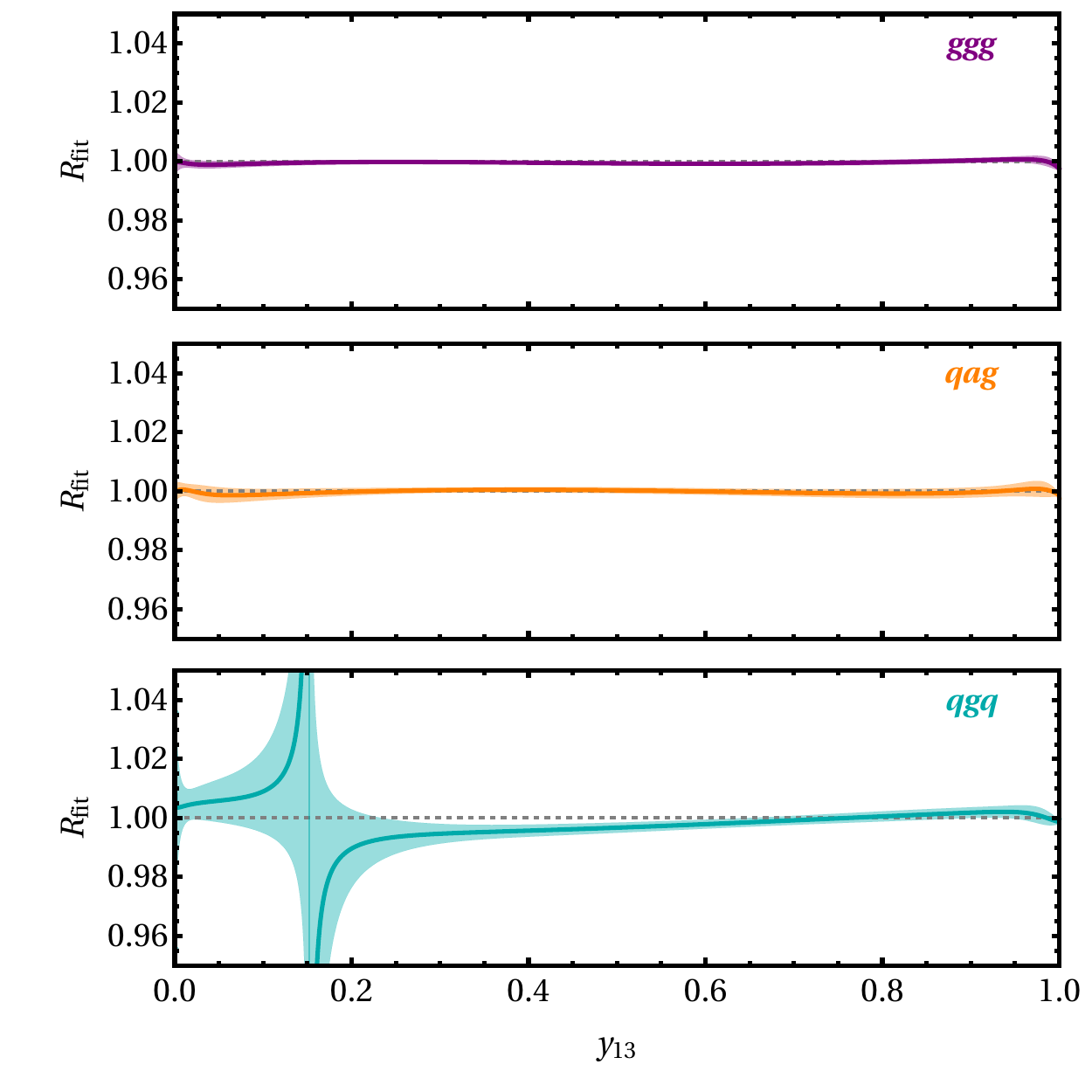}\includegraphics[scale=0.68]{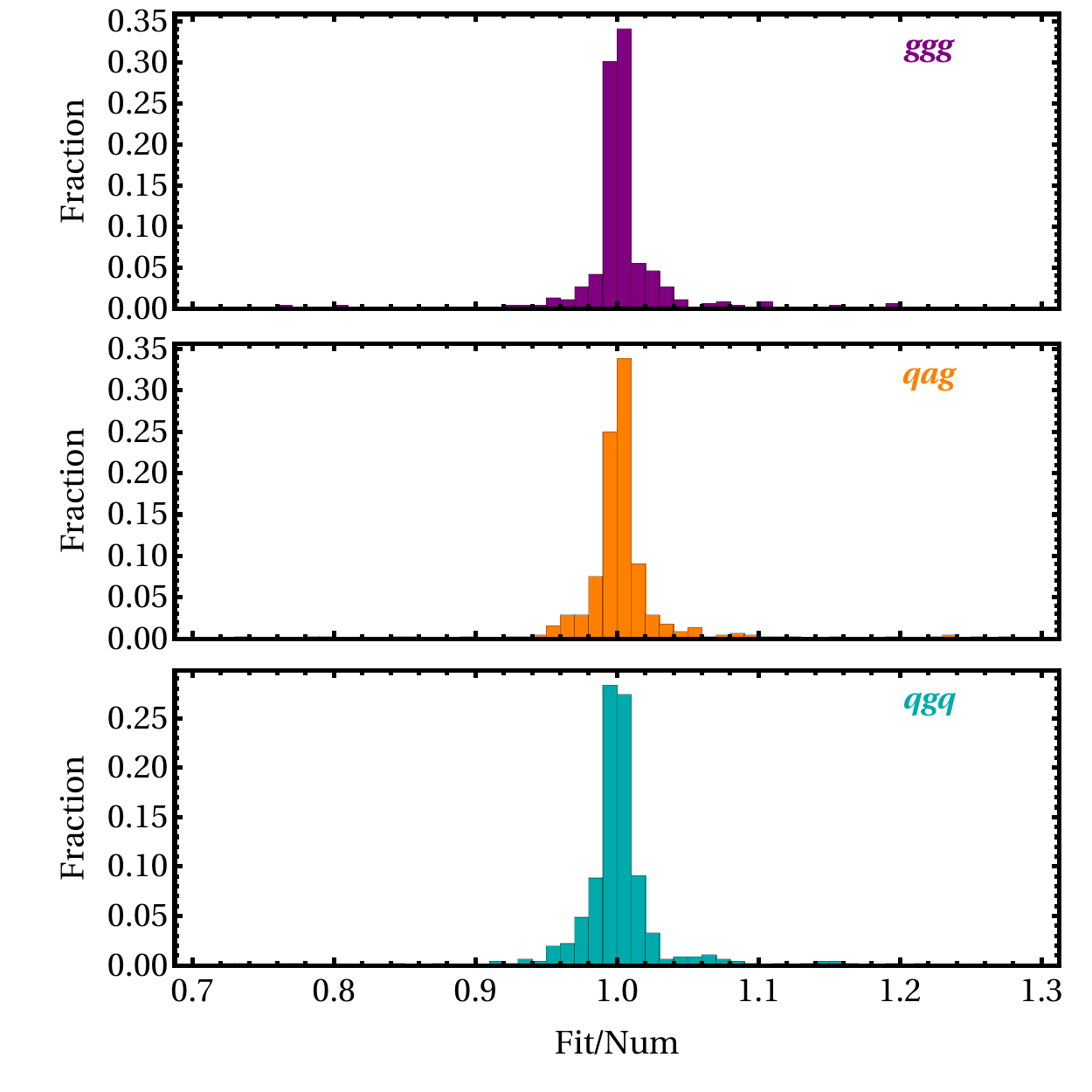}}
\vspace{0.2cm}
\caption{Left panel: ratio between the LHC limit of the fits of Eq.~\eqref{fitformgenkin} and the fits of Eq.~\eqref{fitform} for the three partonic channels. The ratio $R_{\rm{fit}}$ has been defined in Eq.~\eqref{ratiogenkin}. Right panel: ratio between predicted values from the fits and obtained numerical results for 450 randomly-generated phase-space points in the generic kinematics case.}
\label{fitratiocheckhisto}
\end{figure}

We can test the validity of the generic fits by ensuring that they correctly reproduce the dedicated LHC fits obtained in Section \ref{1jetresults} when choosing $y_{12} = 1$ and $y_{23} = 1-y_{13}$ in Eq.~\eqref{fitformgenkin}. We therefore define the following ratio 
\begin{eqnarray} \label{ratiogenkin}
R_{\rm{fit}}(y_{13}) = \frac{C^{\rm{ab}}_{-1}(y_{13})+C^{\rm{gen}}_{-1,\text{fit}}(1,y_{13},1-y_{13})}{C^{\rm{ab}}_{-1}(y_{13})+C_{-1,\rm{fit}}(y_{13})} \, .
\end{eqnarray}
In the above equation $C^{\rm{ab}}_{-1}(y_{13})$ represents the $\delta({\cal{T}}_1)$ coefficient of the Abelian part of the soft function (evaluated for LHC kinematics). We have added it to the numerator and denominator such that $R_{\rm{fit}}$ compares the total NNLO $\delta({\cal{T}}_1)$  coefficients using the two different fits.  This combination is the only one that is relevant for phenomenological applications. In addition, since $C_{-1}$ vanishes for certain values of $y_{13}$, the sum of Abelian and non-Abelian pieces helps to ensure that the ratio is not dominated by the regions in which one of the fits is close to zero. The ratios of the two fits are shown in the left panel of Fig.~\ref{fitratiocheckhisto}. By inspecting Fig.~\ref{fitratiocheckhisto} we see that for the $ggg$ and $q\bar{q}g$ channels the generic fit reproduces the dedicated LHC fit to better than 1\%. On the other hand, the $qgq$ channel is poorly behaved in the region $y_{13} \sim 0.15$. This is due to the vanishing of both the Abelian and the non-Abelian pieces in this region, which causes large sensitivity to fitting uncertainties. Away from this region the general fit does a good job (within 1\%) at reproducing the dedicated LHC fit.

Additionally, as in the LHC case, we can test the fits by generating random phase-space points and comparing the numerical results of the non-Abelian piece with the predicted values from the corresponding fits. In this case we perform the comparison by generating 450 phase-space points and plotting the ratios between fit and numerical values as histograms. The results are shown in the right panel of Fig.~\ref{fitratiocheckhisto}. We observe that for all three channels the histograms are well-centered around the value of $1$. In particular, the values obtained with the fit lie within $2\%$ of a dedicated calculation of a configuration for $74\%$, $75\%$, and $74\%$ of the total number of random phase-space points respectively ($ggg$, $q\bar{q}g$, and $qgq$), and within $5\%$ for $87\%$, $88\%$, and $88\%$ of the phase-space points. Similarly to the LHC case, the agreement between fitted and numerical values is better for points with smaller MC uncertainties and coefficients that are not close to zero. We therefore believe that the fits of Eq.~\eqref{fitformgenkin}, which are valid for non-LHC kinematics, could be successfully used, for instance, in $ep$ and $e^+e^-$ applications. 

%% file: conclu.tex
In this paper we have presented a calculation of the next-to-next-to leading
order (NNLO) 1-jettiness soft function.  The soft function is a component
part of NNLO calculations which employ slicing methods based on the
$N$-jettiness global event shape variable.  In particular, this function is a
required piece of the calculation of differential $pp \rightarrow X + j$ type
processes at NNLO, in which $X$ represents a color singlet, for instance a
single vector boson. 

Our calculation bears the traditional hallmarks of NNLO calculations in
regards to its complexity in unresolved limits. In order to deal with
these issues we have employed a numerical approach which uses sector
decomposition of the relevant phase-space integrals to disentangle
overlapping singularities present at this order. In order to ensure
the correctness of our results we have implemented two completely
independent computational codes, and validated them against one
another.  We have further validated our results by recomputing the
known 0-jettiness and 1-jettiness results at NNLO and NLO accuracy
respectively. As a final validation of our results we have checked the
known analytic pieces of the 1-jettiness soft function at NNLO that
can be derived from renormalization group arguments.  The primary
result of our calculation is a numerical determination of the
$\delta({\cal {T}}_1)$ endpoint contribution, which cannot be
deduced from the RGE's alone. We have computed this contribution for
LHC kinematics (scattering processes with two back-to-back partons and one final-state jet at Born level)
and for generic kinematics (three colored partons at leading order).
In order to disseminate our results we
have produced polynomial fits to the results of our numerical
integration for both configurations. We have additionally checked both fits using randomly-generated phase-space points,
and find excellent agreement with our Monte Carlo
output.

Our fits represent the first such results presented in the literature. A
previous calculation of the $1$-jettiness soft function at NNLO for LHC kinematics has been published \cite{Boughezal:2015eha}.
It does not contain the information required
to implement the results in a standalone Monte Carlo program.
We therefore believe our calculation, and the corresponding numerical
fits, will be useful for those interested in applying jettiness-slicing and
subtraction methods to NNLO calculations where three colored particles are present.
Further applications of this method are certainly possible
in future, for instance to analyze a variety of global event shape definitions
or to increase the number of partonic scatters under consideration, such as in the calculation of
the 2-jettiness soft function at NNLO. We leave such applications to a
future study.

%% file: acknowledgments.tex
R.K.E would like to acknowledge the hospitality of the
Fermilab theory group, where this work was initiated,
and SUNY (Buffalo) where it was completed.
C.W. and R.M. are supported by a National Science Foundation
CAREER award PHY-1652066. Support provided by the Center for Computational Research at the University at Buffalo.
This manuscript has been authored by Fermi Research
Alliance, LLC under Contract No. DE-AC02-07CH11359 with the U.S. Department of
Energy, Office of Science, Office of High Energy Physics. The U.S. Government retains
and the publisher, by accepting the article for publication, acknowledges that the U.S.
Government retains a non-exclusive, paid-up, irrevocable, world-wide license to publish
or reproduce the published form of this manuscript, or allow others to do so, for U.S.
Government purposes.

%% file: Notation.tex
%\section{Notation}

%\subsection{Epsilontica}
%\beq
%\frac{(e^{\gamma_E})^\ep}{\Gamma(1-\ep)}=  1- \ep^2 \frac{\pi^2}{12}- \ep^3 \frac{\zeta(3) }{3}+ \ep^4 \frac{\pi^4}{1440}+O(\ep^5)
%\eeq
%Legendre duplication formula
%\beq
%\frac{1}{4^\ep \sqrt{\pi}} \frac{\Gamma(1-\ep)}{\Gamma(1-2 \ep)} = \frac{1}{\Gamma(\frac{1}{2}-\ep)}
%\eeq

\section{Dimensional regularization}

A $d$-dimensional Euclidean integral may be written as,
\begin{eqnarray} \label{x21}
\int d^d \kappa \; f(\kappa^2) & =& \int d |\kappa| \; f(\kappa^2) \;
|\kappa|^{d-1}
\sin^{d-2} \theta_{d-1} \sin^{d-3} \theta_{d-2} \ldots  \nonumber \\
&& \times \sin \theta_{2} \; d \theta_{d-1} d \theta_{d-2}
\ldots d \theta_{2} d \theta_{1}.
\end{eqnarray}
The range of the angular integrals is
$ 0 \leq \theta_i \leq \pi $ except for $ 0 \leq \theta_1 \leq 2\pi $.
Eq.~(\ref{x21}) is best proved by induction. Assuming that it is true for
a $d$-dimensional integral, in $(d+1)$ dimensions we can write,
\beqn
\int \; d^{d+1} \kappa &=& \int \; d \kappa_{d+1}
\; d^{d} \kappa \\
&=& \int  \; d \kappa_{d+1} \; d |\kappa| \; |\kappa|^{d-1}
\sin^{d-2} \theta_{d-1} \sin^{d-3} \theta_{d-2} \ldots
\sin \theta_{2} \; d \theta_{d-1} d \theta_{d-2}
\ldots d \theta_{2} d \theta_{1}\nonumber \\
\eeqn
The $d$-dimensional length, $\kappa$, can be written
in terms of the $(d+1)$-dimensional length, $\rho$, as
\beqn
\kappa_{d+1}&=&\rho \cos \theta_d \nn \\
|\kappa|&=&\rho \sin  \theta_d \, .
\eeqn
Changing variables to $\rho$ and $\theta_d$ we recover
the $(d+1)$-dimensional version of Eq.~(\ref{x21}).

For our particular case at hand we have $d \to d-2$:
\begin{eqnarray} 
\int d^{d-2} \kappa  & =& \frac{1}{2} \int d |\kappa^2| \; |\kappa^2|^{\frac{(d-4)}{2}} d \Omega_{d-2} \\
\int d \Omega_{d-2} & =& \int \; \sin^{d-4} \theta_{d-3} \sin^{d-5} \theta_{d-4} \ldots  \nonumber \\
&& \times \sin \theta_{2} \; d \theta_{d-3} d \theta_{d-4} 
\ldots d \theta_{2} d \theta_{1}. 
\end{eqnarray}
A $(d-2)$-dimensional vector can be written as,
\beq
(\cos\theta_{d-3} \hat{n}_{\ep} ;\sin\theta_{d-3} \sin\theta_{d-4} ,\sin\theta_{d-3} \cos\theta_{d-4}) \, ,
\eeq
where the components before the semi-colon are the extra-dimensional pieces 
and the vector $n_\ep$ is a unit vector in the first extra dimension.
We arrive at this frame by setting $\theta_{i}=\pi/2$ for $i=1,d-5$.
Let us choose vectors in the transverse plane given by
\beqn
q_{1\,\perp}&=&q_{1\,\perp}(0;\sin(\phiu),\cos(\phiu)) \nonumber \\
q_{2\,\perp}&=&q_{2\,\perp}(\cos(\phid)\hat{n}_\ep;\sin(\phid) \sin(\beta),\sin(\phid) \cos(\beta))
\eeqn
so that for $q_1$ we have set $\theta_{d-3}=\pi/2$, $\theta_{d-4}=\phiu$ 
and for $q_2$ we set $\theta_{d-3}=\phid$, $\theta_{d-4}=\beta$.
The integral over unconstrained angles is,
\beq
\int_0^{2\pi} d\theta_1 \int_0^\pi \sin^{d-6} \theta_{d-5} \ldots \sin \theta_{2}
 \; d \theta_{d-5} \ldots d\theta_{2} 
 = 2\pi \, \frac{(\sqrt\pi)^{d-6}}{\Gamma(\frac{d-4}{2})} \,
\eeq
so that, after integrating over them, we get the following expression 
\beqn
d\Omega_{d-2} &=& -\frac{2 \ep}{\pi^{\ep} \, \Gamma(1-\ep)}
\int_0^\pi d\phid \sin^{-2 \ep} \phid 
\int_0^{\pi} d \beta \sin^{-1-2\ep}\beta \nonumber \\
&=& \frac{2\pi^{1-\ep}}{\Gamma(1-\ep)}
\int_0^\pi \frac{d \phid}{N_\phi} \sin^{-2 \ep} \phid 
\int_0^{\pi} \frac{d \beta}{N_\beta} \sin^{-1-2\ep}\beta \,,
\label{eq:PS2angles}
\eeqn
where
\beq 
N_\beta = -\frac{1}{\ep} \sqrt{\pi} \frac{\Gamma(1-\ep)}{\Gamma(\frac{1}{2}-\ep)},\;\;\;
N_\phi =4^\ep \pi \frac{\Gamma(1-2\ep)}{\Gamma(1-\ep)^2} = \sqrt{\pi} \frac{\Gamma(\frac{1}{2}-\ep)}{\Gamma(1-\ep)}.
\eeq
By performing the integration over $\beta$ we get
\beqn
d\Omega_{d-2}& = & 2 \sqrt{\pi} \frac{1} {\pi^\ep \, \Gamma(\frac{1}{2} -\ep)}  \int_0^{\pi} d \phid \sin^{-2 \ep}\phid \nonumber \\
&=& \frac{2}{(4 \pi)^\ep} \frac{\Gamma(1-\ep)} {\Gamma(1-2 \ep)}  \int_0^{\pi} d \phid \sin^{-2 \ep}\phid \nonumber \\
&=& \frac{2\pi^{1-\ep}} {\Gamma(1-\ep)}  \int_0^{\pi} \frac{d \phid}{N_\phi} \sin^{-2 \ep}\phid.
\label{eq:PS1angles}
\eeqn
Using the standard result,
\beq
\int_0^\pi d\phi \, \sin^d\phi = \sqrt{\pi} \, \frac{\Gamma(\frac{d+1}{2})}{\Gamma(\frac{d+2}{2})} \, ,
\eeq
the total volume of angular integration is 
\beq
\Omega_{d-2} = \frac{2 \pi}{\Gamma(1-\ep) \pi^\ep} \, .
\eeq

%% file: Rotation.tex
\section{Rotational invariance of the solid angle integral measure}
\label{sec:rotation}

Perform an angle-radius decomposition of the coordinates
\beqn
x_1 &=& r \cos\phid \nonumber \\
x_2 &=& r \sin\phid \cos\beta \nonumber \\
x_\ep &=& r \sin\phid \sin\beta 
\eeqn
where $x_\ep$ is a coordinate in the transverse plane beyond the usual two.
Hence
\beq
dx_1 \, dx_2 \, dx_\ep = r^2 \, dr \, \sin\phid \, d\phid \, d\beta \, .
\eeq
Now perform a rotation about the $\ep$-axis by an angle $\phiu$:
\beqn
x_1^\prime &=&r (\cos\beta\, \sin\phiu\, \sin\phid + \cos\phiu\, \cos\phid\,) \nonumber\\
x_2^\prime &=&r (\cos\beta\, \cos\phiu\, \sin\phid - \sin\phiu\, \cos\phid\,) \nonumber\\
x_\ep^\prime &=&r (\sin\beta\, \sin\phid\,) \, .
\eeqn
These coordinates can also be parametrized by introducing new angles,
\beqn
x_1^\prime &=& r \cos\phiud \nonumber \\
x_2^\prime &=& r \sin\phiud \cos\betaud \\
x_\ep^\prime &=& r \sin\phiud \sin\betaud \nonumber \, .
\eeqn
Comparing the two parametrizations, we note that
\beqn
\label{eq:anglerelation12}
\cos\phiud &=& \cos\phiu\, \cos\phid + \cos\beta\, \sin\phiu\, \sin\phid\, , \\
\sin\phiud \sin\betaud &=& \sin\phid \sin\beta
\eeqn
Calculating the solid element, we see that 
\beqn
d x_1^\prime \, d x_2^\prime \, d x_3^\prime &=& r^2 dr d\phiud \, \sin\phiud \, d \betaud \\
&\equiv& r^2 dr d\phid \, \sin\phid\, d \beta \, ,
\eeqn
as can be shown by explicitly calculating the Jacobian for the transformation.
However, after the change of
variables $\{\phid,\beta\} \to \{\phiud,\betaud\}$, we must be sure to express all 
dependence on $\phid,\beta$ in terms of $\phiud$ and $\betaud$,
\beq
\cos\phid=\cos\phiu \cos\phiud-\sin\phiu \sin\phiud\cos\betaud.
\label{eq:anglerelation}
\eeq

%% file: MEs.tex
\section{Double-real matrix elements}
\label{DRME}
Here we list the expressions for the matrix elements $\cJ^{I}_{ij},\cJ^{II}_{ij},\cJ^{III}_{ij}$,
defined in Eqs.~(\ref{CJIdefn}), (\ref{CJIIdefn}), and~(\ref{CJIIIdefn}), that must be integrated to
evaluate the double-real contributions.  The expressions depend on the case specified by the measurement function, which determines the choice of the Sudakov
directions.  For $\cJ^{III}_{ij}$ it is useful to perform a further decomposition to aid the
numerical integration of these contributions, with the division of terms depending on the
case at hand. We have,
\beq
\cJ^{III}_{ij} = \cJ^{IIIa}_{ij}+\cJ^{IIIb}_{ij}+\cJ^{IIIc}_{ij}+\cJ^{IIId}_{ij},
\eeq
for cases 1 and 4, while the sum only runs over the $a$ and $b$ terms for cases 2 and 3.

We note that in the eventual evaluation of these matrix elements we must restore an overall
factor,
\beq
\MF = g^4 S_\ep^2 \mu^{4 \ep} .
\label{MFdefn}
\eeq
according to Eqs.~(\ref{CUdefn}) and~(\ref{CTdefn}).

\subsection{Case 1}
\label{MEcase1}

Using the relations in Eq.~(\ref{Denominators1}) and the transformation of Eq.~(\ref{q1Dq2_transformed}) we have,
\beqn 
\cJ^{I}_{ij}
% &=& 
%\frac{(-2) [\hp_i.q_1 \, \hp_j.q_2-\hp_j.q_1 \, \hp_i.q_2]^2}
%     {q_1.q_2^2 \; \hp_i.(q_1+q_2)^2 \; \hp_j.(q_1+q_2)^2} \nonumber \\
  &=&-8 \, \frac{\hs_{ij}^2}{\tauN^4} \, \frac{s^2 t^2}{[s-t]^2} \,
\frac{[(\sqrt{s}-\sqrt{t})^2 +4 \lambda \sqrt{st}]^2}{[\xi t+(1-\xi)s]^2} 
\label{eq:MEcase1I} \\
\cJ^{II}_{ij}
% &=&\frac{\hs_{ij}}{ q_1.q_2 \; \hp_i.(q_1+q_2) \; \hp_j.(q_1+q_2)} \nonumber \\
&=&8 \, \frac{\hs_{ij}^2}{\tauN^4} \frac{s^2 t^2}{[s-t]^2} \frac{1}{\xi [1-\xi]} 
\frac{[(\sqrt{s}-\sqrt{t})^2 +4 \lambda \sqrt{st}]}{[\xi t+(1-\xi)s]} \\
\cJ^{IIIa}_{ij}&=&8 \, \frac{\hs_{ij}^2}{\tauN^4} \frac{s t}{\xi^2\; (1-\xi)^2}\\
\cJ^{IIIb}_{ij}&=&8 \, \frac{\hs_{ij}^2}{\tauN^4} 
 \frac{st(s+t)\left[2st-(s+t)\sqrt{st}(1-2\lambda)\right]}
 {\xi(1-\xi)(\xi t+(1-\xi)s)(s-t)^2} \\
\cJ^{IIIc}_{ij}&=&-8 \, \frac{\hs_{ij}^2}{\tauN^4}
     \frac{2 st}{ \xi^2 [1-\xi]^2}  \Big(
\frac{\big(t\; \theta(s-t)+ s\; \theta(t-s)\big )\; [(\sqrt{s}-\sqrt{t})^2+4 \lambda \sqrt{st}]}{(s-t)^2}\Big) \nonumber \\
\\
\cJ^{IIId}_{ij}&=&-8 \, \frac{\hs_{ij}^2}{\tauN^4}
\frac{st}{\xi^2 [1-\xi]^2} \;
\Big( \frac{(\theta(s-t)-\theta(t-s)) \; [(\sqrt{s}-\sqrt{t})^2+4 \lambda \sqrt{st}]}{(s-t)}\Big) \, . 
\eeqn
We note that since $\cJ^{IIIa}_{ij}$ does not depend on $\lambda$ it may be
treated using the phase-space measure in Eq.~(\ref{Twophasespacecase1untransformed}).
All other contributions require the use of Eq.~(\ref{Twophasespacecase1transformed})
and the further partitioning indicated in Eq.~(\ref{eq:stpartition}).   

\subsection{Case 2}
\label{MEcase2}

Using the relations in Eq~(\ref{Denominators2}), the matrix elements are given by,
\beqn 
\cJ^{I}_{ij}
% &=& \frac{(-2) [\hp_i.q_1 \, \hp_j.q_2-\hp_j.q_1 \, \hp_i.q_2]^2}
%     {q_1.q_2^2 \; \hp_i.(q_1+q_2)^2 \; \hp_j.(q_1+q_2)^2} \nonumber \\
  &=&-8 \, \frac{\hs_{ij}^2}{\tauN^4} \, 
    \frac{s^2 \; t^2 \; (1-st)^2} {[(1-\sqrt{st})^2+4 \zud\sqrt{st}]^2 \; [\xi t+(1-\xi)]^2 \; [\xi+(1-\xi) s]^2} \\
\cJ^{II}_{ij}
% &=&
% \frac{\hs_{ij}}{ q_1.q_2 \; \hp_i.(q_1+q_2) \; \hp_j.(q_1+q_2)} \nonumber \\
&=&8 \, \frac{\hs_{ij}^2}{\tauN^4} \frac{s^2 t^2}{\xi(1-\xi)} 
\frac{1}{[(1-\sqrt{st})^2+4 \zud\sqrt{st}] \; [\xi t+(1-\xi)] \; [\xi+(1-\xi) s]} \\
\cJ^{IIIa}_{ij}
&=&8 \, \frac{\hs_{ij}^2}{\tauN^4} \frac{st}{\xi^2(1-\xi)^2}
 \left[ 1 - \frac{1+st}{(1-\sqrt{st})^2+4 \zud\sqrt{st}}
 \right] \\
\cJ^{IIIb}_{ij}
&=& 4 \, \frac{\hs_{ij}^2}{\tauN^4} \frac{st}{\xi(1-\xi)}
 \frac{1+st}{[\xi+(1-\xi) s] \, [\xi t+(1-\xi)]}
 \left[ \frac{1+st}{(1-\sqrt{st})^2+4 \zud\sqrt{st}} - 1
 \right] \nn \, . \\
\eeqn
All contributions may be evaluated using the phase-space parametrization given in
Eq.~(\ref{Twophasespacecase2untransformed}).

\subsection{Case 3}
\label{MEcase3}

Using the relations in Eq~(\ref{Denominators3}), the matrix elements are given by,
\beqn 
\cJ^{I}_{ij}
% &=& 
%\frac{(-2) [\hp_i.q_1 \, \hp_j.q_2-\hp_j.q_1 \, \hp_i.q_2]^2}
%     {q_1.q_2^2 \; \hp_i.(q_1+q_2)^2 \; \hp_j.(q_1+q_2)^2} 
 &=& -8 \, \frac{\hs_{ik}^2}{\tauN^4} 
 \frac{s^2 \; t^4}{[(1-\sqrt{st})^2+4 \zud\sqrt{st}]^2} \nonumber \\
 &&\times \frac{[A_{ik,j}(s,\phiu)-A_{ki,j}(t,\phid) s]^2}
 {[\xi t A_{ik,j}(s,\phiu)+(1-\xi) s A_{ki,j}(t,\phid)]^2 \; [\xi t+(1-\xi)]^2 } \\
\cJ^{II}_{ij}
% &=&\frac{\hs_{ij}}{ q_1.q_2 \; \hp_i.(q_1+q_2) \; \hp_j.(q_1+q_2)} 
&=&8 \, \frac{\hs_{ij}\hs_{ik}}{\tauN^4} \frac{s^2 t^3}{\xi(1-\xi) \; [\xi t+(1-\xi)]} \nonumber \\
&&\times \frac{1}{[(1-\sqrt{st})^2+4 \zud\sqrt{st}] \; [\xi t A_{ik,j}(s,\phiu)+(1-\xi) s A_{ki,j}(t,\phid)]} \\
\cJ^{IIIa}_{ij}&=&8 \,  \frac{\hs_{ij}\hs_{ik}}{\tauN^4} \frac{s t^2}{\xi^2(1-\xi)^2} \frac{1}{A_{ik,j}(s,\phiu)\; A_{ki,j}(t,\phid)} \nonumber \\
&&\times 
 \left[\frac{\hs_{ij}}{\hs_{ik}}-\frac{\Big(A_{ik,j}(s,\phiu)+s A_{ki,j}(t,\phid)\Big)}{((1-\sqrt{st})^2+4\zud \sqrt{st})}\right]  \\
\cJ^{IIIb}_{ij}&=&4 \,  \frac{\hs_{ij}\hs_{ik}}{\tauN^4}
 \frac{s t^3}{\xi(1-\xi) \, (\xi t+(1-\xi)) \, (\xi t A_{ik,j}(s,\phiu)+(1-\xi) s A_{ki,j}(t,\phid)} \nonumber \\
&&\times \Bigg\{ \frac{1}{[(1-\sqrt{st})^2+4 \zud \sqrt{st}]} 
     \left[\sqrt{\frac{A_{ik,j}(s,\phiu)}{A_{ki,j}(t,\phid)}}
                            +s \sqrt{\frac{A_{ki,j}(t,\phid)}{A_{ik,j}(s,\phiu)}}\right]^2 \nonumber \\       
&& \quad - \frac{\hs_{ij}}{\hs_{ik}} \Bigg(\frac{1}{A_{ki,j}(t,\phid)}+\frac{s}{A_{ik,j}(s,\phiu)}\Bigg)
\Bigg\} \, .
\eeqn
All contributions may be evaluated using the phase-space parametrization given in
Eq.~(\ref{Twophasespacecase3untransformed}).

\subsection{Case 4}
\label{MEcase4}

\beqn 
\cJ^{I}_{ij}&=&
 -8 \, \frac{\hs_{ik}^2}{\tauN^4} \, \frac{s^4 t^4}{(s-t)^4}
 \Bigg[\frac{( A_{ki,j}(s,\phi_{1})- A_{ki,j}(t,\phi_{2}))\; [(\sqrt{s}-\sqrt{t})^2 +4 \lambda \sqrt{st}]}
 {[\xi t A_{ki,j}(s,\phi_{1})+(1-\xi) s A_{ki,j}(t,\phi_{2})]\;[\xi t+(1-\xi)s]}\Bigg]^2 \nn \\ && \\
\label{MsqII-case4}
\cJ^{II}_{ij}&=& 8 \, \frac{\hs_{ij} \hs_{ik} }{\tauN^4} \frac{s^3 t^3}{(s-t)^2 \, \xi (1-\xi)}
 \frac{[(\sqrt{s}-\sqrt{t})^2 +4 \lambda \sqrt{st}]}
 {[\xi t A_{ki,j}(s,\phi_{1})+(1-\xi) s A_{ki,j}(t,\phi_{2})][\xi t+(1-\xi)s]} \nn \\ && \\
\cJ^{IIIa}_{ij}&=&8\, \frac{\hs_{ij}^2}{\tauN^4}
         \frac{s^2 t^2}{\xi^2(1-\xi)^2 \, A_{ki,j}(s,\phiu) A_{ki,j}(t,\phid)} \\
\cJ^{IIIb}_{ij}&=&-8\, \frac{\hs_{ij} \hs_{ik}}{\tauN^4}\frac{s^2 t^2}{\xi^2 (1-\xi)^2} 
  \frac{\big((\sqrt{s}-\sqrt{t})^2+4 \lambda \sqrt{st}\big)}{(s-t)^2} 
   \Bigg[\frac{1}{A_{ki,j}(s,\phiu)}+\frac{1}{A_{ki,j}(t,\phid)} \Bigg] \nn \\ && \\
\cJ^{IIIc}_{ij}&=&4\, \frac{\hs_{ij} \hs_{ik}}{\tauN^4}\frac{s^3\; t^3}{\xi (1-\xi)}
\frac{((\sqrt{s}-\sqrt{t})^2+4 \lambda \sqrt{st})}{(s-t)^2} 
   \left[2+\frac{A_{ki,j}(s,\phiu)}{ A_{ki,j}(t,\phid)}+\frac{A_{ki,j}(t,\phid)}{A_{ki,j}(s,\phiu)}\right] \nonumber \\
   &&\times\frac{1}{\big(\xi t A_{ki,j}(s,\phiu)+(1-\xi) s A_{ki,j}(t,\phid)\big)\; (\xi t+(1-\xi) s)} \\
\cJ^{IIId}_{ij}&=&-4\, \frac{\hs_{ij}^2}{\tauN^4}\frac{s^3 t^3}{\xi (1-\xi)}
         \left[\frac{1}{A_{ki,j}(s,\phiu)}+\frac{1}{A_{ki,j}(t,\phid)}\right] \nonumber \\
         &&\times \frac{1}{\big(\xi t A_{ki,j}(s,\phiu)+(1-\xi) s A_{ki,j}(t,\phid)\big)\; (\xi t+(1-\xi) s)} \, .
\eeqn
The contibutions corresponding to $\cJ^{IIIa}_{ij}$ and $\cJ^{IIId}_{ij}$ do not depend on $\lambda$ and therefore may be
treated using the phase-space measure in Eq.~(\ref{Twophasespacecase4untransformed}).
All other contributions require the use of both Eq.~(\ref{Twophasespacecase4transformed})
and the partitioning of Eq.~(\ref{eq:stpartition}).

%% file: Integrals.tex
\section{Results for double-real integrals}
\label{sec:RRintegrals}

Results for the ${\cal O}(\ep)$ coefficient of each of the basic integrals entering the double-real emission calculation, at a
sample phase-space point, are given in Table~\ref{tableintegrals}.
These coefficients are the ones that enter the calculation of the endpoint contribution to the NNLO soft function and may
be useful to the reader interested in reproducing the results of our calculation.

\begin{table}[h]
\centering
\begin{tabular}{|l|l|r|}
\hline
Case & Integral & Result \\
\hline
        C1I $(i,j)$  & $I^{(2),ij,I    }_{ii}$  & $     8.083 \pm    0.001 $ \\
       C1II $(i,j)$  & $I^{(2),ij,II   }_{ii}$  & $   -17.909 \pm    0.004 $ \\
     C1IIIa $(i,j)$  & $I^{(2),ij,IIIa }_{ii}$  & $    36.374 \pm    0.005 $ \\
     C1IIIb $(i,j)$  & $I^{(2),ij,IIIb }_{ii}$  & $   -38.598 \pm    0.013 $ \\
     C1IIIc $(i,j)$  & $I^{(2),ij,IIIc }_{ii}$  & $    98.983 \pm    0.014 $ \\
     C1IIId $(i,j)$  & $I^{(2),ij,IIId }_{ii}$  & $   -67.556 \pm    0.022 $ \\
      C1III $(i,j)$  & $I^{(2),ij,III  }_{ii}$  & $    29.203 \pm    0.029 $ \\
 \hline
        C1I $(j,i)$  & $I^{(2),ij,I    }_{jj}$  & $     3.538 \pm    0.002 $ \\
       C1II $(j,i)$  & $I^{(2),ij,II   }_{jj}$  & $   -10.015 \pm    0.006 $ \\
     C1IIIa $(j,i)$  & $I^{(2),ij,IIIa }_{jj}$  & $    11.693 \pm    0.007 $ \\
     C1IIIb $(j,i)$  & $I^{(2),ij,IIIb }_{jj}$  & $   -24.216 \pm    0.017 $ \\
     C1IIIc $(j,i)$  & $I^{(2),ij,IIIc }_{jj}$  & $    53.675 \pm    0.027 $ \\
     C1IIId $(j,i)$  & $I^{(2),ij,IIId }_{jj}$  & $   -28.305 \pm    0.077 $ \\
      C1III $(j,i)$  & $I^{(2),ij,III  }_{jj}$  & $    12.847 \pm    0.084 $ \\
 \hline
        C2I $(i,j)$  & $I^{(2),ij,I    }_{ij}$  & $    -3.395 \pm    0.002 $ \\
       C2II $(i,j)$  & $I^{(2),ij,II   }_{ij}$  & $     9.648 \pm    0.003 $ \\
     C2IIIa $(i,j)$  & $I^{(2),ij,IIIa }_{ij}$  & $   -13.245 \pm    0.028 $ \\
     C2IIIb $(i,j)$  & $I^{(2),ij,IIIb }_{ij}$  & $    19.332 \pm    0.015 $ \\
      C2III $(i,j)$  & $I^{(2),ij,III  }_{ij}$  & $     6.087 \pm    0.032 $ \\
 \hline
% These are just a check of the symmetry
%        C2I $(j,i)$  & $I^{(2),ij,I    }_{ji}$  & $    -3.398 \pm    0.002 $ \\
%       C2II $(j,i)$  & $I^{(2),ij,II   }_{ji}$  & $     9.653 \pm    0.004 $ \\
%     C2IIIa $(j,i)$  & $I^{(2),ij,IIIa }_{ji}$  & $   -13.242 \pm    0.028 $ \\
%     C2IIIb $(j,i)$  & $I^{(2),ij,IIIb }_{ji}$  & $    19.364 \pm    0.016 $ \\
%      C2III $(j,i)$  & $I^{(2),ij,III  }_{ji}$  & $     6.122 \pm    0.032 $ \\
% \hline
        C3I $(i,j)$  & $I^{(2),ij,I    }_{ik}$  & $    -2.287 \pm    0.002 $ \\
       C3II $(i,j)$  & $I^{(2),ij,II   }_{ik}$  & $     7.213 \pm    0.003 $ \\
     C3IIIa $(i,j)$  & $I^{(2),ij,IIIa }_{ik}$  & $   -12.019 \pm    0.025 $ \\
     C3IIIb $(i,j)$  & $I^{(2),ij,IIIb }_{ik}$  & $    13.072 \pm    0.011 $ \\
      C3III $(i,j)$  & $I^{(2),ij,III  }_{ik}$  & $     1.053 \pm    0.027 $ \\
 \hline
        C3I $(j,i)$  & $I^{(2),ij,I    }_{jk}$  & $     2.244 \pm    0.002 $ \\
       C3II $(j,i)$  & $I^{(2),ij,II   }_{jk}$  & $    -4.272 \pm    0.005 $ \\
     C3IIIa $(j,i)$  & $I^{(2),ij,IIIa }_{jk}$  & $    15.539 \pm    0.034 $ \\
     C3IIIb $(j,i)$  & $I^{(2),ij,IIIb }_{jk}$  & $    -0.963 \pm    0.016 $ \\
      C3III $(j,i)$  & $I^{(2),ij,III  }_{jk}$  & $    14.576 \pm    0.037 $ \\
 \hline
        C4I $(i,j)$  & $I^{(2),ij,I    }_{kk}$  & $     8.263 \pm    0.002 $ \\
       C4II $(i,j)$  & $I^{(2),ij,II   }_{kk}$  & $   -27.980 \pm    0.006 $ \\
     C4IIIa $(i,j)$  & $I^{(2),ij,IIIa }_{kk}$  & $   -36.003 \pm    0.005 $ \\
     C4IIIb $(i,j)$  & $I^{(2),ij,IIIb }_{kk}$  & $   305.627 \pm    0.035 $ \\
     C4IIIc $(i,j)$  & $I^{(2),ij,IIIc }_{kk}$  & $   -55.278 \pm    0.012 $ \\
     C4IIId $(i,j)$  & $I^{(2),ij,IIId }_{kk}$  & $     2.965 \pm    0.002 $ \\
      C4III $(i,j)$  & $I^{(2),ij,III  }_{kk}$  & $   217.312 \pm    0.037 $ \\
 \hline
\end{tabular}
\caption{Results for all the double-real integrals at the point $y_{13} = 0.9$ with $i=1$, $j=2$, and $k=3$.
The table shows the ${\cal O}(\ep)$ coefficient of each integral,
$I^{(2),ij}$, according to the normalization indicated in
Eq.~(\ref{eq:Idefinition}). For each case C$x$III ($x \in \{1,2,3,4\}$) the result presented corresponds to the sum of the individual pieces according to the subdivision shown in Sections \ref{MEcase1}--\ref{MEcase4} respectively.}
\label{tableintegrals}
\end{table}

%% file: Results0j.tex
\section{Analytic result for $0$-jettiness at NLO}
\label{sec:Results0j}

For LHC kinematics, the $0$-jettiness case refers to processes with two initial-state colored partons and
no final-state jets (plus any non-colored final-state particle). The only allowed leading-order
configurations are therefore either a pair of gluons or a quark-antiquark pair. For the purposes of our
calculation, this means that $i,j \in \{1,2\}$ and that we need not consider the phase-space sector $F^{ij}_k$. Moreover, there is no $\theta$-function involving an angle left in the problem.
As discussed in Section~\ref{sec:color}, color conservation means that the
overall color factor, $C = - \bT_1 \cdot \bT_2$, is given by either $C=C_F$ for a quark-antiquark pair or $C=C_A$ for a pair of gluons. 
The $0$-jettiness soft function at NLO is thus given by Eq.~\eqref{eq:soft1} after setting $F^{ij}_k \to 0$.  The phase-space integrals can then be performed analytically so that the soft function reads,
\bal
\tS^{(1)} &= 4 \, C \,\frac{e^{\gamma_E\eps}}{\eps \,\Gamma(1-\eps)}
  \, \frac{1}{\mathcal{T}_0} \, \Big[\frac{\mathcal{T}_0}{\mu \sqrt{\hs_{12}}}\Big]^{-2\ep} \label{resultsoftnlonoexp} \\
&= 4 \, C \, \frac{1}{\mathcal{T}_0} \, \Big[\frac{\mathcal{T}_0}{\mu}\Big]^{-2\ep}
 \bigg[\frac{1}{\eps} + L_{12} + \eps \Big(\frac{L^2_{12}}{2}-\frac{\pi^2}{12}\Big) 
  +\eps^2 \Big(\frac{L^3_{12}}{6}-\frac{\pi^2}{12}L_{12}-\frac{\zeta_3}{3}\Big) +\mathcal{O}(\eps^3)\bigg] \, , \notag
\end{align}
where $L_{12} = \log y_{12}$. The result in Eq.~\eqref{resultsoftnlonoexp} agrees with Eq.~(52) of
Ref.~\cite{Jouttenus:2011wh} after taking into account the different notation and overall normalization.
By performing the expansion of Eq.~\eqref{deltaplus} on the result in Eq.~\eqref{resultsoftnlonoexp},
and keeping the physical $\mathcal{O}(\eps^0)$ term, we find:
\bal \label{resultsoftnloeps0}
\tS^{(1)} &= C \bigg[ \left(\zeta_2-L_{12}^2\right) \delta(\mathcal{T}_0) + 4 L_{12} \, \mathcal{L}_0(\mathcal{T}_0) -8 \, \mathcal{L}_1(\mathcal{T}_0) \bigg] \, .
\end{align}
For color-singlet production at the LHC the two initial-state partons are back-to-back so that $y_{12}=1$ ($L_{12} = 0$) and
the soft function simplifies to,
\bal
\tS^{(1)} = C \left[\zeta_2 \,\delta(\mathcal{T}_0) -8 \,\mathcal{L}_1(\mathcal{T}_0) \right] \,.
\end{align}
This result agrees with the literature (c.f.~Eq.~(173) of Ref.~\cite{Fleming:2007xt} and Eq.~(2.23) of
Ref.~\cite{Berger:2010xi}). The more general result of Eq.~\eqref{resultsoftnloeps0} can be used for $e p$ or $e^+ e^-$ processes with
respectively one or two final-state jets.\\